\documentclass[
  preprint,
  superscriptaddress,
  amsmath,amssymb,
  aps,
  prfluids
]{revtex4-2}

\usepackage{supertabular}
\usepackage{booktabs}
\usepackage{graphicx}
\usepackage{dcolumn}
\usepackage{bm}
\usepackage{siunitx}

\newcommand{\vect}[1]{\mathbf{#1}}
\newcommand{\pderiv}[2]{\frac{\partial #1}{\partial #2}}
\newcommand{\dd}{\mathrm{d}}

\begin{document}

\title{Diffusion–compaction coupling controls pore-pressure dynamics in granular–fluid flows}

\author{Eric C.P.~Breard}
\email{eric.breard@ed.ac.uk}
\affiliation{School of GeoSciences, The University of Edinburgh, Edinburgh, UK}
\affiliation{Department of Earth Sciences, University of Oregon, Eugene, OR, USA}

\author{Claudia Elijas~Parra}
\thanks{These authors contributed equally to this work.}
\affiliation{School of GeoSciences, The University of Edinburgh, Edinburgh, UK}

\author{Mattia de'~Michieli Vitturi}
\thanks{These authors contributed equally to this work.}
\affiliation{Istituto Nazionale di Geofisica e Vulcanologia (INGV), Sezione di Pisa, Pisa, Italy}

\begin{abstract}
Excess pore pressure in granular--fluid mixtures can transiently suppress frictional contacts and dramatically enhance flow mobility, yet its evolution is commonly modeled using constant effective diffusivities. Here we show that the apparent diffusivity is not intrinsic but emerges from the coupling between pore-pressure diffusion and granular compaction.

Starting from two-phase mass conservation for a deformable, gas-saturated granular assembly, we derive an evolution equation for excess pore pressure that captures deformation of the granular skeleton. In the thin-flow, small-excess-pressure limit, this reduces to a one-dimensional diffusion--compaction equation with a time-dependent source term controlled by porosity changes.

A modal analysis yields a reduced basal equation that separates diffusive drainage from compaction-driven forcing and identifies the corresponding timescales. This framework introduces a dimensionless source-to-diffusion ratio, $\Psi_0$, which governs the competition between these processes and collapses effective diffusivities obtained from high-resolution two-fluid simulations over nearly two orders of magnitude in bed height. This scaling implies that the apparent diffusivity, and thus flow mobility, is not intrinsic but depends on flow thickness through the competition between diffusion and compaction.

Incorporating this physics into a depth-averaged model demonstrates that the resulting closure reproduces the thickness dependence of pore-pressure decay and runout observed in experiments. These results provide a physically grounded description of pore-pressure evolution in granular--fluid flows and clarify how diffusion--compaction coupling controls their mobility.
\end{abstract}


\maketitle
% \linenumbers  % disabled for arXiv

\section{\label{sec:intro}Introduction}

In geophysical settings, gravity-driven granular flows such as volcanic pyroclastic density currents (PDCs) and debris flows can attain unexpectedly long runout distances because of excess pore-fluid pressure. These mixtures of solid particles and interstitial fluid (gas or water) may develop high pore-fluid pressures that temporarily reduce intergranular friction, thereby partially fluidizing the material and enhancing its mobility \citep{Iverson1997}. In debris flows, for example, sudden compaction of a water-saturated sediment mass can generate pore pressures exceeding hydrostatic equilibrium when permeability is low, liquefying the flow's interior \citep{Iverson1997}. Likewise, concentrated pyroclastic density currents have been observed to propagate as a basal grain-rich avalanche cushioned by pressurized gas, distinct from the dilute turbulent ash cloud above. The presence of excess pressurized pore fluid in such concentrated granular currents is now recognized as a key mechanism for reduced frictional resistance and extreme mobility \citep{Lube2020}.

A fundamental driver of this behaviour is the coupling between granular bulk volume change (dilatation or compaction) and pore-fluid pressure diffusion. As a granular flow deforms, particles rearrange and either create void space (dilatation) or collapse voids (compaction). Compaction tends to squeeze the pore fluid (liquid or gas), elevating local pore pressure if the fluid cannot escape quickly, whereas dilatation produces suction. This dilatancy–pore-pressure feedback, central to soil mechanics and debris-flow dynamics \citep{Iverson1997}, has since been demonstrated in controlled experiments and numerical models. Two-phase continuum simulations have shown that internal pore pressure is dynamically modulated by intermittent episodes of dilatation and compaction within a moving granular mass (known as pore pressure feedback, \citet{Iverson2005}), explaining transient basal pressure signals observed in small-to-large-scale experiments of granular media \citep{Breard2016, Lube2019, iverson2011positive}.

The competition between pore-pressure generation (from compaction) and dissipation (via Darcy-type diffusion) gives rise to characteristic timescales. If diffusion is slow relative to deformation, excess pressures are sustained and the material remains highly mobile; if diffusion is rapid, the flow quickly re-solidifies as pore pressure equilibrates. \citet{Iverson2014} formalized this interplay by introducing two key timescales---one for downslope advection and one for pore-pressure relaxation---and showed that their ratio, along with the material's initial dilative tendency, governs whether a surge remains fluidized or quickly transitions to a frictional regime. This framework generalizes earlier shallow-flow models by \citet{Iverson2001}, which treated pore-pressure evolution as an advection–diffusion process within a depth-averaged granular mixture.

Several material properties and initial conditions control the effectiveness of this diffusion–compaction coupling. Grain-size distribution strongly influence the pore-pressure diffusion rate: fine-grained, polydiperse size particle distributions generate low-permeability beds that dissipate pressure slowly, sustaining fluidization over longer timescales \citep{Montserrat2012, Roche_2010_pore, Roche2012, Breard2019b}. Laboratory studies of initially fluidized granular columns show that thicker beds retain elevated pore pressures far longer than thinner ones, consistent with a diffusive timescale $t_d \sim H^2 / D$. However, the inferred diffusivities still vary systematically with initial height, suggesting that $D$ is not a constant material property in these systems \citep{Montserrat2012, Roche_2012_depositional}. Laboratory studies of initially fluidized granular columns show that thicker beds retain elevated pore pressures far longer than thinner ones, consistent with a diffusive timescale $t_d \sim H^2 / D$. However, the inferred diffusivities still vary systematically with initial height, suggesting that $D$ is not a constant material property in these systems \citep{Montserrat2012}. 

Continuum modeling of dam-break configurations with incompressible solvers—where the solid volume fraction is assumed constant—exhibits a similar inconsistency: reproducing the observed runout distances requires prescribing an effective diffusivity that depends on the initial bed dimensions. In such models, high-aspect-ratio columns tend to lose pore pressure more rapidly during collapse, whereas shorter columns remain pressurized over much of their evolution, producing nonlinear dependencies between runout and initial column height \citep{Aravena_2021_influence}. 

Taken together, these observations indicate that a classical Darcy description with constant diffusivity is often inadequate. In static defluidization experiments, the apparent diffusion coefficient increases with bed height \citep{Montserrat2012, Roche_2012_depositional}, and theoretical estimates are frequently an order of magnitude larger than values obtained through inverse modeling.

A key limitation of most analyses is the implicit assumption of constant porosity, despite the fact that even small variations in porosity or permeability can strongly influence pore-pressure dissipation rates. When grain-size distributions evolve, for example through clast comminution, both porosity and permeability may change significantly, making compaction and mixture compressibility central to the dynamics. In pyroclastic density currents, where fragmentation continuously generates fines and broadens the particle-size distribution, these effects can promote fragmentation-induced fluidization and sustained pore-pressure support \citep{Breard2023}.

Furthermore, if pore pressure dissipates through gas diffusion as described by Darcy's law, the escaping gas necessarily reduces the gas volume fraction, implying that porosity cannot remain constant under such conditions.

Capturing these coupled processes in predictive models remains a major challenge. High-fidelity multiphase simulations solve separate conservation equations for the solid and fluid phases, allowing spatial and temporal variations in pressure and volume fraction to emerge naturally. Such models reproduce fine-scale behaviours, including basal underpressures and overpressures beneath a propagating flow front, and the modulation of basal friction by dilatancy \citep{Breard2019}. \citet{Aravena_2021_influence} used depth-resolved two-phase simulations of collapsing fluidized columns and showed that realistic dynamics could be obtained by adjusting an effective pore-pressure diffusion parameter consistent with static-bed measurements \citep{Montserrat2012}. These simulations confirm an early high-mobility phase followed by a deposition phase as pore pressures dissipate, but also demonstrate that mis-specifying diffusivity can lead to substantial run-out errors.

In parallel, reduced-order (depth-averaged) models have been developed to efficiently incorporate dilatancy effects. Building on the Coulomb mixture theory of \citet{Iverson1997, Iverson2001}, \citet{Iverson2014} introduced a depth-averaged framework that couples mass and momentum conservation with an evolving solid volume fraction and basal pore pressure, successfully reproducing the dynamics of debris-flows. Similarly, \citet{Bouchut2016} developed a two-layer shallow model incorporating dilatancy, while \citet{GarresDiaz2020} extended this approach to multilayer two-phase debris flows with a vertical pore-pressure structure. These reduced-order models are far more computationally efficient than full multiphase simulations but require careful closure relations and validation.

Given this background, the present study leverages both high-resolution two-phase simulations and depth-averaged modeling to investigate pore-pressure diffusion and compaction in granular flows. By comparing detailed simulations with 1D modeling and applying insights to simplified models, we clarify how diffusion–compaction coupling deviates from classical Darcy-type behaviour and explore the implications for geophysical hazard-scale modeling. Ultimately, this integration of rigorous simulation and reduced-order theory advances our understanding of how excess pore pressure can transiently transform a granular mass from a frictional pile into a fluidized, mobile flow, and under what conditions that transformation is most effective.

% Renamed duplicate label sec:intro to sec:twofluid
\section{\label{sec:twofluid}Two-Fluid modeling Methods}
\subsection{The solid stress description}
Following Srivastava and Sundaresan (as implemented in MFIX), the frictional contribution is activated when the gas volume fraction falls below a threshold and is otherwise set to zero. The frictional shear viscosity for solid phase $m$ is
\begin{equation}
\mu^{\mathrm{fric}}_{m} =
\begin{cases}
\dfrac{\sqrt{2}\,P_f \sin\varphi}{\sqrt{\Theta_m/d_m^2 + S_{mij}S_{mji}}}
\left[
\,\nu_{SS} - (\nu_{SS}-1)\left(\dfrac{P_f}{P_c}\right)^{\frac{1}{\,\nu_{SS}-1\,}}
\right]\!\left(\dfrac{\varepsilon_m}{\varepsilon_s}\right),
& \varepsilon_g < 1-\varepsilon_f^{\min}, \\[9pt]
0, & \varepsilon_g \ge 1-\varepsilon_f^{\min},
\end{cases}
\label{eq:mfix-mu-fric}
\end{equation}
where $\nu_{SS}$ is an exponent controlling the shape of the frictional yield surface and the nonlinear dependence of the stress on the ratio $P_f/P_c$, with distinct values for dilatation and compaction. $P_f$ is the frictional pressure of the solid phase (a bulk quantity shared across all solid phases), and $P_c$ is the critical-state pressure. $\varphi$ is the internal friction angle, $d_m$ the particle diameter, $\Theta_m$ the granular temperature of phase $m$, and $S_{mij}$ is the deviatoric strain-rate tensor defined as
\[
S_{mij} = \tfrac{1}{2}\left(\partial_i v_{mj} + \partial_j v_{mi}\right) - \tfrac{1}{3}(\nabla \cdot \mathbf{v}_m)\delta_{ij},
\]
with $S_{mij}S_{mji}$ its second invariant. $\varepsilon_m$ is the volume fraction of phase $m$, and $\varepsilon_s=\sum_{m=1}^{M}\varepsilon_m$ is the total solids volume fraction.

The exponent $N$ depends on whether the granular assembly is dilating or compacting:
\begin{equation}
\nu_{SS} =
\begin{cases}
\dfrac{\sqrt{3}}{2\sin\varphi}, & \nabla \cdot \mathbf{v}_m \ge 0 \quad \text{(dilatation)},\\[6pt]
1.03, & \nabla \cdot \mathbf{v}_m < 0 \quad \text{(compaction)}.
\end{cases}
\label{eq:mfix-N}
\end{equation}

The frictional pressure ratio is given by
\begin{equation}
\frac{P_f}{P_c}
=
\left(
1 - \frac{\nabla \cdot \mathbf{v}_m}{N \sqrt{2}\sin\varphi \,\sqrt{S_{mij}S_{mji} + \Theta_m/d_m^2}}
\right)^{\,N-1}
\left(\frac{\varepsilon_m}{\varepsilon_s}\right),
\label{eq:mfix-PfPc}
\end{equation}
where $\nabla \cdot \mathbf{v}_m = D_{mkk}$ is the volumetric strain rate.

The critical-state pressure $P_c = P_c(\varepsilon_g)$ is defined as
\begin{equation}
P_c =
\begin{cases}
0, & \varepsilon_g \ge 1-\varepsilon_f^{\min},\\[4pt]
F_r\;
\left(\dfrac{(1-\varepsilon_g)-\varepsilon_f^{\min}}{\varepsilon_g-\varepsilon^\ast}\right)^{\!r}
\,(\varepsilon_g-\varepsilon^\ast)^{-s}\;
\mathcal{L}(\varepsilon_g),
& \varepsilon^\ast+\delta \le \varepsilon_g < 1-\varepsilon_f^{\min},\\[8pt]
\mathcal{L}(\varepsilon_g),
& \varepsilon_g < \varepsilon^\ast+\delta,
\end{cases}
\label{eq:mfix-Pc}
\end{equation}
where $\varepsilon^\ast$ is the gas volume fraction at maximum packing, $\delta$ is a small regularization parameter, $F_r$, $r$, and $s$ are empirical constants, and $\mathcal{L}(\varepsilon_g)$ is a smoothing (limiter) function ensuring numerical regularity as $\varepsilon_g \to \varepsilon^\ast$. This formulation implicitly assumes that compressibility arises from changes in packing rather than grain deformation.

The linearization term is
\begin{equation}
L_{\varepsilon_g+\delta}(\varepsilon_g)
=
\mathrm{Fr}\left[
\frac{(1-\varepsilon^\ast-\delta-\varepsilon_f^{\min})^{r}}{\delta^{s}}
+\frac{r\,[1-\varepsilon^\ast-\delta-\varepsilon_f^{\min}]^{\,r-1}}{\delta^{s}}
+\frac{s\,[1-\varepsilon^\ast-\delta-\varepsilon_f^{\min}]^{\,r}}{\delta^{\,s-1}}
\right](\varepsilon^\ast+\delta-\varepsilon_g).
\label{eq:mfix-L}
\end{equation}

The frictional solids pressure $P_m^{\mathrm{fric}}$ contributes to the total solids-phase pressure; the frictional shear viscosity $\mu_m^{\mathrm{fric}}$ adds to the total solids viscosity, and the frictional second viscosity $\lambda_m^{\mathrm{fric}}$ contributes analogously to the second viscosity (as per the MFIX-TFM formulation).

\begin{table}[h]
\centering
\renewcommand{\arraystretch}{1.1}
\caption{\label{tab:mfix-constants}Constants used with the Srivastava friction model in MFIX.}
\begin{tabular}{lcl}
\hline
Parameter & Value & Notes \\
\hline
Restitution coefficient, $e$ & 0.9 & measured \\
Internal friction angle, $\varphi$ & $26^{\circ}$ & measured \\
Minimum solids fraction, $\varepsilon_f^{\min}$ & 0.5 & Srivastava and Sundaresan \\
$\mathrm{Fr}$ & 0.5 & Srivastava and Sundaresan (pressure scale) \\
$r$ & 2.0 & Srivastava and Sundaresan \\
$s$ & 5.0 & Srivastava and Sundaresan \\
$\delta$ & 0.01 & Srivastava and Sundaresan \\
$\varepsilon^\ast$ & 0.38 & close-packing \\
\hline
\end{tabular}
\end{table}

\subsection{\label{sec:gidaspow}Gidaspow Drag Model}

The Gidaspow drag model combines the Ergun correlation for dense regimes and the Wen and Yu correlation for dilute regimes, providing a continuous description of gas--solid momentum exchange across a wide range of void fractions.
In this model, the interphase momentum exchange coefficient $\beta'_{gm}$ depends on the local gas volume fraction $\varepsilon_g$ (or equivalently, the solids concentration $\phi_s=1-\varepsilon_g$) and is defined piecewise as:
\begin{equation}
\beta'_{gm} =
\begin{cases}
\beta'^{\,\mathrm{WenYu}}_{gm}, & \varepsilon_g \ge 0.8, \\[6pt]
\beta'^{\,\mathrm{Ergun}}_{gm}, & \varepsilon_g < 0.8,
\end{cases}
\label{eq:gidaspow_piecewise}
\end{equation}
where the Ergun correlation applies to dense (packed-bed) conditions and the Wen--Yu correlation to more dilute suspensions.

For the dense regime ($\varepsilon_g < 0.8$), the Ergun form is used:
\begin{equation}
\beta'^{\,\mathrm{Ergun}}_{gm}
=
\frac{150\,(1-\varepsilon_g)^2\,\mu_g}{\varepsilon_g\,d_m^2}
\;+\;
1.75\,\rho_g\,(1-\varepsilon_g)\,
\frac{|\mathbf{U}_g - \mathbf{U}_m|}{d_m},
\label{eq:gidaspow_ergun}
\end{equation}
where $\mu_g$ and $\rho_g$ are the gas viscosity and density, $d_m$ is the particle diameter of phase $m$, and $U_{gi}-U_{mi}$ is the local relative velocity between gas and solid phases. The first term represents viscous drag, and the second inertial drag due to form resistance.

For the dilute regime ($\varepsilon_g \ge 0.8$), the Wen--Yu correlation is applied:
\begin{equation}
\beta'^{\,\mathrm{WenYu}}_{gm}
=
\frac{3}{4}\,C_D\,\frac{\rho_g\,\varepsilon_g\,(1-\varepsilon_g)}{d_m}
\,|\mathbf{U}_g-\mathbf{U}_m|,
\label{eq:gidaspow_wenyu}
\end{equation}
where the drag coefficient $C_D$ is a function of the particle Reynolds number,
\begin{equation}
\mathrm{Re}_p = \frac{\rho_g\,\varepsilon_g\,|\mathbf{U}_g-\mathbf{U}_m|\,d_m}{\mu_g},
\qquad
C_D =
\begin{cases}
\dfrac{24}{\mathrm{Re}_p}\left(1 + 0.15\,\mathrm{Re}_p^{0.687}\right), & \mathrm{Re}_p \le 1000,\\[6pt]
0.44, & \mathrm{Re}_p > 1000.
\end{cases}
\label{eq:Cd_WenYu}
\end{equation}

The transition at $\varepsilon_g=0.8$ defines a piecewise switch between the packed-bed and fluidized-flow regimes.
This model reproduces the Ergun limit for dense beds and approaches the isolated-particle drag behaviour embedded in the Wen--Yu correlation as $\varepsilon_g\to1$.
It is widely used in MFIX-TFM and other two-fluid solvers because of its robustness and its ability to handle a broad range of granular concentrations in gas--particle flows.

We employ the Eulerian two-fluid model (TFM) as implemented in MFIX, in which the gas and particulate phases are treated as interpenetrating continua coupled through closure relations for interphase momentum exchange and granular stresses. In particular, interphase drag is described using the Gidaspow model, which combines the Ergun and Wen--Yu correlations across dense and dilute regimes.

\begin{figure}[ht]
    \centering
    \includegraphics[width=\linewidth]{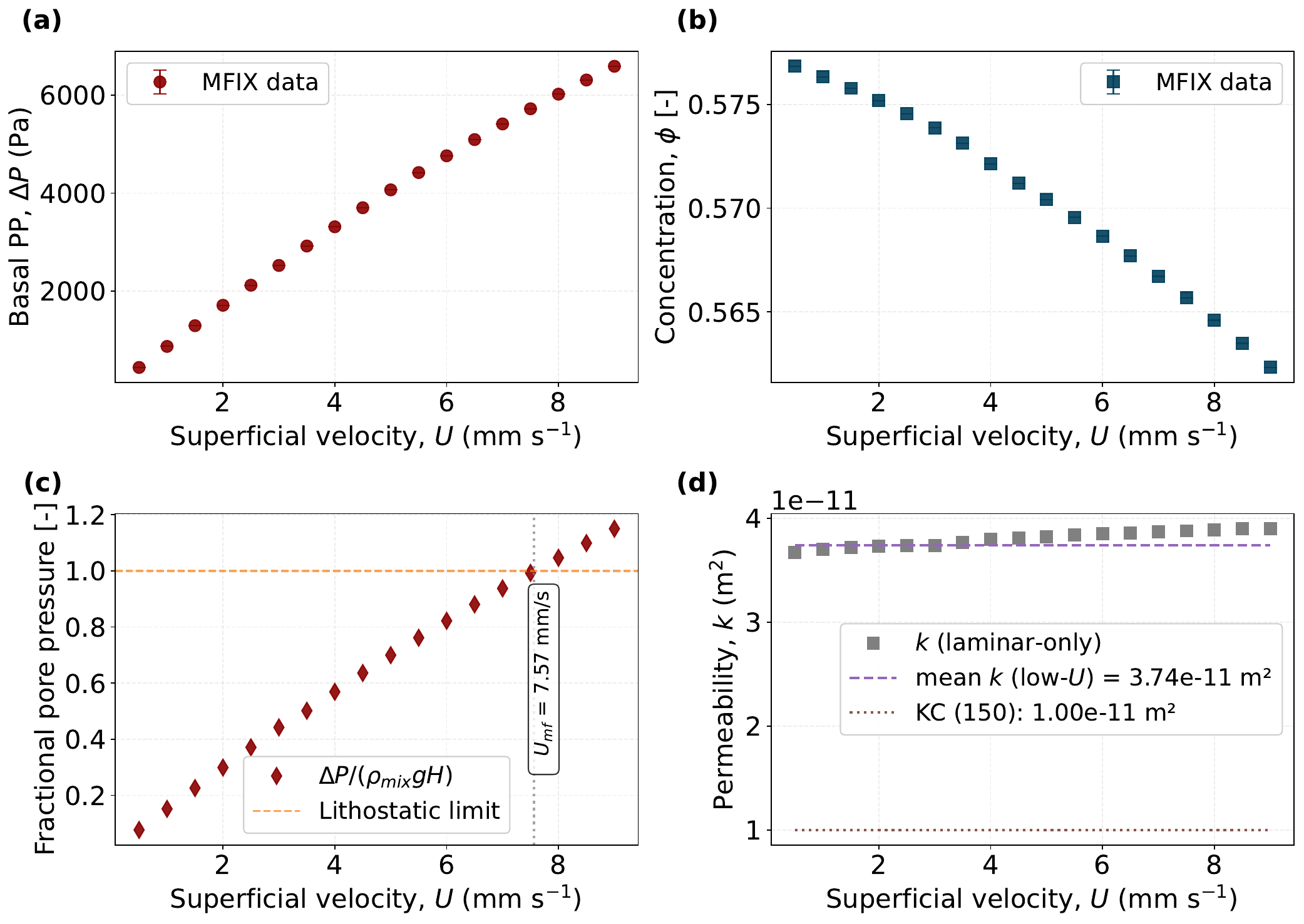}
    \caption{%
    One-dimensional MFIX-TFM simulations of a \SI{0.4}{m} thick bed of glass beads ($d_{\mathrm{eq}} = \SI{75}{\micro\meter}$) in air for various superficial velocities.
    (a) Basal pore pressure as a function of time.
    (b) Depth-averaged concentration of the bed.
    (c) Fractional excess pore pressure.
    (d) Permeability estimates using the Ergun equation. Each data point represents a single simulation at steady state.
    }
    \label{fig:permeability}
\end{figure}

We test the MFIX--TFM framework against the experimental results of Roche et al.\ (2010) and find that it reproduces the measured fluidization behaviour with good fidelity: the predicted minimum fluidization velocity $U_{\mathrm{mf}}$ is only about 8\% lower than in the experiments (Figure~\ref{fig:permeability}). The simulations also capture the slight but measurable dilatation of the mixture as the superficial velocity increases, corresponding to a reduction in solids packing of roughly 2--4\%.

Because this study focuses on pore-pressure evolution in the regime $U \le U_{\mathrm{mf}}$, we restrict the simulations to one dimension, thereby excluding bubble formation and the associated plateau in excess pore pressure observed at higher velocities.

The granular material used in both the experiments and simulations consists of spherical glass beads with equivalent diameter $d_{\mathrm{eq}} = \SI{75}{\micro\metre}$ and density $\rho_s = \SI{2500}{\kilogram\per\cubic\metre}$. Under ambient air conditions at $20~^\circ\mathrm{C}$ ($\rho_f = \SI{1.2}{\kilogram\per\cubic\metre}$, $\mu_g = \SI{1.8e-5}{\pascal\second}$), the corresponding Archimedes number,
\begin{equation}
\mathrm{Ar} = \frac{g\,d_{\mathrm{eq}}^{3}\,\rho_f\,(\rho_s - \rho_f)}{\mu_g^{2}},
\label{eq:archimedes}
\end{equation}
is $\mathrm{Ar} \simeq 3.8\times10^{1}$. This value places the system within the Geldart Group~A regime ($\mathrm{Ar}\lesssim 20{-}100$), which is characterized by smooth, homogeneous expansion of the bed prior to bubbling. Such beds fluidize uniformly and compact continuously when gas flow ceases, making them particularly well suited for studying pore-pressure diffusion without interference from bubble dynamics.

As shown in Figure~\ref{fig:permeability}(d), the linear (viscous) term of the Ergun equation predicts a permeability nearly three times higher than the Carman--Kozeny estimate. In the laminar, pre-bubbling regime applicable here, the Ergun pressure--velocity relationship reduces to
\begin{equation}
\frac{\Delta P}{L}
\simeq
150\,\frac{(1-\varepsilon)^2}{\varepsilon^3}\,
\frac{\mu_g\,U}{d_{\mathrm{eq}}^2},
\label{eq:ergun_linear}
\end{equation}
where $\Delta P$ is the pressure drop across a bed of length $L$, $\varepsilon$ the porosity, $U$ the superficial gas velocity, $\mu_g$ the gas viscosity, and $d_{\mathrm{eq}}$ is an equivalent particle diameter representing the surface-area-controlled length scale of the granular assembly (e.g.\ the Sauter mean diameter $d_{32}$ for polydisperse systems).

This highlights that the effective permeability is controlled by the evolving surface-area-to-volume ratio of the mixture rather than a single characteristic grain size, and therefore cannot, in general, be treated as constant in polydisperse or fragmenting granular flows.

The quadratic inertial term of the full Ergun equation,
\[
1.75\,\frac{(1-\varepsilon)}{\varepsilon^3}\,
\frac{\rho_f\,U^2}{d_{\mathrm{eq}}},
\]
is negligible under these low-Reynolds-number conditions ($\mathrm{Re}_p \approx 0.04$ at $U_{\mathrm{mf}}$) for $U \le U_{\mathrm{mf}}$. In this viscous limit, the pressure gradient follows Darcy’s law, $\Delta P / L = \mu_g U / k$, yielding the equivalent permeability
\begin{equation}
k_{\mathrm{Ergun}} = \frac{\varepsilon^3 d_{\mathrm{eq}}^2}{150(1-\varepsilon)^2}.
\label{eq:ergun_perm}
\end{equation}
This is the expression used to compute the theoretical permeability values plotted in Figure~\ref{fig:permeability}(d).

These simulations confirm that the bed operates in a strictly laminar, Darcy-type regime, with negligible inertial contributions and porosities well within the Geldart~A range. Under such conditions, the pressure field is governed primarily by the coupled diffusion and mechanical compaction of the gas--particle mixture. To capture this interaction quantitatively, we now derive the governing equations describing the evolution of pore pressure in a deformable, compressible granular medium.

\section{\label{sec:governing}Governing Equations of the Compressible Granular Medium}

The behaviour observed in both experiments and simulations---diffusive pore-pressure decay coupled with small but finite changes in porosity---indicates that the gas and solid phases remain dynamically coupled even in the laminar, pre-bubbling regime. To describe this interaction quantitatively, we consider the general two-phase formulation governing a deformable, fluid-saturated granular medium.

We denote by $\varepsilon$ the porosity (gas volume fraction), $\phi = 1-\varepsilon$ the
solids concentration, $\rho_s$ and $\rho_f$ the mass densities of the solid grains and
interstitial fluid, and $\mathbf{u}_s$ and $\mathbf{u}_f$ their respective velocities.
Following the work of \citet{Goren2010}, the evolution of each phase is described by
local conservation of mass:
\begin{equation}
\frac{\partial(\phi\,\rho_s)}{\partial t}
+ \nabla \cdot \bigl(\phi\,\rho_s\,\mathbf{u}_s\bigr) = 0,
\label{eq:mass_s}
\end{equation}
\begin{equation}
\frac{\partial(\varepsilon\,\rho_f)}{\partial t}
+ \nabla \cdot \bigl(\varepsilon\,\rho_f\,\mathbf{u}_f\bigr) = 0,
\label{eq:mass_f}
\end{equation}
where the relative velocity between phases corresponds to the Darcy flux per unit
porosity,
\begin{equation}
\varepsilon\,(\mathbf{u}_f - \mathbf{u}_s)
= -\,\frac{k}{\mu_g}\,\nabla P,
\label{eq:darcy}
\end{equation}
with $k$ the intrinsic permeability and $P$ the excess (over
hydrostatic) pore pressure.

The gas density is related to pressure through the ideal-gas equation of state,
\begin{equation}
\rho_f = \frac{P_{\mathrm{abs}}}{R\,T},
\label{eq:rho_f}
\end{equation}
where $P_{\mathrm{abs}}$ is the absolute pressure, $T$ the gas temperature, and $R$ the specific gas constant.
The gas compressibility is defined as $\beta = (1/\rho_f)(\partial \rho_f / \partial P_{\mathrm{abs}})_{T}$, which for an isothermal ideal gas gives $\beta = 1/P_{\mathrm{abs}}$.

\medskip
Assuming (i) negligible grain compressibility relative to the fluid, (ii) constant $\rho_s$, and (iii) negligible fluid inertia, combination of equations~\eqref{eq:mass_s}--\eqref{eq:rho_f} yields the governing equation for the evolution of the excess pore pressure $P$ (cf.\ Eq.~(5) of \citet{Goren2010}):
\begin{equation}
\beta \,\varepsilon \,\frac{\partial P}{\partial t}
= \nabla \cdot 
\left[(1+\beta P)\frac{k}{\mu_g}\nabla P\right]
- (1+\beta P)\,\nabla \cdot \mathbf{u}_s
- \beta \,\varepsilon \,\mathbf{u}_s \cdot \nabla P.
\label{eq:governing_general}
\end{equation}
The first term on the right-hand side represents pore-pressure diffusion governed by Darcy flow, while the second and third terms describe mechanical compaction/dilatation (through $\nabla\!\cdot \mathbf{u}_s$) and the gradient part of the Lagrangian derivative (advection of pressure by the solid skeleton), respectively.

Following the non-dimensional analysis in \citet{Goren2010}, the grain Deborah number is defined as the ratio of the pore-pressure diffusion timescale across a single grain to the deformation timescale:
\begin{equation}
\mathrm{De}_d = \frac{t_d}{t_0} = \frac{d^2/D}{t_0},
\end{equation}
where $d$ is the grain diameter, $D = k/(\mu_g\beta\varepsilon)$ is the pressure diffusivity, and $t_0$ is the characteristic deformation time. For our system with $d = 75~\mu$m glass beads in air at ambient conditions, taking $k \approx 10^{-11}$~m$^2$ (Carman--Kozeny), $\mu_g = 1.8 \times 10^{-5}$~Pa$\cdot$s, $\beta = 10^{-5}$~Pa$^{-1}$, and $\varepsilon = 0.4$, we obtain $D \approx 1.4 \times 10^{-2}$~m$^2$/s. For typical compaction rates with $t_0 \sim 1$~s, the grain diffusion time is $t_d = d^2/D \approx 4 \times 10^{-7}$~s, yielding $\mathrm{De}_d \approx 4 \times 10^{-7} \ll 1$. In this limit, pore pressure equilibrates across individual grains essentially instantaneously compared to the rate of skeleton deformation, and the advective term $\beta\varepsilon\mathbf{u}_s\!\cdot\nabla P$ in Eq.~\eqref{eq:governing_general} is negligible. We therefore obtain
\begin{equation}
\beta \,\varepsilon \,\frac{\partial P}{\partial t}
= \nabla \cdot 
\left[(1+\beta P)\frac{k}{\mu_g}\nabla P\right]
- (1+\beta P)\,\nabla \cdot \mathbf{u}_s .
\label{eq:R10}
\end{equation}

It is often convenient to express the forcing via \emph{solids} concentration rather than $\nabla\!\cdot\mathbf{u}_s$. From solid mass conservation with constant $\rho_s$,
\begin{equation}
\phi\,\nabla\!\cdot\mathbf{u}_s \;=\; -\frac{\partial \phi}{\partial t} - \mathbf{u}_s\!\cdot\nabla \phi .
\label{eq:div_us_phi}
\end{equation}
Substituting \eqref{eq:div_us_phi} into \eqref{eq:R10} and recalling $\phi=1-\varepsilon$ yields a form that unifies ``thick'' (3-D) and ``thin'' closures:
\begin{equation}
\beta\,\varepsilon\,\frac{\partial P}{\partial t}
=
\nabla \cdot \!\left[(1+\beta P)\frac{k}{\mu_g}\nabla P\right]
+
\frac{(1+\beta P)}{\phi}\left(\frac{\partial \phi}{\partial t}+\mathbf{u}_s\!\cdot\nabla \phi\right).
\label{eq:full_general}
\end{equation}

\subsection*{The case of thin-flow (small-excess-pressure) approximation}

When applied to realistic configurations, Eq.~\eqref{eq:full_general} captures the full coupling between pore-pressure diffusion and time-dependent compaction or dilatation of the granular matrix. In many practical situations, however, the flow thickness is small compared to its lateral extent and the excess pressures remain weak, so that spatial gradients of porosity and nonlinear gas-compressibility effects can be neglected. Under these assumptions, the equation simplifies considerably and can be expressed in one dimension as a diffusion equation with a time-dependent source term that accounts for bed compaction.

While the full Eq.~\eqref{eq:full_general} remains valid for any three-dimensional configuration, one can reduce its complexity under the ``thin-flow approximation.'' For small excess pressures ($\beta P \ll 1$) and spatially uniform porosity $\varepsilon(t)$, such that advection of $\varepsilon$ is negligible, Eq.~\eqref{eq:full_general} reduces in one dimension (vertical $z$) to
\begin{equation}
\frac{\partial P}{\partial t}
= D\,\frac{\partial^2 P}{\partial z^2} \;-\; S(t),
\qquad
D = \frac{k}{\mu_g\,\beta\,\varepsilon},
\qquad
S(t) = \frac{1}{\beta\,\varepsilon\,(1-\varepsilon)}\,\frac{\partial \varepsilon}{\partial t}.
\label{eq:pde}
\end{equation}
Here $\partial \varepsilon/\partial t$ is purely temporal because the porosity is assumed spatially uniform, i.e.\ $\nabla \varepsilon = 0$.

When the skeleton compacts ($\mathrm{d}\varepsilon/\mathrm{d}t<0$), $-S(t)>0$ and the source term raises $P$; for dilatation ($\mathrm{d}\varepsilon/\mathrm{d}t>0$), $-S(t)<0$ and $P$ decreases.

This reduced form provides a direct framework for analysing how temporal changes in porosity influence the evolution of pore pressure, while retaining the essential competition between diffusion and mechanical compaction.

\section{\label{sec:1dsolver}1D numerical solver}

To quantify the interplay between pore-pressure diffusion and compaction, we develop a one-dimensional numerical solver that integrates the simplified diffusion--compaction equation~\eqref{eq:pde} for a porous bed with spatially uniform but time-varying porosity~$\varepsilon(t)$. This formulation enables direct comparison with analytical solutions and with porosity histories extracted from two-fluid simulations.

Starting from the general diffusion--compaction expression, the governing equation reads:
\begin{equation}
\frac{\partial P}{\partial t}
=
\frac{1}{\mu_g\,\beta\,\varepsilon}\,
\nabla\!\cdot\!\big[(1+\beta P)\,k\,\nabla P\big]
\;-\;
\frac{(1+\beta P)}{\beta\,\varepsilon\,(1-\varepsilon)}\,\frac{\partial \varepsilon}{\partial t}.
\label{eq:vector_depthavg}
\end{equation}

Expanding Eq.~\eqref{eq:vector_depthavg} in one spatial dimension ($z$ vertical) gives:
\begin{equation}
\boxed{%
\frac{\partial P}{\partial t} =
\frac{k}{\mu_g \beta \varepsilon}\!\left[(1+\beta P)\frac{\partial^2 P}{\partial z^2}
+ \beta \!\left(\frac{\partial P}{\partial z}\right)^{\!2}\right]
- \frac{(1+\beta P)}{\beta \varepsilon (1-\varepsilon)}\frac{\partial \varepsilon}{\partial t}}
\label{eq:governing}
\end{equation}
where $k=k(\varepsilon)$ is the intrinsic permeability, spatially uniform but evolving with $\varepsilon(t)$ according to the Carman--Kozeny relation:
\begin{equation}
k(\varepsilon) = \frac{\varepsilon^3\,d_{\mathrm{eq}}^2}{150(1-\varepsilon)^2},
\label{eq:carman_kozeny}
\end{equation}
with $d_{\mathrm{eq}}=d_{32}\psi$ the equivalent particle diameter defined from the Sauter mean size $d_{32}$ and sphericity~$\psi$ (Breard et al. 2019).

\medskip
Equation~\eqref{eq:governing} expresses the nonlinear balance between pressure diffusion and the volumetric source term linked to the bed’s compaction rate~$\partial_t\varepsilon$.

\subsection*{Small-excess-pressure limit ($\beta P \ll 1$)}

For small excess pressures, $\beta P \ll 1$, Eq.~\eqref{eq:governing} reduces to a linear, non-homogeneous diffusion equation:
\begin{equation}
\frac{\partial P}{\partial t} =
\frac{k}{\mu_g \beta \varepsilon}\frac{\partial^2 P}{\partial z^2}
- \frac{1}{\beta \varepsilon (1-\varepsilon)}\frac{\partial \varepsilon}{\partial t}.
\label{eq:linear}
\end{equation}
This reduced form provides a convenient framework for analysing pore-pressure decay in experiments where the gas remains nearly isothermal and the compaction amplitude is small.

\subsection*{Boundary and initial conditions}

We consider a vertical column of height $H(t)$, which may evolve due to compaction or dilatation, with mixed Neumann--Dirichlet boundary conditions:
\begin{align}
\left.\frac{\partial P}{\partial z}\right|_{z=0} &= 0,
&
P(H,t) &= 0,
&
P(z,0) &= P_{b0}\!\left(1-\frac{z}{H_0}\right).
\end{align}
The no-flux condition at $z=0$ represents an impermeable base, while $P(H,t)=0$ enforces atmospheric pressure at the top surface.

\subsection{\label{sec:numerical}Numerical implementation}

To solve Eq.~\eqref{eq:governing}, we employ a finite-difference scheme that updates the pressure field in time for a bed with spatially uniform porosity~$\varepsilon(t)$, evolving solely as a function of time. The permeability $k(\varepsilon)$ follows the Carman--Kozeny relation~\eqref{eq:carman_kozeny}, with $d_{\mathrm{eq}} = 75~\text{microns}$ for the glass beads used in the reference experiments.

\subsubsection*{Spatial discretization and moving boundary}

The solver operates on a fixed Eulerian grid with uniform vertical spacing $\Delta z = \SI{5}{mm}$. The computational domain extends from $z=0$ (base) to $z_{\mathrm{max}}$, chosen to exceed the maximum expected bed height. As the bed compacts or dilates, its height $H(t)$ evolves according to solids mass conservation:
\begin{equation}
H(t) = H_0\,\frac{\phi_0}{\phi(t)},
\label{eq:mass_conservation}
\end{equation}
where $H_0$ and $\phi_0$ are the initial bed height and solids concentration, and $\phi = 1-\varepsilon$.

At each timestep $t_j$, only the grid points within the active domain $[0, H(t_j)]$ are updated. The number of active nodes is
\begin{equation}
N_{\mathrm{active}}(t_j) =
\left\lceil \frac{H(t_j) + \Delta z}{\Delta z} \right\rceil,
\end{equation}
and nodes above $z = H(t_j)$ are masked as undefined. This fixed-grid approach efficiently represents the moving free surface without remeshing, preserving numerical stability while accurately tracking compaction-driven height changes.

\subsubsection{\label{sec:temporal}Temporal discretization and stability}

We use an explicit forward Euler scheme for time integration. At interior nodes $i \in [1, N_{\mathrm{active}}-2]$, the pressure is advanced as:
\begin{equation}
P_{i}^{n+1} = P_{i}^{n} + \Delta t \left[ D_{\mathrm{loc}} \left( \delta^2 P_i^n + \beta_{\mathrm{loc}} \left( P_i^n \delta^2 P_i^n + (\delta P_i^n)^2 \right) \right) + S_i^n \right]
\label{eq:explicit_update}
\end{equation}
where:
\begin{itemize}
\item $D_{\mathrm{loc}} = k(t_n)/(\mu_g \beta_{\mathrm{loc}} \varepsilon(t_n))$ is the local pressure diffusivity
\item $\beta_{\mathrm{loc}} = 1/(P_{\mathrm{amb}} + P_i^n)$ accounts for variable gas compressibility (ideal gas)
\item $\delta^2 P_i^n = (P_{i-1}^n - 2P_i^n + P_{i+1}^n)/(\Delta z)^2$ is the second-order centered difference
\item $\delta P_i^n = (P_{i+1}^n - P_{i-1}^n)/(2\Delta z)$ is the first-order centered difference
\item $S_i^n = -\frac{\partial \varepsilon}{\partial t}\Big|_{t_n} \cdot \frac{1+\beta_{\mathrm{loc}} P_i^n}{\beta_{\mathrm{loc}} \varepsilon(t_n)(1-\varepsilon(t_n))}$ is the compaction/dilatation source term
\end{itemize}

The use of locally varying $\beta_{\mathrm{loc}}$ based on absolute pressure $P_{\mathrm{abs}} = P_{\mathrm{amb}} + P$ corrects for the spatial variation in gas compressibility. For a 1~m thick bed with bulk density $\rho_{\mathrm{bulk}} \approx 1500~\mathrm{kg\,m^{-3}}$, the hydrostatic excess pressure at the base is approximately $\Delta P/P_{\mathrm{amb}} \approx 14.6\%$. Using a constant $\beta$ would introduce a comparable error in the local diffusivity and nonlinear factors.

To ensure numerical stability, the timestep is constrained by the diffusive CFL condition:
\begin{equation}
\Delta t = 0.5 \frac{(\Delta z)^2}{D_0}
\label{eq:timestep_stability}
\end{equation}
where $D_0 = k(\varepsilon_0)/(\mu_g \beta_0 \varepsilon_0)$ is the initial diffusivity and $\beta_0 = 1/P_{\mathrm{amb}}$. This choice corresponds to a safety factor of $0.4$ times the classical von Neumann stability limit $(\Delta z)^2/(2D)$ for the linear diffusion equation. For typical initial conditions ($\varepsilon_0 = 0.6$, $d_{\mathrm{eq}} = 75~\mu$m, $\Delta z = 5$~mm), this yields $\Delta t \approx 53~\mu$s, requiring $\mathcal{O}(10^4)$ to $\mathcal{O}(10^5)$ timesteps for simulations spanning several seconds. The solver is accelerated using Numba just-in-time compilation to maintain reasonable computational costs.

\subsubsection{\label{sec:bc}Boundary conditions}

At the base ($z=0$), a no-flux (Neumann) condition is imposed:
\begin{equation}
\left.\frac{\partial P}{\partial z}\right|_{z=0} = 0
\quad \Rightarrow \quad
P_0^{n+1} = P_1^{n+1}
\label{eq:bc_base}
\end{equation}
This represents an impermeable lower boundary, appropriate for gas-fluidized beds on a distributor plate or granular flows on solid substrates.

At the top ($z=H(t)$), atmospheric pressure is enforced (Dirichlet condition):
\begin{equation}
P(H(t), t) = 0
\quad \Rightarrow \quad
P_{N_{\mathrm{active}}-1}^{n+1} = 0
\label{eq:bc_top}
\end{equation}
As the bed height changes, the index $N_{\mathrm{active}}-1$ tracks the uppermost grid point within the bed, effectively implementing a moving Dirichlet boundary on the fixed Eulerian mesh. This approach naturally handles both compaction (decreasing $H$) and dilatation (increasing $H$) without requiring interpolation or remapping of the pressure field.

\subsubsection{\label{sec:input}Input data and source term}

The solver accepts time-dependent porosity $\varepsilon(t)$ and its time derivative $\partial\varepsilon/\partial t$ as input, typically extracted from depth-averaged concentration profiles in the CFD simulations. When $\varepsilon(t)$ is provided as discrete data points, $\partial\varepsilon/\partial t$ is computed using a smoothed numerical derivative (Savitzky-Golay filter with 9-point window and second-order polynomial) to reduce sensitivity to noise. The compaction source term in equation~\eqref{eq:explicit_update} is then evaluated pointwise at each timestep.

This numerical solver provides a flexible tool for studying pressure dynamics under realistic, time-varying porosity profiles while retaining the ability to compare with analytical Fourier series solutions when $\varepsilon(t)$ is prescribed analytically (e.g., linear or exponential compaction laws).

\subsection*{Validation against the pure-diffusion limit}

Before analysing cases with time-dependent porosity, we first disable the compaction source term in Eq.~\eqref{eq:governing} (i.e.\ $\partial \varepsilon / \partial t = 0$) so that the problem reduces to classical one-dimensional pore-pressure diffusion in a fixed bed. In the MFIX-TFM simulations, this configuration is enforced by suppressing compaction of the granular skeleton, such that the solids volume fraction remains constant and the solids velocity field is identically zero.

In this setting, the basal pressure is expected to decay exponentially with a rate controlled solely by the diffusivity $D = k/(\mu_g \beta \varepsilon)$ and the bed height through the fundamental mode $alpha_m = \pi^{2}D/(4H^{2})$.

\begin{figure}[ht]
    \centering
    \includegraphics[width=0.6\textwidth]{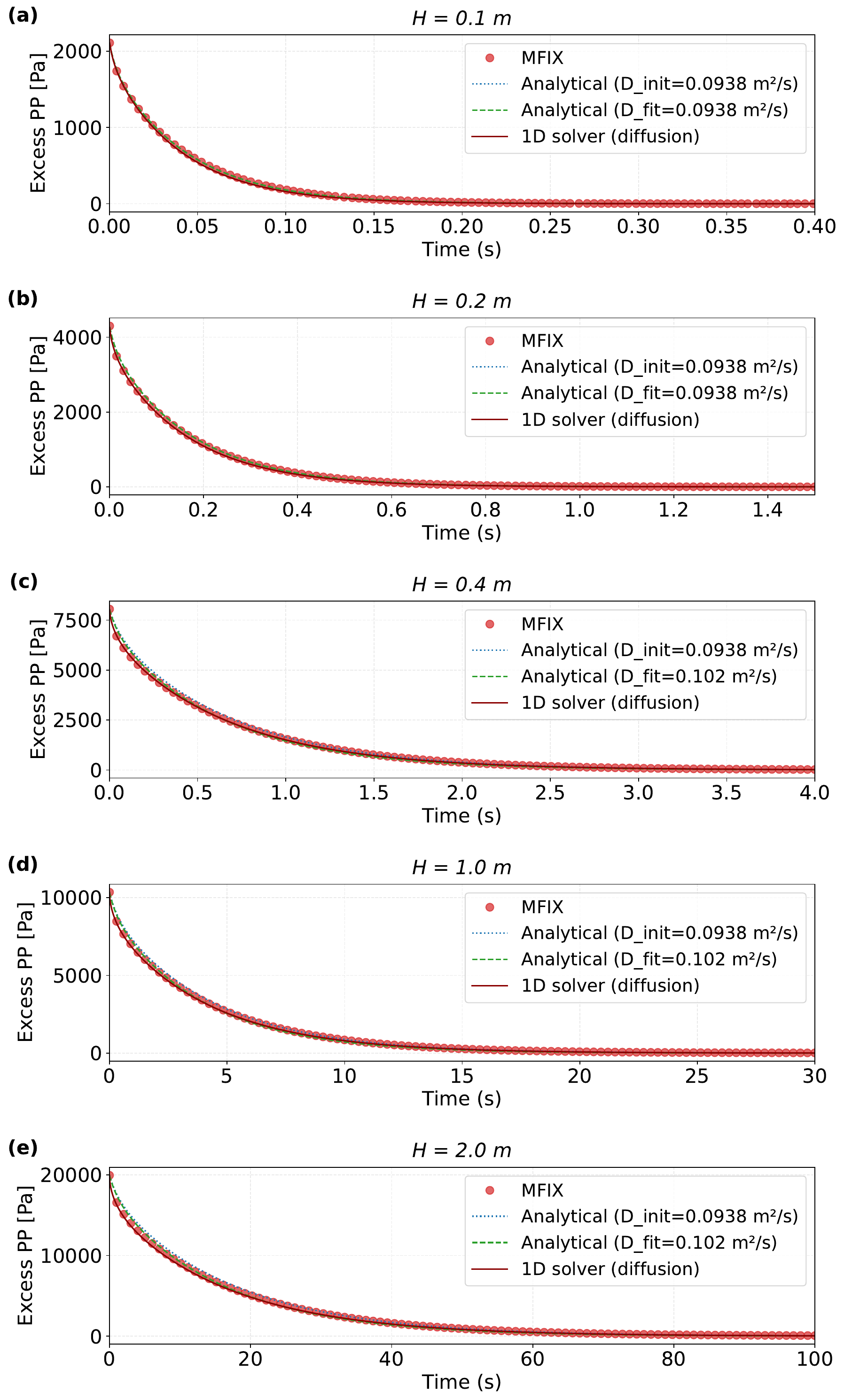}
    \caption{Comparison between the 1-D diffusion-only solver (no dilatancy, $\partial \varepsilon/\partial t = 0$) and MFIX-TFM simulations for the same bed height and permeability. Basal excess pore pressure decays exponentially with the fundamental mode predicted from $D = k/(\mu_g \beta \varepsilon)$, demonstrating that the 1-D model reproduces the pure-diffusion limit.}
    \label{fig:combined_excess_solver}
\end{figure}

Figure~\ref{fig:combined_excess_solver} compares the 1-D finite-difference solution with MFIX-TFM simulations for this no-dilatancy configuration. The agreement is excellent over the full drainage window, confirming that (i) the numerical scheme correctly captures the mixed Neumann--Dirichlet boundary conditions and (ii) deviations observed in the full problem arise from the compaction/dilatation term rather than from numerical artefacts.

\subsection*{Validation against MFIX-TFM porosity histories}

To assess how sensitive the 1-D diffusion–compaction solver is to the prescribed porosity history, we drove it with several versions of the time-dependent bed porosity extracted from the MFIX–TFM runs. The raw depth-averaged porosity from CFD contains small temporal oscillations that are dynamically irrelevant but that appear in the source term
\(
S(t) = [\beta\,\varepsilon(1-\varepsilon)]^{-1}\,\mathrm{d}\varepsilon/\mathrm{d}t
\)
as high-frequency noise. We therefore fitted the MFIX-TFM depth-averaged porosity with a simple exponential relaxation law
\[
\varepsilon(t) \approx \varepsilon_\infty + \bigl(\varepsilon_0 - \varepsilon_\infty\bigr)\,\exp(-t/\tau),
\]
and used both the fitted \(\varepsilon(t)\) and its analytic time derivative \(\mathrm{d}\varepsilon/\mathrm{d}t\) to force the 1-D solver. This removes numerical jitter from the source term while keeping the correct compaction amplitude and timescale.

\begin{figure}[htbp]
    \centering
    \includegraphics[width=0.9\linewidth, height=0.80\textheight, keepaspectratio]{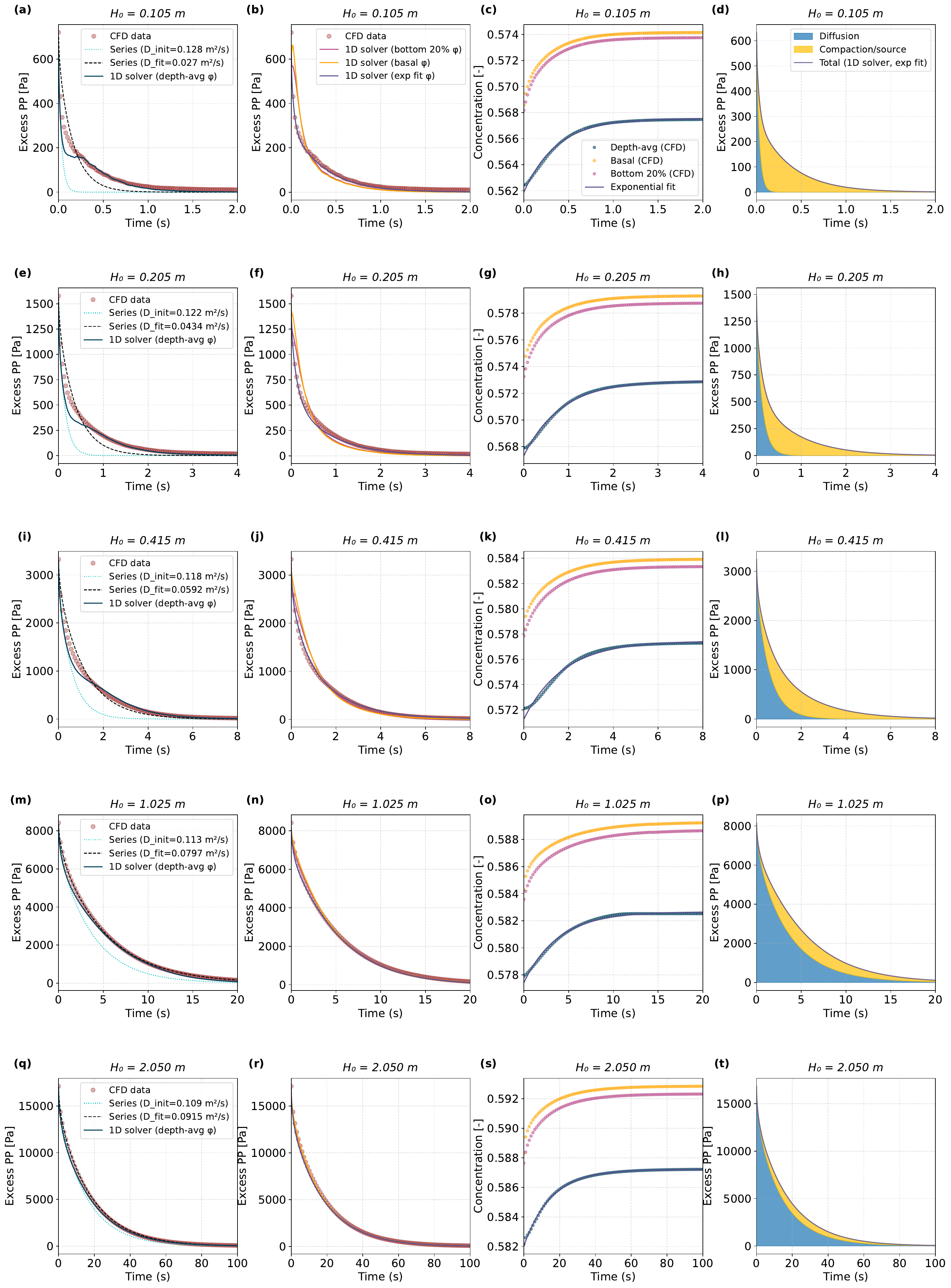}
    \caption{Comparison between MFIX–TFM basal excess pore pressure and reduced 1-D models for several bed heights. Each row shows (a) MFIX-TFM data with Fourier-series reconstructions using \(D_{\text{pred}}\) and the fitted \(D_{\text{fit}}\); (b) 1-D solver forced with different MFIX-derived porosity histories (depth-averaged, basal, bottom 20\%, and exponential fit); (c) corresponding concentration/porosity signals from MFIX-TFM together with the exponential fit used to clean the source term; and (d) separation of the 1-D solution into diffusion-only and compaction/source contributions, whose sum matches the total pressure. The excellent overlap obtained when the solver is driven by the exponential porosity demonstrates that a simple two-parameter compaction law is sufficient to reproduce the MFIX-TFM pressure decay across bed thicknesses.}
    \label{fig:fourpanel}
\end{figure}

Figure~\ref{fig:fourpanel} shows, for representative bed thicknesses, (i) the MFIX-TFM basal excess-pressure decay, (ii) the Fourier-series prediction using the permeability-based diffusivity \(D_{\text{pred}}\), (iii) the same series refitted with an effective diffusivity \(D_{\text{fit}}\), and (iv) the 1-D solver forced by the exponential porosity history. For the latter, we tested various representations of the concentration to determine the most appropriate description.

For all beds, including thinner beds where compaction contributes a large fraction of the total pressure signal, the exponential-forced 1-D model reproduces the MFIX-TFM curves almost exactly, confirming that the small mismatch seen with the raw CFD porosity was due to noise in \(\mathrm{d}\varepsilon/\mathrm{d}t\) rather than to missing physics.

The decomposition of the pressure signal in panel~(Fig.~3d) further shows that the total pressure signal is well represented as the sum of a purely diffusive contribution and a compaction/source contribution, consistent with the basal ODE derived in Sec.~\ref{sec:basal_ode_timescales}.

\section{\label{sec:basal_ode_timescales}Single-mode reduction and diagnostic timescales}

In order to relate the fully coupled diffusion–compaction problem to quantities that can be measured or extracted from simulations (basal pressure, porosity history, bed height), we project the 1-D equation onto the eigenmodes defined by the mixed Neumann–Dirichlet boundary conditions. This leads to a basal ordinary differential equation (ODE) that makes explicit the competition between drainage and compaction. From this ODE, natural diffusive and compaction timescales emerge, together with a dimensionless source–to–diffusion ratio.

\subsection{Modal decomposition and basal ODE}

We consider the one-dimensional diffusion–compaction equation
\begin{equation}
\frac{\partial P}{\partial t}
= D\,\frac{\partial^2 P}{\partial z^2} \;-\; S(t),
\label{eq:modal_pde}
\end{equation}
in a bed of thickness $H$, subject to a no-flux (Neumann) condition at the base and a fixed (Dirichlet) pressure at the top.

This boundary-value problem admits the separated spatial eigenfunctions
\begin{equation}
  \psi_k(z) = \cos\!\Bigl[\bigl(\tfrac{\pi}{2} + k\pi\bigr)\,\tfrac{z}{H}\Bigr],
  \qquad k = 0,1,2,\dots
\end{equation}
with corresponding eigenvalues
\begin{equation}
  \lambda_k = \bigl(\tfrac{\pi}{2} + k\pi\bigr)^{2}.
\end{equation}
These $\psi_k$ satisfy $\partial_z \psi_k|_{z=0}=0$ (impermeable base) and $\psi_k(H)=0$ (venting top), which is the same pair of boundary conditions enforced in the numerical solver.

Any excess pore-pressure field $P(z,t)$ that is compatible with these boundaries can therefore be expanded as
\begin{equation}
  P(z,t) = \sum_{k=0}^{\infty} P_k(t)\,\psi_k(z)
         = \sum_{k=0}^{\infty} P_k(t)\,
           \cos\!\Bigl[\bigl(\tfrac{\pi}{2} + k\pi\bigr)\,\tfrac{z}{H}\Bigr].
  \label{eq:Nd-series}
\end{equation}
Evaluating \eqref{eq:Nd-series} at the base $z=0$ gives the basal pressure directly as a modal sum
\begin{equation}
  P_b(t) \equiv P(0,t)
  = \sum_{k=0}^{\infty} P_k(t),
  \label{eq:Pb-modal-sum}
\end{equation}
because $\psi_k(0)=1$ for all $k$.

In the absence of compaction forcing, each modal amplitude obeys
\begin{equation}
  \frac{\dd P_k}{\dd t} + \frac{D}{H^{2}}\,\lambda_k\,P_k = 0,
  \qquad k=0,1,2,\dots,
  \label{eq:modal-ode-diffusion}
\end{equation}
so that
\begin{equation}
  P_k(t) = P_{0}^{\,k}\,
  \exp\!\Bigl[-\,\lambda_k\,D\,\tfrac{t}{H^{2}}\Bigr],
  \label{eq:modal-solution-diffusion}
\end{equation}
where $P_{0}^{\,k}$ are the modal coefficients of the \emph{initial} pressure profile.

For the linear initial profile used throughout this work,
\begin{equation}
  P(z,0) = P_{b0}\Bigl(1 - \frac{z}{H}\Bigr),
  \label{eq:linear-initial}
\end{equation}
the projection onto the Neumann–Dirichlet eigenfunctions can be done analytically and yields
\begin{equation}
  \boxed{%
  P_{0}^{\,k}
  = \frac{8\,P_{b0}}{\pi^{2}\,(1+2k)^{2}}, \qquad k=0,1,2,\dots}
  \label{eq:modal-coeff-linear}
\end{equation}
The basal signal is reconstructed by summing exponentials with eigenvalues $\lambda_k=(\tfrac{\pi}{2}+k\pi)^2$ and amplitudes given by \eqref{eq:modal-coeff-linear}.

Substituting \eqref{eq:modal-solution-diffusion} and \eqref{eq:modal-coeff-linear} into \eqref{eq:Pb-modal-sum}, the diffusion-only basal pressure is
\begin{equation}
  P_b(t)
  = \sum_{k=0}^{\infty}
    \frac{8\,P_{b0}}{\pi^{2}\,(1+2k)^{2}}\,
    \exp\!\Bigl[-\,\bigl(\tfrac{\pi}{2}+k\pi\bigr)^{2}\,\frac{D}{H^{2}}\,t\Bigr].
  \label{eq:Pb-multi-mode}
\end{equation}
Because the eigenvalues grow like $(1+2k)^{2}$, the higher modes decay very quickly; truncating \eqref{eq:Pb-multi-mode} to the first one or a few modes already reproduces the MFIX-TFM basal signal with good accuracy, which is why the code uses $n=10$ modes by default.

\begin{figure}[htbp]
    \centering
    \includegraphics[width=1.0\linewidth]{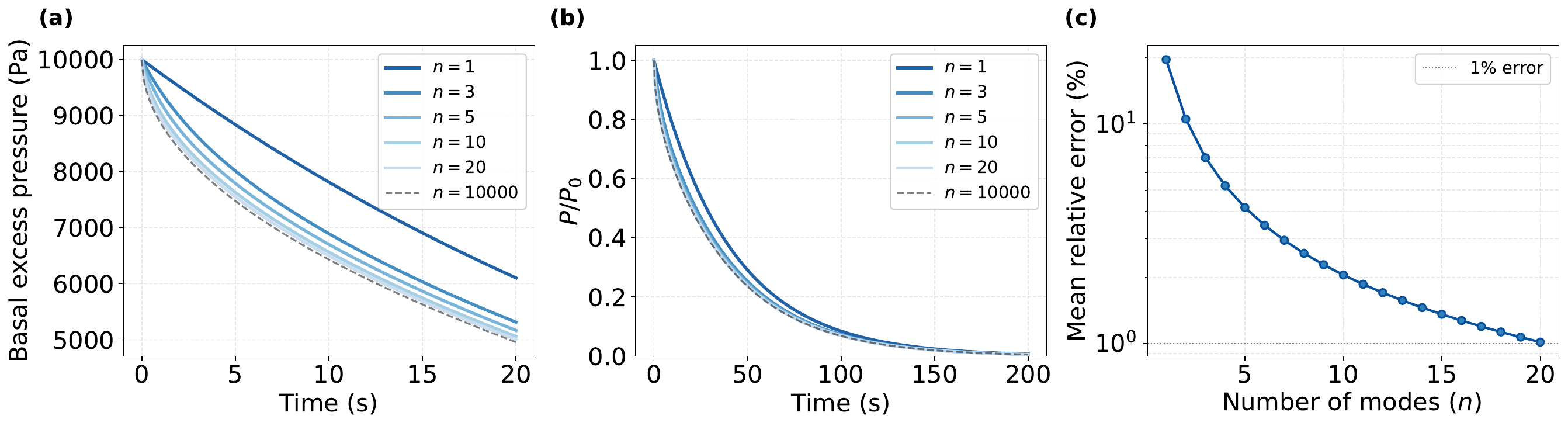}
    \caption{%
    Convergence of the Fourier-series reconstruction of the basal excess pressure for a \SI{1.0}{m} bed with $P_{b0}=\SI{10}{kPa}$ and $D=\SI{0.01}{m^2\,s^{-1}}$. 
    The modal coefficients are normalised so that $P_b(t{=}0)=P_{b0}$. 
    The reference curve ($n=10^4$) is shown in black; successive truncations ($n=1$–$20$) rapidly approach it, with mean relative error below 2\% for $n\ge5$ over the short-time interval. 
    }
    \label{fig:fourier_convergence}
\end{figure}

To verify the accuracy and convergence of the truncated Neumann–Dirichlet expansion, we computed the Fourier-series reconstruction of the basal pressure for increasing mode numbers ($n=1$–$20$) and compared it against a high-resolution reference solution using $n=10^4$ modes. The modal coefficients $P_0^k$ were normalised so that the reconstructed basal pressure at $t=0$ exactly matches the prescribed $P_{b0}$, ensuring consistency of the initial condition. As shown in Fig.~\ref{fig:fourier_convergence}, the series converges rapidly: fewer than ten modes are sufficient to reproduce the basal pressure decay within 2\% of the reference solution over the entire short-time window. This justifies the use of $n=10$ modes in the subsequent analyses of effective diffusivity.

When compaction is present, the depth-averaged porosity $\varepsilon(t)$ acts as a source or sink of excess pore pressure and each mode acquires a source term. At the level of the basal signal—and keeping the leading mode—this reduces to the familiar single-mode ODE
\begin{equation}
  \frac{\dd P_b}{\dd t}
  = -\,\alpha_m\,P_b \;-\; R\,\frac{\dd \varepsilon}{\dd t},
  \qquad
  \alpha_m = \frac{\pi^{2}D}{4H^{2}},
  \qquad
  R = \frac{1}{\beta\,\varepsilon\,(1-\varepsilon)},
  \label{eq:basal-ode-updated}
\end{equation}
which is the same structure obtained previously, but now we understand $P_b(t)$ as the $k=0$ member of the Neumann–Dirichlet series whose initial amplitude is fixed by \eqref{eq:modal-coeff-linear}. The first term in \eqref{eq:basal-ode-updated} describes exponential drainage of pore pressure through the bed; the second term accounts for the mechanically driven pressure change due to time-varying porosity.

As a useful benchmark, setting $\dd \varepsilon/\dd t = 0$ in \eqref{eq:basal-ode-updated} gives the pure-diffusion limit
\begin{equation}
  P_b(t) = P_{b0}\,\exp(-\alpha_m t),
  \qquad
  \alpha_m = \frac{\pi^{2}D}{4H^{2}},
\end{equation}
which is exactly the $k=0$ term of \eqref{eq:Pb-multi-mode}. Any deviation from this single exponential in the MFIX-TFM data is therefore attributed to either (i) the higher Neumann–Dirichlet modes in \eqref{eq:Pb-multi-mode}, or (ii) the compaction forcing represented by $R\,\dot{\varepsilon}$ in \eqref{eq:basal-ode-updated}.

\subsection{Characteristic diffusion and compaction timescales}

Starting from Eq.~\eqref{eq:basal-ode-updated},
\begin{equation}
\frac{\mathrm{d}P_b}{\mathrm{d}t}
= -\,\alpha_m\,P_b \;-\; R\,\frac{\mathrm{d}\varepsilon}{\mathrm{d}t},
\end{equation}
we can write the general solution using an integrating factor:
\begin{equation}
P_b(t)
= P_{b0}\,e^{-\alpha_m t}
- e^{-\alpha_m t}\!
\int_{0}^{t} e^{\alpha_m \tau}\,
R(\tau)\,\frac{\mathrm{d}\varepsilon}{\mathrm{d}\tau}\,
\mathrm{d}\tau.
\label{eq:Pbt_integral_combined}
\end{equation}
This expression reveals that basal pressure at any time is the superposition of pure diffusion (first term) and compaction forcing (integral term).

Two natural timescales follow:

\begin{enumerate}
\item \textbf{Diffusive timescale} (drainage):
\begin{equation}
T_{\mathrm{diff}}
= \frac{1}{\alpha_m}
= \frac{4H^{2}}{\pi^{2}D},
\label{eq:Tdiff_combined}
\end{equation}
which is the time over which pressure would decay if diffusion acted alone.

\item \textbf{Compaction (dilatancy) timescale}:
\begin{equation}
T_{c}
= \frac{\varepsilon(1-\varepsilon)}{-\,\mathrm{d}\varepsilon/\mathrm{d}t},
\label{eq:Tc_combined}
\end{equation}
which is the time over which porosity changes by an amount comparable to $\varepsilon(1-\varepsilon)$. For $\mathrm{d}\varepsilon/\mathrm{d}t<0$ (compaction), $T_c>0$ and measures the duration of pressure injection; for dilatation, $T_c$ characterizes pressure relaxation.
\end{enumerate}

\subsection{Diffusion–to–compaction ratio and source metric}

The competition between drainage and compaction is captured by the ratio of these two timescales, which we identify with an instantaneous Deborah number:
\begin{equation}
\mathrm{De}
= \frac{T_{\mathrm{diff}}}{T_{c}}
= \frac{1}{\alpha_m}\,
\frac{-\,\mathrm{d}\varepsilon/\mathrm{d}t}{\varepsilon(1-\varepsilon)}.
\label{eq:De_combined}
\end{equation}
Large $\mathrm{De}$ indicates that porosity is changing fast compared to diffusion (compaction-dominated), while small $\mathrm{De}$ corresponds to classical diffusion-dominated decay.

However, the basal ODE~\eqref{eq:basal-ode-updated} shows that the influence of the source term is inversely proportional to the instantaneous pressure. This motivates the definition of the source–to–diffusion ratio
\begin{equation}
\boxed{%
\Psi(t)
\;\equiv\;
\frac{\big|R(t)\,\dot{\varepsilon}(t)\big|}{\alpha_m(t)\,P_b(t)}
=
\frac{1}{\beta\,P_b(t)}\,
\frac{1}{\alpha_m(t)}\,
\frac{|\dot{\varepsilon}(t)|}{\varepsilon(t)\,[1-\varepsilon(t)]}
}
\label{eq:Psi_combined}
\end{equation}
so that, using Eq.~\eqref{eq:De_combined},
\begin{equation}
\boxed{%
\Psi(t) = \frac{\mathrm{De}(t)}{\beta\,P_b(t)} }.
\end{equation}
This form is convenient because $\Psi(t)$ can be evaluated directly from data or simulations: it needs only the measured basal pressure, the porosity and its rate of change, and the bed thickness (to get $\alpha_m$). Using initial values,
\begin{equation}
\boxed{%
\Psi_0
=\frac{|R_0\,\dot{\varepsilon}_0|}{\alpha_{m,0}\,P_{b0}}
=\frac{1}{\beta\,\varepsilon_0(1-\varepsilon_0)}\,\frac{|\dot{\varepsilon}_0|}{\alpha_{m,0}\,P_{b0}}
=\frac{\mathrm{De}_0}{\beta\,P_{b0}},
\qquad
\alpha_{m,0}=\frac{\pi^2 D_0}{4H_0^2},
}
\end{equation}
which we later use to collapse fitted diffusivities across bed heights.

\section{\label{sec:effective_diffusivity}Effective diffusivity accounting for compaction}

In depth-averaged flow models of granular systems, the pore pressure evolution is often simplified by assuming a pure diffusion process governed by a single effective diffusivity. This approach reduces computational complexity and provides physical insight into the characteristic timescales controlling flow mobility. However, when the granular bed undergoes significant compaction or dilatation, the source term arising from porosity changes can substantially alter the pressure evolution, effectively modifying the apparent diffusion rate. Our objective is to develop a predictive correction to the Darcy diffusivity that accounts for compaction effects in an a-priori manner, enabling accurate depth-averaged modeling without resolving the full coupled diffusion-compaction dynamics.

\subsection{\label{sec:fitted_diffusivity}Fitted versus predicted diffusivity}

We define the fitted diffusivity $D_{\text{fit}}$ as the effective constant-diffusivity value that reproduces the observed exponential decay of basal excess pressure in each simulation:
\begin{equation}
P_b(t) \simeq P_{b0}\,e^{-\alpha_{m,\text{fit}} t},
\qquad
D_{\text{fit}} = \frac{4H_0^2}{\pi^2}\,\alpha_{m,\text{fit}}.
\label{eq:Dfit_def_final}
\end{equation}
Here, $D_{\mathrm{fit}}$ should be interpreted as an apparent relaxation diffusivity inferred from the decay of basal excess pressure, rather than as the intrinsic Darcy diffusivity governing gas transport alone. Our goal is to predict $D_{\mathrm{fit}}$ from the initial conditions only: bed height $H_0$, porosity $\varepsilon_0$, and basal excess pressure $P_{b0}$.

For comparison, the purely diffusive coefficient expected from Darcy's law and the Carman--Kozeny permeability relation is
\begin{equation}
D_{\text{pred}} = \frac{k(\varepsilon_0)}{\mu_g\,\beta_0\,\varepsilon_0},
\qquad
k(\varepsilon_0) = \frac{\varepsilon_0^3 d_{\mathrm{eq}}^2}{150(1-\varepsilon_0)^2},
\label{eq:Dpred_final}
\end{equation}

The subscript 0 indicates the \emph{a priori} initial values at $t=0$. The ratio $D_{\text{fit}}/D_{\text{pred}}$ quantifies how much the compaction dynamics retard the apparent diffusivity relative to the pure diffusion prediction.

While this definition is expressed in terms of initial conditions for the purpose of analysing transient column experiments, it is not directly intended for use in depth-averaged models. In such models, diffusivity should be evaluated from instantaneous local variables (e.g.\ $\varepsilon$, $P$, and $H$). The present formulation instead provides a framework to identify how these instantaneous quantities control the effective diffusivity through the underlying compaction–diffusion coupling.

\subsection{\label{sec:timescales}Characteristic diffusion and compaction timescales}

For the fundamental diffusion mode, the characteristic diffusive timescale is
\begin{equation}
T_{\mathrm{diff}} = \frac{4 H_0^2}{\pi^2 D_{\mathrm{pred}}}.
\end{equation}
If diffusion acted alone, the basal excess pressure would decay exponentially as
\[
P_b(t) = P_{b0}\,\exp(-t/T_{\mathrm{diff}}),
\]
with $T_{\mathrm{diff}}$ setting the duration of the drainage process observed in the MFIX-TFM simulations.

\begin{figure}[htbp]
    \centering
    \includegraphics[width=0.6\linewidth]{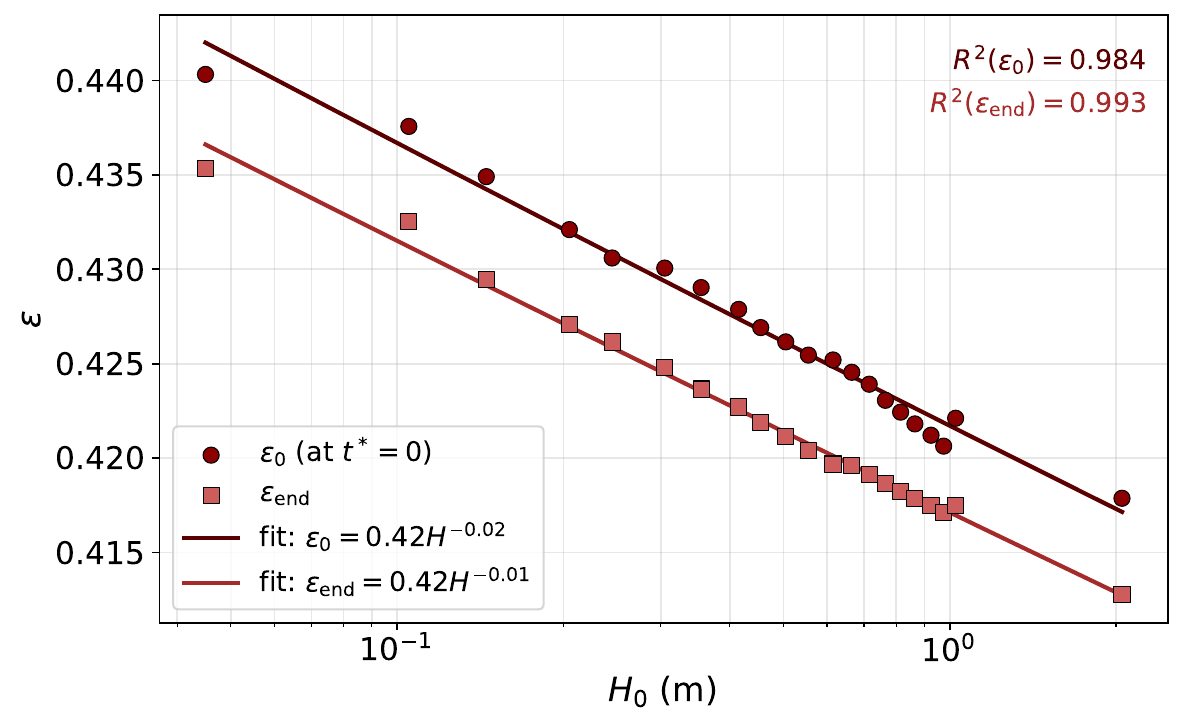}
    \caption{Porosity change $\Delta\varepsilon = \varepsilon(t^*=H_0) - \varepsilon_{\mathrm{end}}$ as a function of initial bed height $H_0$ from MFIX-TFM simulations. Thin beds compact appreciably during drainage, while thick beds remain nearly rigid. This behaviour motivates the empirical scaling in Eq.~\eqref{eq:Delta_eps_rule}.}
    \label{fig:delta_eps_vs_H}
\end{figure}

The characteristic compaction timescale is defined from the initial porosity evolution rate as
\begin{equation}
T_c = \frac{\varepsilon_0(1-\varepsilon_0)}{-\,\mathrm{d}\varepsilon/\mathrm{d}t\vert_{t=0}}.
\label{eq:Tc_final}
\end{equation}
This represents the time required for the porosity to change by a fractional amount $\varepsilon_0(1-\varepsilon_0)$ at the initial compaction rate. To evaluate this \emph{a priori}, we first quantify how the compaction amplitude depends on bed thickness. Figure~\ref{fig:delta_eps_vs_H} shows the porosity near the onset of drainage, $\varepsilon_{\mathrm{start}}$, and the porosity at the end of drainage, $\varepsilon_{\mathrm{end}}$, extracted from the MFIX-TFM simulations as a function of the initial bed height $H_0$. Both series follow nearly linear trends with $H_0$ ($R^2 \approx 0.99$), indicating that thicker beds start and finish at slightly lower porosity than thinner ones, but with an almost constant offset between the two curves. This implies that the net porosity loss
\[
\Delta\varepsilon = \varepsilon_{\mathrm{start}} - \varepsilon_{\mathrm{end}}
\]
depends only weakly on $H_0$, i.e.\ thin and thick beds compact by roughly the same small amount over the drainage interval (Fig.~\ref{fig:delta_eps_vs_H}).
Guided by these data, we approximate $\Delta\varepsilon$ by the observed MFIX-TFM drop (capped so it cannot exceed the available porosity):
\begin{equation}
\Delta\varepsilon \approx \min\bigl(\Delta\varepsilon_{\text{MFIX}},\, \varepsilon_0 - \varepsilon_{\min}\bigr),
\label{eq:Delta_eps_rule}
\end{equation}
with $\varepsilon_{\min} \simeq 0.41$. Substituting this estimate of the total porosity change into the definition of the compaction timescale gives the \emph{a priori} relation
\begin{equation}
T_c \approx \frac{\varepsilon_0(1-\varepsilon_0)}{\Delta\varepsilon}\,T_{\mathrm{diff}},
\label{eq:Tc_estimate}
\end{equation}
which links the effective duration of compaction to the MFIX-measured porosity drop. Because $\Delta\varepsilon$ varies only weakly with $H_0$, most of the height dependence in the diffusion–compaction model still enters through $T_{\mathrm{diff}}$.

\subsection{\label{sec:psi_definition}Source-to-diffusion ratio and collapse parameter}

The relative influence of compaction on pore-pressure diffusion can be quantified by a single dimensionless number comparing the strength of the compaction source term to the characteristic diffusive flux. We define this \emph{source-to-diffusion ratio} as
\begin{equation}
\Psi_0 = 
\frac{1}{\beta_0 P_{b0}}\,
\frac{\Delta\varepsilon}{\varepsilon_0(1-\varepsilon_0)},
\label{eq:Psi0_final}
\end{equation}
where $\Delta\varepsilon$ is the total porosity change between the start and end of drainage, $\varepsilon_0$ is the initial porosity, $\beta_0 = 1/P_{\mathrm{amb}}$ is the gas compressibility at ambient pressure, and $P_{b0}$ is the initial basal overpressure.

Physically, $\Psi_0$ expresses how strongly the volumetric compaction source perturbs the diffusion-driven relaxation of pore pressure. When $\Psi_0 \ll 1$, the system behaves as a purely diffusive bed with negligible deformation. When $\Psi_0 \gtrsim 1$, compaction contributes comparably to or dominates over diffusion, leading to a marked reduction of effective diffusivity. This single parameter successfully collapses the MFIX-derived diffusivity data across all tested bed heights (0.045--2~m) and initial pressures, as shown in Table~\ref{tab:psi_algebraic}. Thin, strongly compacting beds exhibit large $\Psi_0$, while thicker beds with small porosity change approach $\Psi_0 \to 0$.

\subsection{\label{sec:psi_algebraic_model}Algebraic correction based on the source–to–diffusion ratio $\Psi_0$}

The effective diffusivity $D_{\text{fit}}$ obtained from the MFIX-TFM pressure–time series is consistently lower than the theoretical diffusivity
\[
D_{\text{pred}} = \frac{k(\varepsilon_0)}{\mu_{\text{air}}\beta_0 \varepsilon_0},
\]
whenever bed compaction occurs during gas drainage. The reduction scales systematically with~$\Psi_0$, providing a natural basis for correction.

To describe this relationship, we evaluated several empirical forms (exponential, rational, and algebraic). The algebraic expression
\begin{equation}
\frac{D_{\text{fit}}}{D_{\text{pred}}} = \frac{1}{1 + c\,\Psi_0^{\,m}},
\label{eq:alg_model}
\end{equation}
provided the best agreement with the simulation data, while satisfying both limiting behaviours:
\[
\frac{D_{\text{fit}}}{D_{\text{pred}}} \to 1 \quad \text{as } \Psi_0 \to 0,
\qquad
\frac{D_{\text{fit}}}{D_{\text{pred}}} \to 0 \quad \text{as } \Psi_0 \to \infty.
\]

A global least-squares fit to all $(\Psi_0, D_{\text{fit}}/D_{\text{pred}})$ pairs from the ten-mode Fourier reconstructions yields
\begin{equation*}
c = 1.5637, \qquad m = 1.0392, \qquad
R^2 = 0.991,\ \text{mean error } = 3.7\%.
\end{equation*}
The algebraic law accurately captures both the gentle attenuation at small~$\Psi_0$ and the strong suppression at large~$\Psi_0$, where the compaction source term dominates the diffusion–compaction dynamics.

We therefore adopt the following $\Psi_0$-based predictive law for the effective diffusivity:
\begin{equation}
\boxed{%
D_{\Psi}
= \frac{D_{\text{pred}}}{1 + 1.5637\,\Psi_0^{\,1.0392}}
}
\label{eq:Dfit_algebraic_final}
\end{equation}
Here $D_{\Psi}$ denotes the \emph{apriori} corrected diffusivity that accounts for the coupling between compaction and diffusion using only the initial bed properties $(H_0, \varepsilon_0, P_{b0})$ and the magnitude of porosity change $\Delta\varepsilon$.  

Table~\ref{tab:psi_algebraic} summarises the predicted and MFIX-fitted diffusivity across the full range of bed heights, including the thick-bed limit ($H_0\approx2$~m) where $D_{\text{fit}}\approx D_{\text{pred}}$. The overall mean deviation is $3.7\%$, with the largest discrepancy ($17\%$) observed for $H_0=0.105$~m, one of the most strongly compacting cases.

\begin{table}[htbp]
\centering
\caption{Predicted versus MFIX-fitted effective diffusivity using the algebraic correction
$D_{\Psi} = D_{\text{pred}}/(1 + 1.5637\,\Psi_0^{1.0392})$ for all simulated bed heights ($n=5$ modes).}
\label{tab:psi_algebraic}
\begin{tabular}{cccccc}
\toprule
$H_0$ & $\Psi_0$ & $D_{\text{pred}}$ & $D_{\Psi}$ (pred.) & $D_{\text{fit}}$ (MFIX) & Error \\
(m)   &          & (m$^2$/s)         & (m$^2$/s)                 & (m$^2$/s)               & (\%)  \\
\midrule
0.045 & 6.400 & $1.307\times10^{-1}$ & $1.111\times10^{-2}$ & $1.068\times10^{-2}$ &  4.0 \\
0.105 & 2.884 & $1.278\times10^{-1}$ & $2.241\times10^{-2}$ & $2.700\times10^{-2}$ & 17.0 \\
0.145 & 1.949 & $1.250\times10^{-1}$ & $3.029\times10^{-2}$ & $2.881\times10^{-2}$ &  5.1 \\
0.205 & 1.299 & $1.222\times10^{-1}$ & $4.003\times10^{-2}$ & $4.336\times10^{-2}$ &  7.7 \\
0.245 & 0.916 & $1.207\times10^{-1}$ & $4.972\times10^{-2}$ & $4.283\times10^{-2}$ & 16.1 \\
0.305 & 0.897 & $1.202\times10^{-1}$ & $5.014\times10^{-2}$ & $5.068\times10^{-2}$ &  1.1 \\
0.355 & 0.773 & $1.192\times10^{-1}$ & $5.424\times10^{-2}$ & $5.479\times10^{-2}$ &  1.0 \\
0.415 & 0.642 & $1.181\times10^{-1}$ & $5.942\times10^{-2}$ & $5.917\times10^{-2}$ &  0.4 \\
0.455 & 0.554 & $1.171\times10^{-1}$ & $6.347\times10^{-2}$ & $6.400\times10^{-2}$ &  0.8 \\
0.505 & 0.494 & $1.164\times10^{-1}$ & $6.649\times10^{-2}$ & $6.361\times10^{-2}$ &  4.5 \\
0.555 & 0.455 & $1.157\times10^{-1}$ & $6.848\times10^{-2}$ & $6.895\times10^{-2}$ &  0.7 \\
0.615 & 0.454 & $1.155\times10^{-1}$ & $6.842\times10^{-2}$ & $6.880\times10^{-2}$ &  0.6 \\
0.665 & 0.375 & $1.149\times10^{-1}$ & $7.346\times10^{-2}$ & $7.461\times10^{-2}$ &  1.5 \\
0.715 & 0.340 & $1.143\times10^{-1}$ & $7.569\times10^{-2}$ & $7.422\times10^{-2}$ &  2.0 \\
0.765 & 0.293 & $1.135\times10^{-1}$ & $7.902\times10^{-2}$ & $8.035\times10^{-2}$ &  1.7 \\
0.815 & 0.264 & $1.129\times10^{-1}$ & $8.110\times10^{-2}$ & $7.994\times10^{-2}$ &  1.4 \\
0.865 & 0.238 & $1.123\times10^{-1}$ & $8.306\times10^{-2}$ & $8.670\times10^{-2}$ &  4.2 \\
0.925 & 0.216 & $1.118\times10^{-1}$ & $8.484\times10^{-2}$ & $8.628\times10^{-2}$ &  1.7 \\
0.975 & 0.195 & $1.113\times10^{-1}$ & $8.652\times10^{-2}$ & $8.587\times10^{-2}$ &  0.8 \\
2.050 & 0.124 & $1.088\times10^{-1}$ & $9.230\times10^{-2}$ & $9.151\times10^{-2}$ &  0.9 \\
\midrule
\multicolumn{6}{l}{Mean error: 3.7\% \quad (median: 1.5\%, max: 17.0\%)} \\
\bottomrule
\end{tabular}
\end{table}

\begin{figure}[htbp]
\centering
\includegraphics[width=1.0\textwidth]{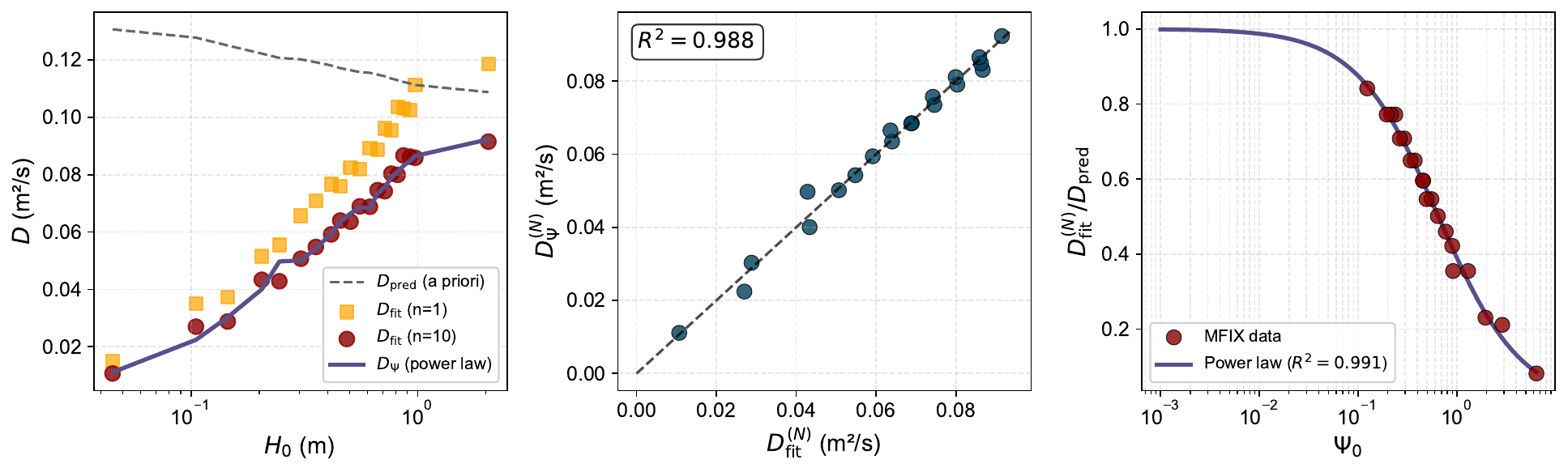}
\caption{$\Psi_0$-based algebraic correction for effective diffusivity. 
(a) Diffusivity versus bed height (logarithmic $x$-axis) showing $D_{\text{pred}}$ (gray dashed), MFIX-fitted $D_{\text{fit}}$ for $n=1$ (orange squares) and $n=10$ (dark red circles), and predictions from Eq.~\eqref{eq:Dfit_algebraic_final} using the $n=10$ fit (purple solid line). 
(b) Parity plot demonstrating excellent correlation ($R^2=0.991$) between predicted $D_{\Psi}$ and MFIX-fitted $D_{\text{fit}}$ diffusivities for $n=10$ modes. 
(c) The correction factor $D_{\text{fit}}/D_{\text{pred}}$ versus $\Psi_0$ on a log scale, showing the algebraic decay. Data points (dark red circles) are from MFIX-TFM simulations with $n=10$ modal reconstruction; the purple curve is the fitted power law $1/(1 + 1.5637\,\Psi_0^{1.0392})$.}
\label{fig:psi_algebraic_model}
\end{figure}

Starting from the single--mode basal ODE, we defined the instantaneous source--to--diffusion ratio $\Psi(t)$ and showed its relation to $\mathrm{De}(t)$ via $\Psi(t)=\mathrm{De}(t)/[\beta\,P_b(t)]$. This provides (i) an \emph{a priori} indicator $\Psi_0$ using the initial state and (ii) a practical, data--driven diagnostic $\Psi(t)$ that uses $H(t)$, $\varepsilon(t)$, $\dot{\varepsilon}(t)$ and $P_b(t)$ directly from the simulation or experiment. The algebraic correction Eq.~\eqref{eq:Dfit_algebraic_final} enables accurate prediction of effective diffusivity across the entire range of bed heights and compaction strengths with a mean error under 4\%.

\section{Modeling geophysical flows using the depth-average approach}

% contribution from Claudia (then we need to move it to a more suitable place in the text)
%\include{imex_pore_pressure}

\subsection{Pore Pressure Dynamics}

The results of the previous Section are not used in the depth-averaged model as a direct algebraic substitution for the diffusivity. Rather, they identify the physical controls that the closure must reproduce locally, namely the competition between drainage, compaction, and porosity-dependent permeability, which we incorporate in the IMEX formulation below.

To model flows where pore fluid pressure significantly affects mobility, we introduce a new prognostic equation for its evolution, inspired by the work of \citet{gueugneau2017effects}. The model is formulated in terms of the \textit{excess pore pressure}, $P_{exc} = P - P_{\text{atm}}$, where $P$ is the absolute pore pressure. Within our depth-averaged framework, the entire vertical profile of the excess pressure is represented by a single parameter: its value at the base of the flow. The prognostic variable that our model solves for is therefore the basal excess pore pressure, denoted as $P_{b,exc} = P_b - P_{\text{atm}}$, where $P_b$ is the absolute basal pressure.

\subsubsection{Physical Formulation and Depth-Averaging}
The evolution of excess pore pressure $P_{\mathrm{exc}}(t,z)$ at any depth within the granular mixture is governed by a 1D vertical diffusion process:
\begin{equation}
    \pderiv{P_{exc}}{t} = D \pderiv{^2 P_{exc}}{z^2}
\end{equation}
where $t$ is the temporal coordinate, $z$ is the vertical coordinate parallel to gravity, and $D$ is the hydraulic diffusivity coefficient, defined as \citet{iverson1997physics}:
\begin{equation}
    D = \frac{k}{\phi_p \mu_g \beta_f}
\end{equation}
Here, $k$ is the hydraulic permeability of the medium, $\phi_p$ is the porosity, $\mu_g $ is the dynamic viscosity of the gas and $\beta_f$ is the compressibility of the fluid, respectively. The hydraulic permeability may be dynamically updated depending on the porosity and the equivalent particle diameter, according to the Carman-Kozeny Eq. (\ref{eq:carman_kozeny}).

To derive a depth-averaged equation suitable for our framework, we approximate the vertical profile of the excess pressure. We assume boundary conditions of zero excess pressure at the free surface ($P_{exc}(h,t)=0$) and zero gas flux through the impermeable bed ($\partial P_{exc}/\partial z |_{z=0}=0$). The first dominant mode of the Fourier series solution that satisfies these conditions provides the following profile approximation:
\begin{equation}
    P_{exc}(z,t) \approx P_{b,exc}(t) \cos\left(\frac{\pi z}{2h}\right)
    \label{eq:pressure_profile}
\end{equation}
where $P_{b,exc}(t)$ is the basal excess pressure at a given time. Integrating the 1D diffusion equation over the flow depth using this profile yields an evolution equation for the depth-averaged quantity, which is consistent with the dissipation term used by \citet{gueugneau2017effects}. The resulting non-conservative (advective) form for the basal excess pore pressure is:
\begin{equation}
    \pderiv{P_{b,exc}}{t} + \vect{u} \cdot \nabla P_{b,exc} = - \left( \frac{\pi}{2h} \right)^2 D P_{b,exc}
    \label{eq:pore_pressure_non_cons_excess}
\end{equation}
The left-hand side represents the advection of the excess pressure field with the flow, while the right-hand side models its dissipation. 

\subsubsection{Mass Loss due to Pore Pressure Dissipation}
The dissipation of excess pore pressure implies a physical loss of gas from the top of the flow. We quantify this mass loss using Darcy's Law, evaluated at the free surface ($z=h$). The volumetric flux of gas leaving the flow per unit area, $v_{\text{loss,gas}}$, depends on the gradient of the \textit{absolute} pressure $P = P_{exc} + P_{\text{atm}}$:
\begin{equation}
    v_{\text{loss,gas}} = -\frac{k}{\mu_g} \left. \pderiv{P}{z} \right|_{z=h} = -\frac{k}{\mu_g} \left. \pderiv{P_{exc}}{z} \right|_{z=h}
\end{equation}
since the atmospheric pressure $P_{\text{atm}}$ is assumed constant. Using the approximated excess pressure profile from Eq. ( \ref{eq:pressure_profile}), the vertical gradient at the surface is:
\begin{equation}
    \left. \pderiv{P_{exc}}{z} \right|_{z=h} = -P_{b,exc}(t) \frac{\pi}{2h} \sin\left(\frac{\pi h}{2h}\right) = -P_{b,exc}(t) \frac{\pi}{2h}
\end{equation}
Substituting this back into the Darcy flux equation yields the gas loss rate:
\begin{equation}\label{EQ:gas_loss}
    v_{\text{loss,gas}} = \frac{k}{\mu_g} \frac{\pi P_{b,exc}}{2h}
\end{equation}
This volumetric loss rate is then used to define a mass flux leaving the system. This mass loss is not only subtracted from the continuity equation, but it is also consistently accounted for in the source terms of the other conservation equations. Specifically, the mass lost carries with it the momentum and energy of the flow at the point of exit (the free surface). Accordingly, the momentum and energy associated with the removed mass are also removed from the system, ensuring physical consistency. Failing to remove the associated momentum, for instance, would lead to an unphysical acceleration of the flow as its mass decreases while its momentum remains unchanged. This comprehensive treatment guarantees that the dissipation of pore pressure is coupled to a consistent loss of mass, momentum, and energy from the system. 

\subsubsection{Effect of Maximum Particle Packing}
The dissipation of pore pressure and the associated gas loss are physically limited by the porosity of the granular mixture. As the solid volume fraction, $\alpha_s = \sum \alpha_{s,i}$, approaches the maximum random close packing fraction, $\alpha_{\text{max}}$, the gas phase loss must decrease and eventually go to zero to guarantee the maximum limit on the solid volume fraction. To account for this physical limit, we introduce an inhibition factor, $f_{\text{inhibit}}$, defined in Eq. (\ref{eq:finhibit}).  Both the dissipation term in Eq. (\ref{eq:pore_pressure_non_cons_excess}) and the gas loss rate in Eq. (\ref{EQ:gas_loss}) are multiplied by this inhibition factor.  As a result, when the flow becomes densely packed ($\alpha_s \to \alpha_{\text{max}}$), pore pressure dissipation and gas loss cease, resulting in a halt in further compaction, and
allowing the model to retain pore pressure. Note that, in this model, solid-phase compaction results from gas-phase loss; the reverse coupling is not included, so compaction does not in turn generate pore pressure.

To represent this effect, we introduce $f_{\mathrm{inhibit}}$ as an ad hoc cutoff function based on the smooth step function $S_n(\alpha)$, Eq.~(\ref{eq:smoothstep}), which provides a gradual transition from unhindered gas loss to strongly inhibited pressure dissipation and the associated suppression of solid-phase compaction. 
\begin{equation}
f_{inhibit} = 1 - S_n(\alpha)
\label{eq:finhibit}
\end{equation}
\begin{equation}
S_n (x)= 
\begin{cases} 
0, & \text{if }  \,\,  x \leq 0 \\[2mm]
x^{n+1} \sum^n_{k=0} \left( ^{n+k}_{k}\right) \left( ^{2n+1}_{n-k}\right)(-x)^k, & \text{if } \,\, 0 \leq x \leq 1 \\[2mm]
1, &\text{if} \,\, 1 \leq x
\end{cases},\,\,\,\, \text{where} \,\, x = \frac{\alpha-\alpha_{tr}}{\alpha_{max} -\alpha_{tr}}
\label{eq:smoothstep}
\end{equation}

\begin{figure}[htbp]
\centering
\includegraphics[width=0.6\textwidth]{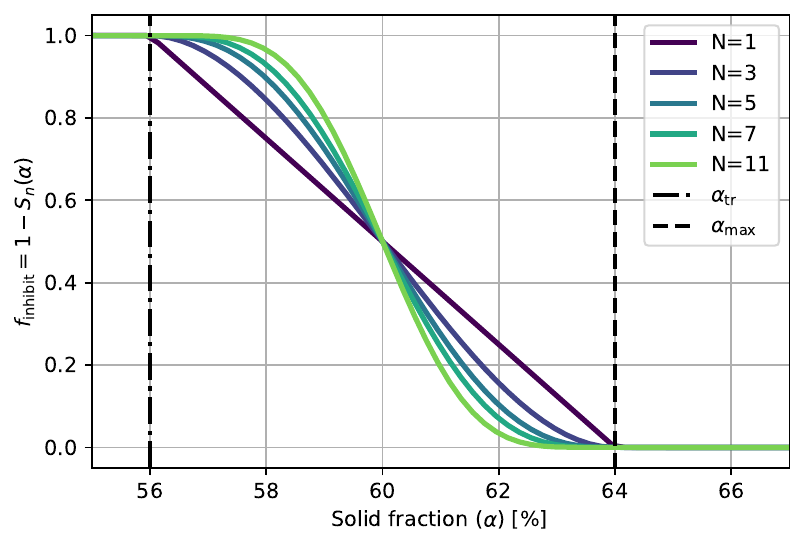}
\caption{Representation of the inhibit factor, $f_{inhibit}$, which accounts for the hindered gas loss and associated pore pressure dissipation as the solid concentration ($\alpha$) approaches the maximum packing fraction ($\alpha_{max}$).  $f_{inhibit}$  is calculated from the smooth step function  $S_n(\alpha)$, Eq. (\ref{eq:finhibit}) and (\ref{eq:smoothstep}), which enables a smooth transition between completely unhindered to completely inhibited gas loss and pressure dissipation. The transition starts at $\alpha_{tr}$ (dash-dotted line) and ends at $\alpha_{max}$ (dashed line). The smoothness of the transition is given by $N$, the degree of the polynomial in $S_n(\alpha)$. }\label{fig:generalised_smoothstep}
\end{figure}

The start and end of the transition are marked by the user-defined parameters $\alpha_{tr}$ and  $\alpha_{max}$ (typically $\approx 0.64$ for spheres), respectively.  The smoothness order parameter of the transition, $n$, defines the degree of the polynomial in $S_n(\alpha)$ as $N = 2n + 1$ (Fig. \ref{fig:generalised_smoothstep}). 

\subsubsection{Conservative Formulation for Numerical Implementation}

As explained by \citet{greenshieldsweller2022}, the advective form of Eq. \eqref{eq:pore_pressure_non_cons_excess} possesses the important mathematical property of \emph{boundedness}. This is a physically desirable characteristic for an intensive quantity like pressure, which represents a state of the material rather than a conserved quantity like mass. However, while physically intuitive, the non-conservative form is not directly suitable for our numerical framework, which is built upon the principle of flux conservation.

The numerical framework adopted in IMEX\_SfloW2D\_v2 (here referred to as 'IMEX') is a finite-volume method that solves equations in conservative form. To integrate the transport of the basal excess pore pressure, $P_{b,exc}$, we must transform the advective equation, Eq.  \eqref{eq:pore_pressure_non_cons_excess}, into its conservative equivalent form. This is achieved by defining a new conservative variable, $Q_p = \rho_m h P_{b,exc} = q_1 P_{b,exc}$, where $q_1$ is the conserved mass per unit area.

The conservation law for this new variable is written as:
\begin{equation}
    \pderiv{Q_p}{t} + \nabla \cdot (\vect{u} Q_p) = S_{Q_p}
    \label{eq:cons_form_Qp}
\end{equation}
To derive the correct form of the source term $S_{Q_p}$, we apply the product rule to the left-hand side of Eq. \ref{eq:cons_form_Qp}:
\begin{align}
    &\pderiv{(q_1 P_{b,exc})}{t} + \nabla \cdot (\vect{u} q_1 P_{b,exc}) \nonumber \\
    &= \left( P_{b,exc} \pderiv{q_1}{t} + q_1 \pderiv{P_{b,exc}}{t} \right) + \left( P_{b,exc} \nabla \cdot (\vect{u} q_1) + q_1 (\vect{u} \cdot \nabla P_{b,exc}) \right) \nonumber \\
    &= P_{b,exc} \left[ \pderiv{q_1}{t} + \nabla \cdot (\vect{u} q_1) \right] + q_1 \left[ \pderiv{P_{b,exc}}{t} + \vect{u} \cdot \nabla P_{b,exc} \right]
\end{align}
The terms in the first bracket represent the mass conservation equation, which equals the total mass source term, $S_m$. The terms in the second bracket correspond to the non-conservative advection-diffusion equation for pore pressure, Eq. (\ref{eq:pore_pressure_non_cons_excess}). Substituting these known relationships, we obtain the expression for the source term $S_{Q_p}$:
\begin{equation}
    S_{Q_p} = P_{b,exc} S_m + q_1 S_{P_{b,exc}}
    \label{eq:source_term_Qp_general}
\end{equation}
where $S_{P_{b,exc}}$ is the source term for the non-conservative pressure equation, i.e., the dissipation term from Eq. \eqref{eq:pore_pressure_non_cons_excess}.

A crucial physical coupling arises because the total mass source term, $S_m$, is itself affected by the pore pressure. Specifically, the dissipation of pore pressure implies a physical loss of gas from the flow's free surface. We can therefore split the mass source term, $S_m$, into two parts: $S_{m, \text{other}}$, which includes erosion, sedimentation, and air entrainment, and $S_{m, \text{pore}}$, the mass flux due to gas escape:
\begin{equation}
    S_m = S_{m, \text{other}} + S_{m, \text{pore}} \quad \text{where} \quad S_{m, \text{pore}} = -f_{\text{inhibit}}\rho_c v_{\text{loss,gas}}
\end{equation}
The volumetric loss rate, $v_{\text{loss,gas}}$, is derived from Darcy's Law as shown previously.

Substituting this decomposition back into Eq. \ref{eq:source_term_Qp_general}, we can explicitly separate the terms related to pore pressure dynamics:
\begin{equation}
    S_{Q_p} = \underbrace{P_{b,exc} S_{m, \text{other}}}_{\text{Advection coupling}} - \underbrace{P_{b,exc} (f_{\text{inhibit}}\rho_c v_{\text{loss,gas}})}_{\text{Self-coupling}} + \underbrace{q_1 \left( - f_{\text{inhibit}}\left( \frac{\pi}{2h} \right)^2 D P_{b,exc} \right)}_{\text{Dissipation}}
    \label{eq:source_term_Qp_detailed}
\end{equation}
This final form of the source term is implemented in the code. The "Advection coupling" and "Self-coupling" terms ensure that mass added or removed by external processes is coupled to the evolution of the conserved pore pressure variable $Q_p$. Conversely, the "Dissipation" term models the decay of the physical pore pressure value itself. This comprehensive and consistent formulation is essential for accurately modeling the coupled evolution of mass and pore pressure within the flow.

\subsubsection{Pore Pressure in Source Terms}
In many scenarios the granular material can be initially fluidized, entering the simulation domain with a high pore pressure. Our model accounts for this by incorporating a specific formulation for the pore pressure within the mass source terms.

Let $\dot{h}$ be the volumetric flux per unit area (with units of velocity, m\,s$^{-1}$) representing the rate at which new material is introduced into the domain, for instance from a basal source. This source introduces mass at a rate of $S_{m,\text{source}} = \rho_{\text{in}}\dot{h}$, where $\rho_{\text{in}}$ is the density of the incoming material. We define a dimensionless parameter, $C_p$, which represents the initial ratio of excess pore pressure to the effective lithostatic pressure of the incoming material.

The basal excess pore pressure of the source material, $P_{b,exc,\text{source}}$, is therefore defined as:
\begin{equation}
    P_{b,exc,\text{source}} = C_p (\rho_{\text{in}} g' h)
\end{equation}
where $h$ is the local flow thickness at the source. This definition allows for a direct physical interpretation of the parameter $C_p$:
\begin{itemize}
    \item $C_p = 0$: The incoming material is unpressurized (dry or fully drained).
    \item $0 < C_p < 1$: The incoming material is partially pressurized.
    \item $C_p = 1$: The incoming material is fully fluidized, with its excess pore pressure exactly balancing its effective weight.
\end{itemize}
Following the conservative formulation, Eq. (\ref{eq:source_term_Qp_general}), the source term for the conserved pore pressure variable, $Q_p$, due to this influx of new material is given by:
\begin{equation}
    S_{Q_p,\text{source}} = P_{b,exc,\text{source}} S_{m,\text{source}} = C_p (\rho_{\text{in}} g' h) (\rho_{\text{in}}\dot{h})
\end{equation}
This term is added to the right-hand side of the conservative pore pressure equation, Eq. (\ref{eq:cons_form_Qp}). This approach provides a robust and physically intuitive way to prescribe the fluidization state of the material entering the simulation domain from a source.

\subsubsection{Coupling of Pore Pressure with Basal Friction}
The primary physical effect of pore pressure is the reduction of inter-particle friction at the base of the flow. This is incorporated into the model via the principle of effective stress. To derive this relationship clearly, we first consider the case of a flow on a horizontal surface before generalizing to complex topography.

On a horizontal surface, the total vertical stress at the bed, $\sigma_{\text{vert}}$, is the sum of the lithostatic pressure exerted by the weight of the material and the ambient atmospheric pressure, $P_{\text{atm}}$:
\begin{equation}
    \sigma_{\text{vert}} = \rho_m g' h + P_{\text{atm}}
\end{equation}
Here, we use the reduced gravity $g' = g (\rho_m - \rho_a)/\rho_m$ to account for the buoyancy of the flow within the ambient fluid. The density of the fluid and the ambient are given by $\rho_m$ and $\rho_a$, respectively. The effective vertical stress, $\sigma'_{\text{vert}}$, is defined as the total stress minus the absolute basal pore pressure, $P_b$:
\begin{equation}
    \sigma'_{\text{vert}} = \sigma_{\text{vert}} - P_b = (\rho_m g' h + P_{\text{atm}}) - P_b
\end{equation}
By rearranging terms, this expression can be conveniently rewritten in terms of the excess pore pressure, which is the prognostic variable evolved by our model:
\begin{equation}
    \sigma'_{\text{vert}} = \rho_m g' h - (P_b - P_{\text{atm}}) = \rho_m g' h - P_{b,exc}
    \label{eq:effective_vertical_stress}
\end{equation}
This formulation highlights a key physical insight: the stress responsible for inter-granular friction depends on the effective weight of the material column reduced by the pore pressure in excess of the atmospheric value.

For flows over inclined and curved terrains, this principle is generalized. The Coulomb friction law states that the frictional shear stress, $\tau_c$, is proportional to the effective normal stress, $\sigma_n'$, which is the component of $\sigma_{vert}'$ acting perpendicular to the bed. Projecting the vertical stress onto the normal direction introduces a factor of $\cos(\alpha)=1/\phi$: 
\begin{equation}
\sigma_{n}' = \frac{\sigma_{vert}'}{\phi}
    \label{eq:effective_normal_stress}
\end{equation}
The magnitude of the friction stress, which acts tangentially (parallel) to the bed surface, is therefore:

\begin{equation}
\tau_c = \mu \sigma_{n}'=  \frac{\mu}{\phi}[\rho_m g' h - P_{b,exc}],
    \label{eq:shear_stress}
\end{equation}where $\mu$ is the coefficient of friction. 

The governing momentum equations are formulated in global Cartesian coordinates for the horizontal velocity components ($u,v$). Consequently, the tangential friction stress $\tau_c$ must be projected from the sloped bed surface back onto the horizontal ($x,y$) plane to obtain the correct source term. This second projection introduces another factor of $\cos(\alpha)=1/\phi$. The final friction source term magnitude, $S_f$ , acting in the horizontal plane is thus:

\begin{equation}
S_f = \frac{\tau_c}{\phi}=  \frac{\mu}{\phi^2}\max(0, \rho_m g' h - P_{b,exc})
    \label{eq:source_term}
\end{equation}
This $1/\phi^2$ factor ensures that both the projection to calculate the normal force and the projection back to the horizontal plane are correctly accounted for. 

\subsubsection{Partially regularised $\mu(I)$ rheology}

The $\mu(I)$ or inertial rheology is a phenomenological model that describes the steady flow of a quasi-monodisperse granular mixture made of small, hard, spherical particles under a finite pressure and shear rate \citep{Jop_2006_constitutive}.  It predicts a one-to-one relationship between coefficient of friction ($\mu$, the ratio between shear to normal stresses acting on the granular material), and inertial number ($I$, a normalised shear rate), as shown in Eq. (\ref{eq:mu_of_I}) and (\ref{eq:inertial_number}).  The formulation of the $\mu(I)$ model renders itself  suitable to be added to numerical methods for fluid flow, and many have been adapted to incorporate non-Newtonian granular viscosity \citep{Lagree_2011_granulacollapse}. 

\begin{equation}
\mu  = \mu_1 + \frac{\mu_2 - \mu_1}{1+I_o/I},
    \label{eq:mu_of_I}
\end{equation}where $\mu_1$ and $\mu_2$ are the asymptotic values of $\mu(I)$ as $I\rightarrow 0$ and $I\rightarrow\infty$, respectively. Note that $\mu_1$ corresponds the value of Coulomb friction. 
\begin{equation}
I = \frac{\dot{\gamma}d}{\sqrt{P_s/\rho_s}},
\label{eq:inertial_number}
\end{equation}where $\dot{\gamma}$ is a constant shear rate, $d$ is the characteristic grain size, $P_s$ is isotropic pressure, and $\rho_s$ is the particle density. 

The $\mu(I)$ model is a local rheology, whereby the stress at a given point depends only on the pressure and shear rate at that point. This implies that the $\mu(I)$ model has no length scale, despite describing systems that clearly have a finite scale dictated by the grain size. In fact, \cite{Barker_2015_illposed} showed that fluid flow  equations that incorporate $\mu(I)$ are ill posed when the inertial number is too high or too low, due to this lack of internal length scale. The ill-posedness results in unbounded growth rates for high-wavenumber perturbations. Consequently, the numerical solution depends inherently on the grid resolution. To address this, \cite{Barker_2017_partial} proposes a partial regularisation of $\mu(I)$ to force  hyperbolicity for all values of $I$, without having to add a length scale to the model. 

\begin{figure}[htbp]
\centering
\includegraphics[width=0.6\textwidth]{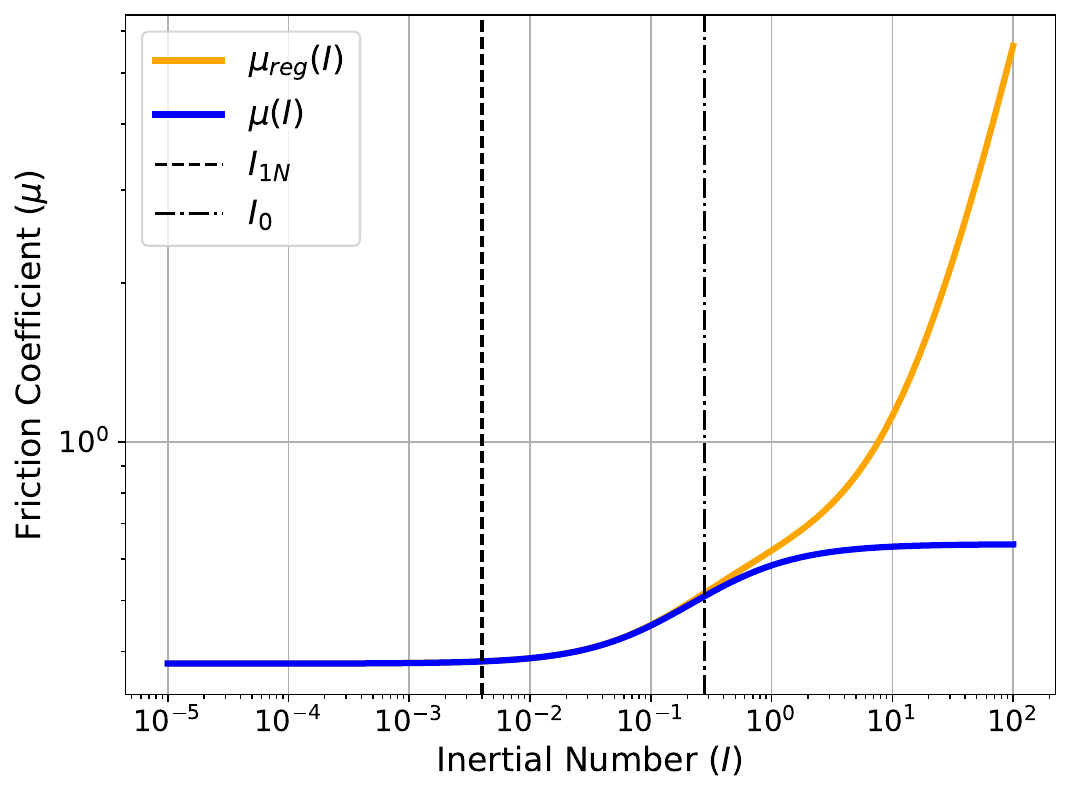}
\caption{The classical $\mu(I)$ model is shown in blue \citep{Jop_2006_constitutive}. The partially regularised $\mu(I)$ proposed by \cite{Barker_2017_partial}, only for $I>I^N_1$, is shown in orange.  $I^N_1$ is the stability neutral number and $I_o$ is the characteristic inertial number. }
\label{fig:regularise_mu}
\end{figure}

We implement the partial regularisation of $\mu(I)$ only  for large values of $I$ (i.e.,  $I>I^N_1$ in Eq. (6.3) of \cite{Barker_2017_partial}, where $I^N_1$ is the stability neutral number), and the standard $\mu (I)$ for $I<I^N_1$, as described in Eq. \eqref{eq:reg_mu_of_I}.
\begin{equation}
\mu_{reg} (I) = \frac{\mu_1I_o + \mu_2I+\mu_{\infty}I^2}{I_o+I},
    \label{eq:reg_mu_of_I}
\end{equation} 
This formulation accounts for the transition in dissipation mechanisms: while at lower values of $I$, energy dissipation is driven predominantly by shearing friction, at intermediate values of $I$, it occurs through binary inelastic collisions, and is captured by $\mu_{\infty}$. Thus, Eq. (\ref{eq:reg_mu_of_I}) ensures that, as $I\rightarrow\infty$,  $\mu(I)$ scales linearly with $I$ ($\mu(I)\rightarrow \mu_{\infty}I$), rather than converging to the constant $\mu_2$ (Fig. \ref{fig:regularise_mu}). Yet, in the limit where dissipation through binary collisions is negligible (i.e., $\mu_{\infty}=0$), Eq. (\ref{eq:reg_mu_of_I}) collapses to the standard $\mu(I)$ form of Eq. (\ref{eq:mu_of_I}).

\subsection{Empirical relation accounting for the reduced diffusivity of thin flows due to compaction limit}

As $\alpha \rightarrow \alpha_{max}$, the $f_{inhibit}$ hinders degassing-driven compaction,  thus retaining pore pressure within the system (Eq. \ref{eq:finhibit}).  The transition to full inhibition is governed by a smooth-step function starting at $\alpha_{tr}$, an empirically obtained parameter (Fig.  \ref{fig:generalised_smoothstep}). Thus, we propose that for optimised values of $\alpha_{tr}$, IMEX can effectively reproduce the MFIX-TFM pore-pressure diffusion curves derived from defluidising static granular columns. To test this, we conducted a grid search for $\alpha_{tr}$ across equivalent initial column heights, setting $N=5$ in  $f_{inhibit}$ and $k\propto f(d,\varepsilon)$ as per (\ref{eq:carman_kozeny}). The grid search was carried out with the quasi-random Satelli sampling method for a  $\alpha_{tr} \in [0.30, 0.60]$, with the goodness of fit evaluated with the residual sum of squares (RSS).

\begin{figure}[htbp]
\centering
\includegraphics[width=1\textwidth]{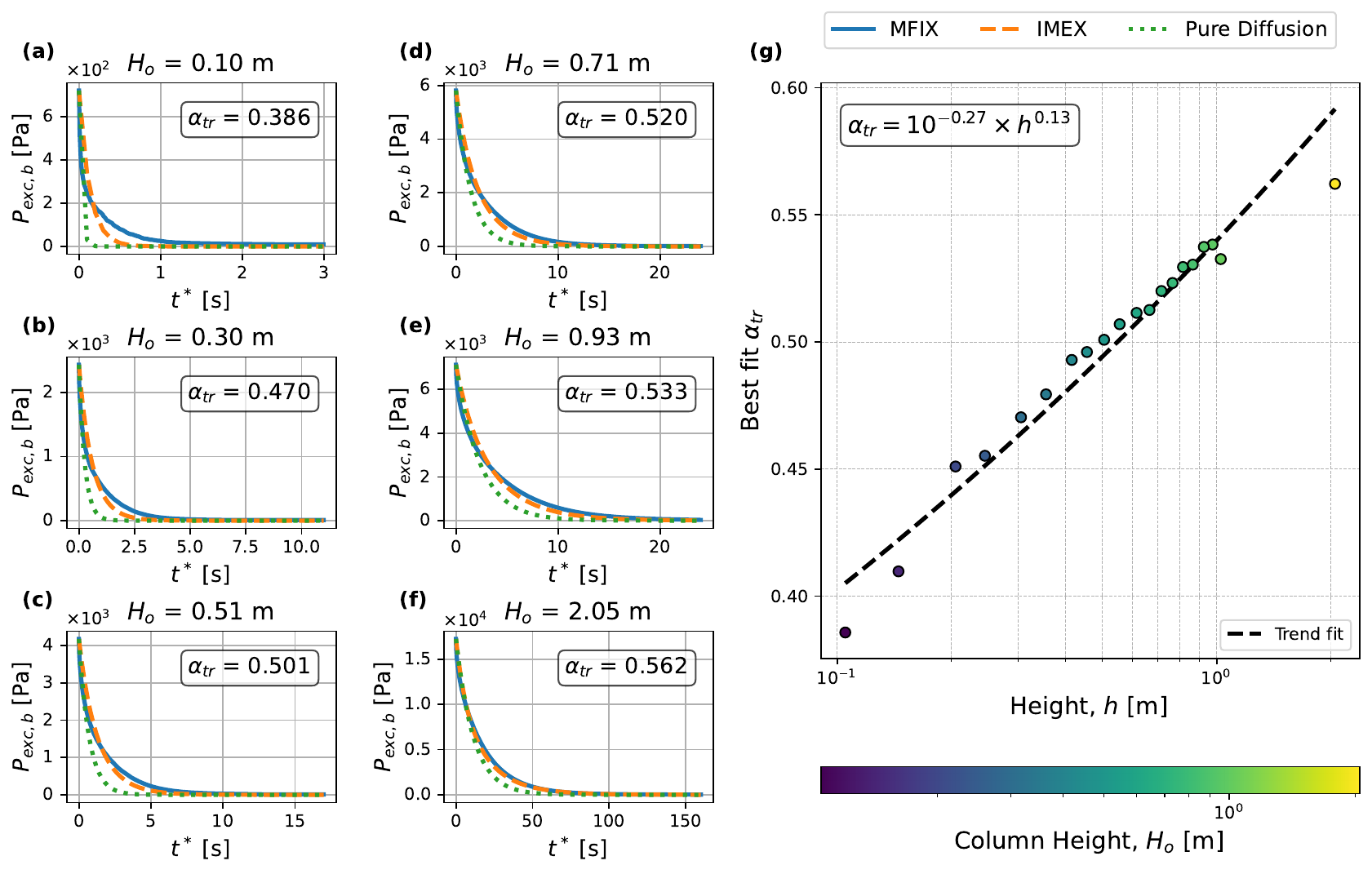}
\caption{(a-f) Diffusion of basal excess pore pressure ($P_{exc,b}$) in a static fluidised column of various heights ($H_o$) with time, where $t^*=0$ is  the time at which the upward supply of air stops. The diffusion yielded by MFIX-TFM simulations  is shown in blue. The green dotted line corresponds to the first mode of the diffusion (only) analytical solution (Eq. (\ref{eq:modal-solution-diffusion})). The diffusion evolution curve yielded from IMEX, accounting for inhibition as the solid fraction approaches maximum packing, is shown in the orange dash line. The $\alpha_{tr}$, as per Eq. (\ref{eq:smoothstep}), to obtain a diffusion curve with the lowest residual sum of squares with respect to the MFIX-TFM diffusion curve is printed in the upper right corner of the subplots. (g) Best-fitting $\alpha_{tr}$ for a range of initial static column heights (dots coloured by $H_o$), alongside the corresponding best-fitting power-law trend (black dashed line), in turn, printed on the upper left corner of the subplot.}
\label{fig:empirical_alpha_height}
\end{figure}

We find that the best-fit values of $\alpha_{tr}$ increase with initial flow height following a power-law trend (Fig. \ref{fig:empirical_alpha_height}.g). Within the present closure, reproducing the MFIX-TFM pressure-decay curves requires a stronger inhibition of degassing in thinner flows at a given solid fraction near $\alpha_{\max}$. This result is consistent with the source-to-diffusion ratio, $\Psi$, as thinner flows have compaction timescales that dominate over diffusion timescales (Fig. \ref{fig:fourpanel}). Although pressure diffusion curves in thinner flows show larger misfit to the MFIX-TFM data compared to thicker flows, they still clearly show hindered diffusion, compared to the diffusion-only model (Fig. \ref{fig:empirical_alpha_height}.(a-f)). 

\begin{figure}[htbp]
\centering
\includegraphics[width=1\textwidth]{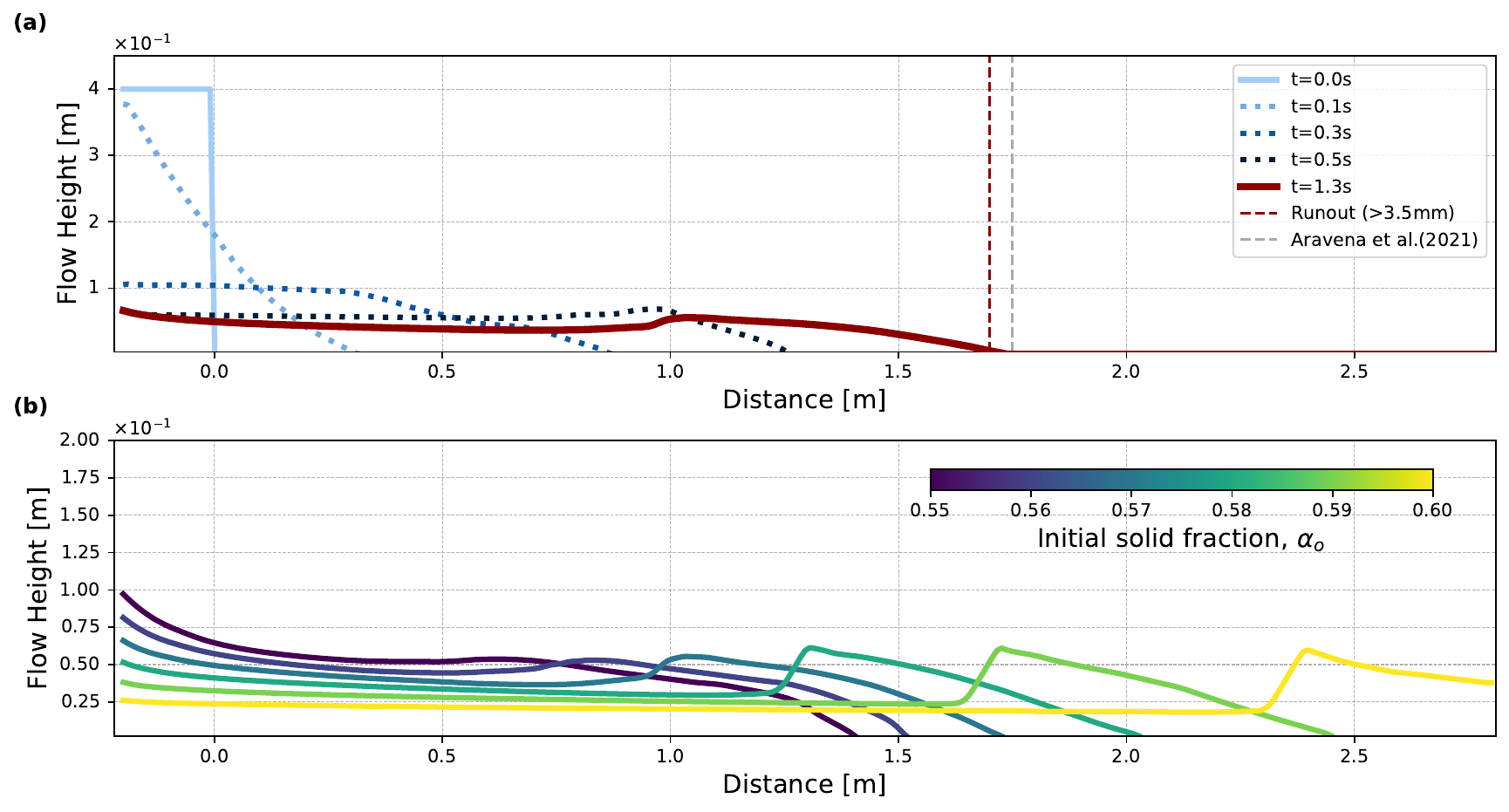}
\caption{IMEX simulations reproducing the dam break experiment presented in \cite{Roche_2010_pore,Roche_2012_depositional,Aravena_2021_influence} for an initial column height of 0.40\,m. These simulations compute $f_{inhibit}$ using the empirical relation $\alpha_{tr}(h)$ shown in Figure \ref{fig:empirical_alpha_height}.g. (a) Flow profiles at different times from the dam break for an initial solid fraction of the granular column of 0.57. The runout distances of the flow of thickness greater than 3.5\,mm yielded by IMEX  and that reported by \cite{Aravena_2021_influence}  for an initial solid fraction of 0.58 are marked by the red and grey dashed lines, respectively. (b) Final flow profiles corresponding to initial granular columns of 0.4\,m height with different solid fractions ($\alpha_o$).}
\label{fig:dambreak_profiles}
\end{figure}

The empirical trend $\alpha_{tr}(h)$ in Fig. \ref{fig:empirical_alpha_height}.g was incorporated in the inhibit factor to implement a flow height dependence in $f_{inhibit}$. We tested the performance of this implementation by reproducing the the dam break experiment presented in \cite{Roche_2010_pore,Roche_2012_depositional,Aravena_2021_influence}. For an initial column height of 0.40\,m, and initial volume fraction of 0.57 (instead of 0.58 in \cite{Aravena_2021_influence}), we are able to match the runout reported by \cite{Aravena_2021_influence}, underestimating it by $\sim 3\%$. We find that the calculated runout distances of the flow ($h>3.5$\,mm)  are very sensitive to the solid volume fraction of the initial granular column (Fig. \ref{fig:dambreak_profiles}). This suggests that the empirical law found in Fig. \ref{fig:empirical_alpha_height}.g. implicitly encompasses a more complex interaction of several other flow parameters.

\subsection{Summary of the Presented Functionalities of IMEX\_SfloW2D\_v2 }
A set of simulations of the dam break experiment in \cite{Aravena_2021_influence} are provided to showcase the effect on flow dynamics of a range of features available in IMEX related to the theory discussed above (Table \ref{tab:demo_models}). 

Simulations I and II do not include interstitial pore pressure, serving as a baseline comparison, and highlighting the difference in final flow profile arising solely from the choice of rheological model (Voellmy-Salm and $\mu(I)$, respectively). The succeeding simulations progressively incorporate various pore pressure–related mechanisms. Simulation III illustrates the introduction of excess pore pressure in the initial granular column, keeping a constant permeability, and an inhibit factor that transitions to complete hindered gas loss at $\alpha_{max}$ as a step-function. Building on this configuration, simulation IV, enables permeability to vary dynamically with porosity, while simulation V models a smooth transition of the inhibit factor at $\alpha_{tr}<\alpha_{max}$. Simulations VI and VII, model a smooth transition of $f_{inhinbit}$ starting at a $\alpha_{tr}$ that depends on flow height according to the empirical formulation in Figure \ref{fig:empirical_alpha_height}.g. Simulation VII additionally incorporates the dynamic variation of permeability with porosity.  
\begin{table}
    \centering
\caption{Dam break simulations presenting different parameter values of the available features in IMEX related to pore pressure dissipation in thin flows. The third column specifies if the initial column has excess pore pressure. The fourth column specifies if the permeability is kept constant (Const.) or varies as a function of porosity (Dyn.). The fifth column indicates what inhibit factor modality is input to the software: step function ('STEP'), smooth function with fixed transition beginning ('STATIC'), and smooth function with transition beginning depending on flow height ('DYNAMIC'). }
\label{tab:demo_models}
\begin{tabular}{ccccc}\toprule
          Simulation&Rheology&  Pore pressure&  Permeability& Inhibit Factor\\\midrule
          I&Voellmy-Salm&  No&  Const.& N/A\\
          II&$\mu (I)$ &  No&  Const.& N/A\\
          III&$\mu (I)$&  Yes&  Const.& 'STEP'\\
          IV&$\mu (I)$&  Yes&  Dyn.& 'STEP'\\
          V&$\mu (I)$&  Yes&  Const.& 'STATIC'\\
          VI&$\mu (I)$&  Yes&  Const.& 'DYNAMIC'\\
          VII&$\mu (I)$&  Yes&  Dyn.& 'DYNAMIC'\\ \bottomrule 
    \end{tabular}
\end{table}

\begin{figure}[htbp]
\centering
\includegraphics[width=1\textwidth]{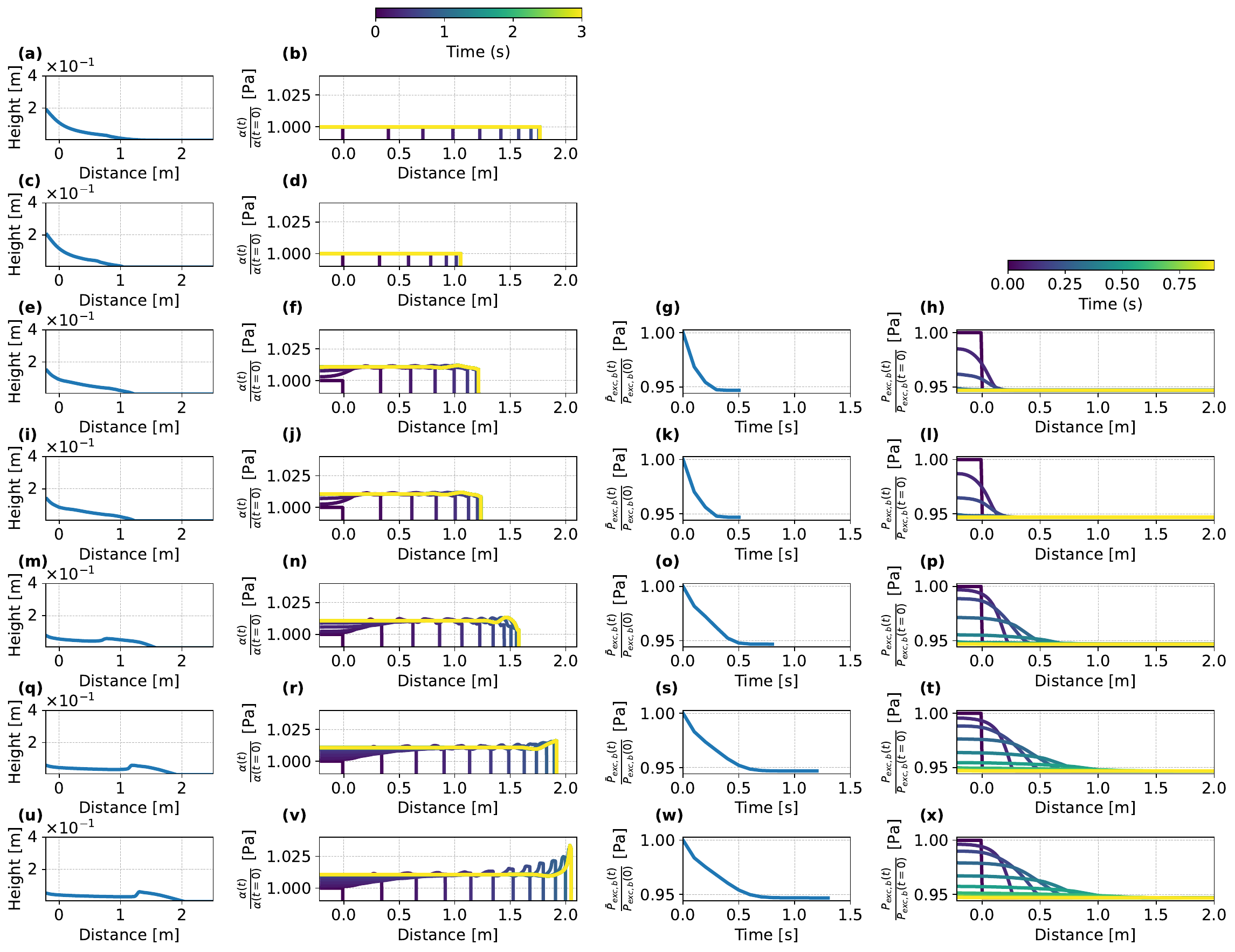}
\caption{Simulations to showcase the effect on flow dynamics of a range of features available in IMEX\_SfloW2D\_v2 related to pore pressure dissipation in thin flows. Each row corresponds to a simulation in Table \ref{tab:demo_models}. The first column shows the flow profile at the last time step. The second column presents the relative solid phase concentration along the flow profile, with respect to the initial granular column $\left( \frac{\alpha(t) }{\alpha (t=0)} \right)$ , for the first 3\,s of simulation. The third column depicts the evolution in time of the average basal excess pore pressure in the flow, with respect to its value at $t=0\,$s $\left( \frac{\bar{P}_{exc,b}(t)}{P_{exc,b}(0)} \right)$. The fourth column presents the basal excess pore pressure along the flow profile, normalized by its value at $t=0\,$s, for the first second of simulation  $\left(\frac{P_{exc,b}(t)}{P_{exc,b}(t=0)} \right)$. Note that simulations I and II in Table \ref{tab:demo_models}do not have excess pore pressure, and thus the corresponding third and fourth subplots here cannot be plotted. }
\label{fig:dambreak_demo_profiles}
\end{figure}

When comparing simulations I and II, we see that by using the $\mu(I)$ rheology the flow attains shorter runout distances (Fig. \ref{fig:dambreak_demo_profiles}.a-d). Although both rheologies yield flow profiles with very similar concave profiles, the Voellmy-Salm rheology produces a longer and thinner flow front. Adding pore pressure to the initial granular column (simulation III) increases the runout and solid concentration throughout the flow with time due to degassing, and produces a less concentrated flow front (Fig. \ref{fig:dambreak_demo_profiles}.e-h). 

Allowing for variable permeability ($k=f(\phi)$, simulation IV), while using a step function degassing inhibition, results in a marginal increase in flow runout distance, with no significant impact on concentration and pressure dissipation timescales (Fig. \ref{fig:dambreak_demo_profiles}.i-l). By contrast, using constant permeability ($k=const.$) with a smooth degassing inhibition function ($\alpha_{tr}=0.45$, simulation V), alters the flow dynamics: the flow profile evolves into a combined concave and convex geometry, and both pressure dissipation timescales and runout distance are extended (Fig. \ref{fig:dambreak_demo_profiles}.m,o,p). Finally, the solid concentration profiles now show a localised band of higher solid fraction behind the flow front (Fig. \ref{fig:dambreak_demo_profiles}.n). 

Using a degassing inhibition dependent on flow height ($f_{inhibit}=f(h)$, simulation VI),  extends dissipation timescales and runout distances further, and results in a gradual increase in solid concentration towards the flow front (Fig. \ref{fig:dambreak_demo_profiles}.q-t). The effect of dynamic permeability is more visible when the inhibit factor's $\alpha_{tr}$ is a function of flow height (simulation VII), leading to both longer runouts, and a sharpening of the solid-fraction profiles towards the flow front (Fig. \ref{fig:dambreak_demo_profiles}.u-x).

To summarise, the use of $\mu(I)$-rheology, instead of Voellmy-Salm, decreases the runout distance and produces a flow with a similar shape but shorter flow front. Introducing excess pore pressure overall enhances runouts, and modifies the concentration evolution and distribution. Fluidisation is sustained for longer periods when employing smooth $f_{inhibit}$ functions that inhibit gas escape at $\alpha < \alpha_{max}$, and results in localised bands of high solid fraction behind the flow front.  Implementing a dynamic permeability has a higher effect on pressure dissipation when used in conjunction of gas escape inhibition mechanisms, implying that the inhibit factor function exerts a higher dominance than permeability changes on flow dynamics. 

\section{Discussion and implications for geophysical flows}
\label{sec:discussion}

\subsection{Thin versus thick beds: reconciling column and dam-break experiments}

A long-standing outcome of laboratory work on fluidised granular columns and dam-break flows is that thin beds drain more slowly, and remain mobile for longer, than standard diffusion theory would suggest. Column-defluidisation experiments and dam-break runs by \citet{Roche_2010_pore}, \citet{Montserrat2012_pore}, and \citet{Roche_2012_depositional} all show that basal pore pressure in relatively thin columns decays more slowly than predicted from $D_{\text{pred}} = k/(\mu_g\beta\varepsilon)$ using Carman--Kozeny permeabilities, whereas thick columns approach the classical diffusion limit. The dam-break simulations of \citet{Aravena_2021_influence} reach similar conclusions: for a given material, beds with smaller initial height retain higher pore pressures over a larger fraction of the runout, and these ``overpressured'' thin flows exhibit anomalously long travel distances compared with thick columns. Building on this, the combined experiments and 2-D incompressible simulations of collapsing, initially fluidised columns with different aspect ratios by \citet{Aravena_2024_runout} show that the effective pore-pressure diffusion coefficient inferred from depth-averaged models with constant $D$ increases with initial column height and can be approximated as a thickness-weighted average of the flow depth during propagation. Interpreted through the present framework, that apparent dependence of $D$ on column geometry is precisely what one would expect if compaction-induced sources and height-dependent permeability are neglected in the closure and absorbed into an effective diffusivity instead.

Our MFIX-TFM two-fluid flow simulations and the diffusion--compaction analysis developed here provide a unified explanation for these trends. In all cases, the diffusivity $D_{\text{fit}}$ inferred from the basal pressure decay is systematically lower than the theoretical Darcy diffusivity $D_{\text{pred}}$, with the departure being largest in thin beds. The single-mode basal ODE~\eqref{eq:basal-ode-updated} shows that this discrepancy is not an artefact of the fitting procedure but a real effect of compaction: as the granular skeleton contracts during drainage, the associated volumetric source term continuously recharges the pore pressure, slowing its relaxation relative to the purely diffusive case. The thinner the bed, the more important this effect becomes, because the drainage time $T_{\mathrm{diff}}\propto H^{2}$ decreases while the compaction timescale $T_{c}$ set by $\dot{\varepsilon}$ remains of order seconds. 

The source--to--diffusion ratio $\Psi_0$ defined in Eq.~\eqref{eq:Psi0_final} provides a convenient way to quantify this competition. Thin beds have small $\Pi_H=\beta_0(1-\varepsilon_0)\rho_s g H_0$ and therefore large $\Psi_0$, whereas thick beds have large $\Pi_H$ and $\Psi_0\to0$. The algebraic correction $D_\Psi/D_{\text{pred}}=1/(1+1.5637\,\Psi_0^{1.0392})$ collapses all MFIX-derived diffusivities across nearly two orders of magnitude in $H_0$, including the thick-bed limit where $D_{\text{fit}}\approx D_{\text{pred}}$. In this view, the apparently distinct behaviours of ``thin'' and ``thick'' flows in the experiments of \citet{Roche_2010_pore,Montserrat2012_pore,Roche_2012_depositional,Aravena_2021_influence} are two ends of a continuous spectrum controlled by $\Psi_0$: thin, strongly compacting beds inhabit the compaction-dominated regime ($\Psi_0\gtrsim1$), whereas thick beds lie in the diffusion-dominated domain ($\Psi_0\ll1$).

This interpretation is consistent with the internal kinematics documented by \citet{Roche_2010_pore} in air--particle gravity currents. In those experiments, thin, initially fluidised columns produce highly mobile currents whose interiors remain weakly supported by gas overpressures, whereas thick columns rapidly lose mobility as pore pressure diffuses away. Our decomposition of the basal pressure into a sum of diffusion and compaction contributions (Fig.~\ref{fig:fourpanel}) shows that the same mechanism operates in the MFIX-TFM columns: in thin beds, the compaction term contributes a large fraction of the basal signal, whereas in thick beds it is negligible and the pressure is governed entirely by diffusion.

\begin{figure}[htbp]
    \centering
    \includegraphics[width=0.7\linewidth]{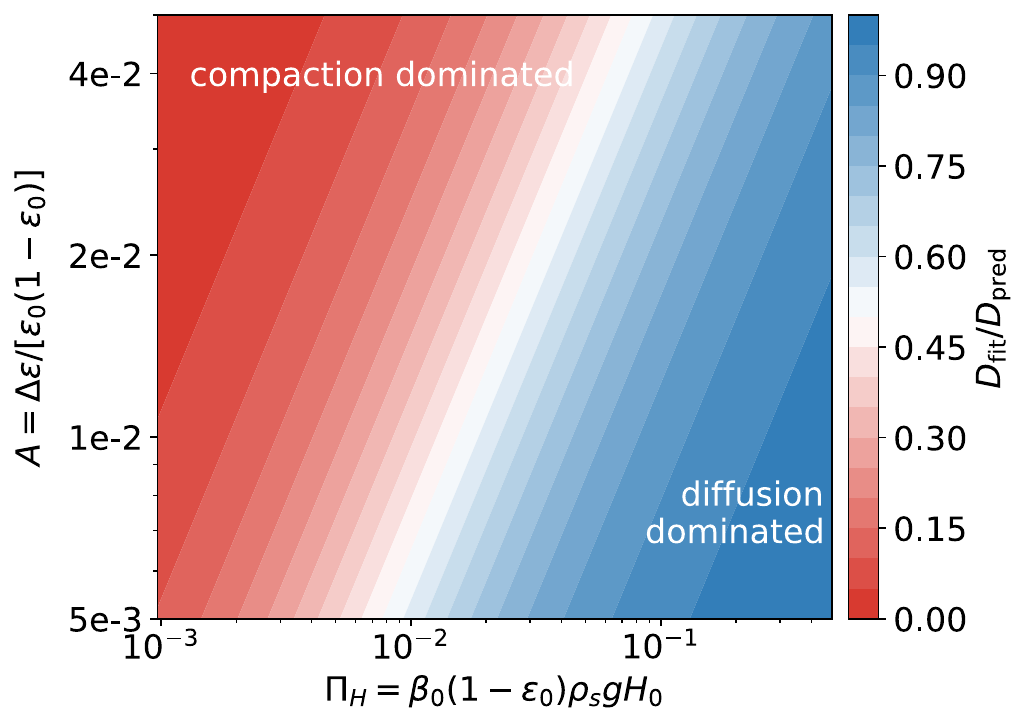}
    \caption{%
    Contour map of the normalized diffusivity ratio $D_{\mathrm{fit}}/D_{\mathrm{pred}}$ for gas--particle beds in air, showing how compaction modifies the effective drainage rate.
    The horizontal axis $\Pi_H = \beta_0(1-\varepsilon_0)\rho_s g H_0$ is a dimensionless measure of bed thickness that captures the pressure loading relative to gas compressibility.
    The vertical axis $A = \Delta\varepsilon/[\varepsilon_0(1-\varepsilon_0)]$ represents the magnitude of porosity change relative to the available pore volume.
    Blue regions correspond to diffusion-dominated conditions where $D_{\mathrm{fit}}\!\approx\!D_{\mathrm{pred}}$, while red zones denote compaction-dominated regimes where drainage is increasingly inhibited.}
    \label{fig:Dfit_Dpred_Psi0_map_air}
\end{figure}

Beds with large $\Pi_H$ (thick, heavily loaded columns) and modest compaction amplitudes $A$ cluster in a blue region where $D_{\mathrm{fit}}/D_{\mathrm{pred}}\approx1$ and the classical rigid-medium approximation is valid (Fig.~\ref{fig:Dfit_Dpred_Psi0_map_air}). Thin beds or columns with large $A$ fall in the red region, where $D_{\mathrm{fit}}$ can be reduced by more than an order of magnitude relative to $D_{\mathrm{pred}}$. The column experiments of \citet{Montserrat2012_pore} and \citet{Breard2019b} occupy precisely this domain: modest but finite compaction amplitudes coupled with small $H_0$ yield $\Psi_0\gtrsim1$ and strongly inhibited drainage. 

Recent experiments by Chupin et al. (2025) indicate that the minimum fluidization velocity ($U_{\mathrm{mf}}$) increases by ~50\% as the bed height rises from 0.2 m to 2 m, despite showing no systematic variation in solids concentration at different degrees of fluidization. All measurements yielded a concentration of 0.54, which is notably lower than their independently determined random close packing value of 0.60. This inconsistency is difficult to reconcile. It implies either unexpectedly strong Janssen-type stress redirection from the cylindrical wall, despite a column diameter more than 1700 particle diameters, or the presence of an unrecognized mechanism that increases drag within the bed. Our MFIX-TFM simulations do not reproduce this behavior; for otherwise comparable conditions, we find that $U_{\mathrm{mf}}$ increases by less than ~10\% when bed thickness increases from 0.2 m to 2 m. Chupin et al. also used a modified Darcy relation to back-calculate solids concentration from defluidization data, which suggests that the concentration decreases from 0.54 at 2 m to roughly 0.4 at 0.2 m, contradicting their direct observation of minimal bed expansion. Taken together, several independent indicators point to non-negligible compaction within the bed as a plausible contributor to many of the reported trends, although compaction alone does not readily explain the mildly non-linear pressure profile measured with bed height. Additional experiments explicitly designed to quantify pore-pressure feedback and its coupling to particle compaction would be particularly valuable for resolving these inconsistencies.
\subsection{Implications for depth-averaged models and IMEX implementations}

Most depth-averaged models used for granular and debris flows retain only a highly simplified representation of pore-pressure effects, typically through a depth-averaged effective friction coefficient that may depend on Froude number or thickness, but not explicitly on the pore-pressure state \citep[e.g.][]{savage1989motion,zhu2020high,xia2018new}. For volcanic flows, many widely used depth-averaged codes treat the mobility of pyroclastic density currents (PDCs) through prescribed basal friction or basal shear stress, sometimes supplemented by heuristic fluidisation factors, but still without a prognostic equation for pore pressure. This approach can reproduce observed runouts only by tuning friction coefficients over wide, case-dependent ranges, and it can easily lead to unphysical behaviour such as nearly inviscid acceleration once friction drops below a threshold, or sudden, grid-dependent arrest when friction is increased.

A smaller set of models incorporates an explicit evolution equation for excess pore pressure. In debris-flow mechanics, \citet{iverson1997physics} and \citet{denlinger2001flow} coupled depth-averaged mass and momentum equations with a Darcy-type pore-pressure diffusion equation, demonstrating that the competition between viscous drainage and material compaction can explain much of the observed variability in runout and deposition. Large-scale flume experiments \citep{iverson2010perfect} confirmed that sustained high pore pressures are required to maintain long, low-friction runouts, and that the loss of fluidisation is controlled by drainage and consolidation dynamics rather than by velocity alone. In the volcanic context, \citet{gueugneau2017effects} extended a Savage--Hutter-type PDC model by adding a depth-averaged pore-pressure equation, calibrated against air--particle experiments inspired by \citet{Roche_2010_pore} and \citet{Montserrat2012_pore}. Their results showed that including pore-pressure evolution greatly improves the ability of depth-averaged models to reproduce both runout and deposit thickness.

The IMEX\_SfloW2D\_v2 model of \citet{de2023imex_sflow2d}, which extends the earlier IMEX formulation of \citet{IMEX2019}, advances this line of work by solving a prognostic equation for the basal excess pore pressure, $P_{b,\mathrm{exc}}$, coupled to the depth-averaged mass and momentum equations. In its original formulation, the dissipation term is obtained from a first-mode approximation of a diffusion equation with fixed diffusivity, while the permeability is updated dynamically as a function of porosity via a Carman--Kozeny relation. Our contribution is to link this depth-averaged pore-pressure closure directly to the physics of diffusion--compaction inferred from MFIX-TFM and column experiments, and to provide an \emph{a priori} correction to the diffusivity that depends only on flow thickness, porosity, and initial pressure. A key strength of IMEX is its numerical scheme, which does not require empirical stopping criteria, in contrast to other depth-averaged models (e.g.\ VolcFlow, TITAN2D) commonly used to simulate geophysical mass flows \citep{NEGLIA2021107146,Karim2011}.

In IMEX, the effective drainage of pore pressure is controlled by a Darcy-type diffusivity based on Carman--Kozeny permeability together with the inhibition factor $f_{\mathrm{inhibit}}(\alpha_s)$, which suppresses gas loss as the solid fraction approaches random close packing. Rather than introducing the correction from Secs.~V and VI into IMEX as an explicit modified diffusivity, we use those results to identify the local processes that the depth-averaged closure must reproduce, namely the interplay between drainage, compaction, and porosity-dependent permeability. The inhibition factor represents the additional hindering of gas escape as $\alpha_s \to \alpha_{\max}$, effectively shutting down pore-pressure dissipation in densely packed regions. By calibrating the transition parameter $\alpha_{tr}(h)$ against the MFIX-TFM defluidisation curves for static columns of different heights (Fig.~\ref{fig:empirical_alpha_height}), we obtain an empirical relation that strengthens inhibition in thin flows and weakens it in thick ones—exactly the trend suggested by the diffusion--compaction analysis and by the experiments and simulations of \citet{Roche_2010_pore,Montserrat2012_pore,Roche_2012_depositional,Aravena_2021_influence,Aravena_2024_runout}. In this way, the MFIX-TFM and column results guide the choice of permeability and inhibition parameters in IMEX, without requiring an explicit prescription of a corrected diffusivity in the code.

The dam-break simulations in Fig.~\ref{fig:dambreak_profiles} illustrate the practical impact of this coupling. When the empirical $\alpha_{tr}(h)$ relation and $f_{\mathrm{inhibit}}$ are used, IMEX reproduces both the magnitude and the timing of basal pressure decay inferred from MFIX-TFM and the dam-break experiments of \citet{Aravena_2021_influence}, including the strong sensitivity of runout distance to the initial solid volume fraction. Flows that start slightly more compact rapidly lose pore pressure, lock up, and come to rest over relatively short distances, whereas more dilute initial conditions maintain high pore pressures over a significant portion of the runout and travel much farther. In a purely friction-based model, these contrasting behaviours would have to be imposed through large, ad-hoc changes in basal friction; with the present pore-pressure closure, they emerge naturally from the interplay between diffusion, compaction, and permeability reduction.

In geophysical applications, this is particularly important because the effective basal friction inferred from volcanic and non-volcanic mass flows can vary greatly between events \citet{zuccarello2025trigger, zhu2020high, Ogburn2017}. The present framework suggests that part of this variability reflects differences in the local competition between pore-pressure diffusion and compaction, which depends on evolving flow thickness, packing, permeability, and excess pore pressure. In the idealized column problem this competition is conveniently summarized by $\Psi_0$, while in depth-averaged models it is represented through a state-dependent inhibition of degassing. Incorporating such physics-based state dependence into hazard models offers a route to reducing the arbitrariness of friction calibration and to linking model parameters more closely to measurable physical properties of the flow and substrate.

\section{Future work}
\label{sec:future_work}

Several extensions of this work follow naturally from the present results.

First, the diffusion--compaction framework has so far been tested for homogeneous, Geldart~A-type beds of nearly monodisperse glass beads in air. Many natural and industrial systems involve strongly polydisperse, angular, or cohesive particles, and pore fluids with very different properties (e.g.\ steam, water, or gas mixtures). Extending the analysis to polydisperse and stratified beds---where permeability, compressibility, and compaction rates may vary with depth---will require coupling the present single-mode reduction to models of vertical segregation and multi-layer diffusion. In such settings, the effective diffusivity may be controlled by a combination of local $\Psi(z)$ and spatially varying permeability, rather than by a single bulk $\Psi_0$.

Second, the present closure assumes Darcy flow with Carman--Kozeny permeability, which is appropriate for the laminar, pre-bubbling regimes considered here. In high-speed, dilute, or highly turbulent flows, non-Darcy effects, inertial drag, and gas compressibility may all become important. Systematic MFIX-TFM or DNS campaigns spanning higher Reynolds, and incorporating realistic equations of state for hot volcanic gases, would allow us to assess how the effective diffusivity and the inhibition factor generalise beyond the current parameter space. 

Third, many granular geophysical flows evolve over complex three-dimensional topographies and interact with erodible or water-saturated substrates. Coupling the present pore-pressure closure to depth-averaged models that include bed erosion, sediment entrainment, and basal liquefaction \citep[e.g.][]{iverson1997physics,denlinger2001flow,xia2018new} is a promising avenue for future work. In such models, the excess pore pressure at the base of the mobile layer would be influenced not only by internal compaction but also by fluid exchange with the substrate, potentially leading to feedbacks between erosion, permeability, and mobility.

Fourth, there is a clear need for laboratory experiments that directly measure compaction and dilatation during gas drainage or rapid deformation. High-speed imaging, X-ray tomography, refractive-index-matched particle tracking, and bed-level sensors (i.e. capacitance sensors) could all provide transient porosity fields with the temporal resolution needed to characterise $\dot{\varepsilon}(t)$ in thin and thick beds. Such measurements would make it possible to build empirical, thickness-dependent compaction laws analogous to those inferred here from MFIX. These laws could then be implemented in IMEX to capture the enhanced influence of compaction in thin flows without relying solely on numerical calibration, thereby strengthening the physical basis of depth-averaged pore-pressure closures.

Finally, the single-mode basal ODE and the diagnostic quantities $\Psi(t)$ and $\mathrm{De}(t)$ provide compact tools for interpreting laboratory and field observations of pore-pressure evolution. Future work could exploit this structure to invert measured basal pressure and thickness histories for effective diffusivities, compaction timescales, or permeability changes, turning pore-pressure records into quantitative constraints on the evolving internal state of geophysical flows. Combining such inversions with high-resolution imaging, seismic, or acoustic observations may help bridge the gap between bench-scale experiments, numerical simulations, and geophysical mass flows in the field.

\section{Conclusion}

In this work, we have combined two-phase flow theory, high-resolution two-fluid simulations, and reduced-order modeling to clarify how diffusion and compaction jointly control pore-pressure dynamics in gas–particle beds and how these effects inform depth-averaged models of geophysical flows. Starting from general mass conservation for a deformable granular skeleton and a compressible pore fluid, we derived a governing equation for excess pore pressure that explicitly couples Darcy-type diffusion, gas compressibility, and the volumetric strain rate of the solid phase. In the thin-flow, small-excess-pressure limit this reduces to a one-dimensional diffusion–compaction equation, which we solved numerically for static, initially fluidized columns and validated against MFIX-TFM simulations over a wide range of bed heights.

A Neumann–Dirichlet modal decomposition shows that the basal pressure can be described by a single ordinary differential equation in which exponential drainage competes with a source term proportional to the porosity rate. This leads naturally to diffusive and compaction timescales and to a source-to-diffusion ratio, $\Psi_0$, that quantifies the strength of compaction forcing relative to drainage. Using MFIX-TFM pressure–time series, we showed that the effective diffusivity inferred from basal decay is systematically smaller than the Darcy–Carman–Kozeny prediction whenever the bed compacts during drainage, with the largest departures in thin columns. A simple algebraic correction, $D_\Psi/D_{\mathrm{pred}} = [1+1.5637\,\Psi_0^{1.0392}]^{-1}$, collapses all fitted diffusivities across nearly two orders of magnitude in bed thickness with a mean error of $3.7\%$.

We then used these results to inform a depth-averaged pore-pressure closure in the IMEX model. The closure consists of a prognostic equation for basal excess pore pressure, a permeability-based diffusivity, and an inhibition factor that suppresses gas loss as the solids fraction approaches maximum packing. Calibrating the inhibition transition against MFIX-TFM defluidization curves yields a thickness-dependent relation that reproduces the stronger drainage inhibition in thin beds. Dam-break simulations demonstrate that this implementation captures both the basal pressure evolution and the strong sensitivity of runout to initial packing documented in laboratory experiments, without resorting to ad-hoc friction tuning.

There is now an opportunity to pair these modeling advances with targeted laboratory measurements of transient compaction and dilatation. Experiments that resolve porosity evolution in space and time would provide direct constraints on $\dot{\varepsilon}(t)$ under controlled drainage and deformation, enabling empirical compaction laws to be derived in the same way that MFIX-TFM informed the inhibition factor. Embedding such experimentally derived compaction closures in IMEX would provide a more robust description of thin-flow behaviour and reduce the reliance on numerical calibration alone.

Taken together, these results provide a physically grounded pathway to represent pore-pressure controlled mobility in depth-averaged models of debris flows, pyroclastic density currents, and other geophysical granular flows. The diffusion–compaction framework, the source metric $\Psi_0$, and the associated diffusivity correction offer practical diagnostics for interpreting experiments and simulations, and a basis for future extensions to polydisperse materials, non-Darcy regimes, and flows interacting with erodible or saturated substrates.

\begin{acknowledgments}
E.C.P.B. was supported by the NERC-IRF (NE/V014242/1), the Leverhulme Trust grant award RPG-2024-294 and the Royal Society (IEC/NSFC/242381) and acknowledges the use of ARCHER2 High Performance Computing. C.E.P's work was supported by the NERC DTP E4.

\textit{Code availability.} The depth-averaged solver IMEX\_SfloW2D\_v2 is open-source and available at \url{https://github.com/demichie/IMEX_SfloW2D_v2}.
The MFIX multiphase solver is available at \url{https://mfix.netl.doe.gov/products/mfix/}.

\textit{Data availability.} The MFIX-TFM simulation data and 1D solver scripts supporting this study are available from the corresponding author upon reasonable request.

\end{acknowledgments}
\appendix

\section{\label{app:symbols}Symbols and Nomenclature}

{\footnotesize

\tablefirsthead{%
\topcaption{Symbols, names, and units used in the manuscript.\label{tab:symbols_units}}\\
\toprule
\textbf{Symbol} & \textbf{Name} & \textbf{Units} \\
\midrule
}

\tablehead{%
\toprule
\textbf{Symbol} & \textbf{Name} & \textbf{Units} \\
\midrule
}

\tabletail{}

\tablelasttail{%
\bottomrule
}

\begin{supertabular}{lll}

%--------------------------------------------------------------------
\multicolumn{3}{l}{\textit{Core phase properties and field variables}}\\
\midrule
\parbox[t]{4.0cm}{$P_{\mathrm{abs}}$} & \parbox[t]{7.5cm}{Absolute (total) gas pressure} & \parbox[t]{1.5cm}{\si{Pa}} \\
\parbox[t]{4.0cm}{$P$} & \parbox[t]{7.5cm}{Excess (over atmospheric) pore pressure} & \parbox[t]{1.5cm}{\si{Pa}} \\
\parbox[t]{4.0cm}{$P_b$} & \parbox[t]{7.5cm}{Basal excess pore pressure} & \parbox[t]{1.5cm}{\si{Pa}} \\
\parbox[t]{4.0cm}{$P_{b0}$} & \parbox[t]{7.5cm}{Initial basal excess pore pressure} & \parbox[t]{1.5cm}{\si{Pa}} \\
\parbox[t]{4.0cm}{$P_{\mathrm{amb}}$} & \parbox[t]{7.5cm}{Ambient (atmospheric) pressure} & \parbox[t]{1.5cm}{\si{Pa}} \\
\parbox[t]{4.0cm}{$\rho_f$} & \parbox[t]{7.5cm}{Gas (fluid) density} & \parbox[t]{1.5cm}{\si{kg.m^{-3}}} \\
\parbox[t]{4.0cm}{$\rho_s$} & \parbox[t]{7.5cm}{Solid grain density} & \parbox[t]{1.5cm}{\si{kg.m^{-3}}} \\
\parbox[t]{4.0cm}{$\varepsilon$} & \parbox[t]{7.5cm}{Porosity (gas volume fraction)} & \parbox[t]{1.5cm}{---} \\
\parbox[t]{4.0cm}{$\varepsilon_0$} & \parbox[t]{7.5cm}{Initial porosity} & \parbox[t]{1.5cm}{---} \\
\parbox[t]{4.0cm}{$\phi = 1-\varepsilon$} & \parbox[t]{7.5cm}{Solids concentration} & \parbox[t]{1.5cm}{---} \\
\parbox[t]{4.0cm}{$\phi_0 = 1-\varepsilon_0$} & \parbox[t]{7.5cm}{Initial solids concentration} & \parbox[t]{1.5cm}{---} \\
\parbox[t]{4.0cm}{$\dot{\varepsilon}$} & \parbox[t]{7.5cm}{Rate of change of porosity ($=\partial\varepsilon/\partial t$)} & \parbox[t]{1.5cm}{\si{s^{-1}}} \\
\parbox[t]{4.0cm}{$\mathbf{u}_f$} & \parbox[t]{7.5cm}{Gas velocity} & \parbox[t]{1.5cm}{\si{m.s^{-1}}} \\
\parbox[t]{4.0cm}{$\mathbf{u}_s$} & \parbox[t]{7.5cm}{Solid-phase velocity} & \parbox[t]{1.5cm}{\si{m.s^{-1}}} \\
\parbox[t]{4.0cm}{$k$} & \parbox[t]{7.5cm}{Intrinsic permeability} & \parbox[t]{1.5cm}{\si{m^{2}}} \\
\parbox[t]{4.0cm}{$\mu_g$} & \parbox[t]{7.5cm}{Dynamic viscosity of gas} & \parbox[t]{1.5cm}{\si{Pa.s}} \\
\parbox[t]{4.0cm}{$\beta$} & \parbox[t]{7.5cm}{Gas compressibility $=(1/\rho_f)(\partial\rho_f/\partial P_{\mathrm{abs}})_T=1/P_{\mathrm{abs}}$} & \parbox[t]{1.5cm}{\si{Pa^{-1}}} \\
\parbox[t]{4.0cm}{$\beta_0$} & \parbox[t]{7.5cm}{Gas compressibility at ambient pressure ($=1/P_{\mathrm{amb}}$)} & \parbox[t]{1.5cm}{\si{Pa^{-1}}} \\
\parbox[t]{4.0cm}{$R$} & \parbox[t]{7.5cm}{Specific gas constant} & \parbox[t]{1.5cm}{\si{J.(kg.K)^{-1}}} \\
\parbox[t]{4.0cm}{$T$} & \parbox[t]{7.5cm}{Gas temperature} & \parbox[t]{1.5cm}{\si{K}} \\
\parbox[t]{4.0cm}{$g$} & \parbox[t]{7.5cm}{Gravitational acceleration} & \parbox[t]{1.5cm}{\si{m.s^{-2}}} \\
\parbox[t]{4.0cm}{$H,\,H(t)$} & \parbox[t]{7.5cm}{Bed/flow height} & \parbox[t]{1.5cm}{\si{m}} \\
\parbox[t]{4.0cm}{$H_0$} & \parbox[t]{7.5cm}{Initial bed height} & \parbox[t]{1.5cm}{\si{m}} \\
\parbox[t]{4.0cm}{$z$} & \parbox[t]{7.5cm}{Vertical coordinate (upward)} & \parbox[t]{1.5cm}{\si{m}} \\
\parbox[t]{4.0cm}{$t$} & \parbox[t]{7.5cm}{Time} & \parbox[t]{1.5cm}{\si{s}} \\

\midrule
\multicolumn{3}{l}{\textit{Particle characteristics and fluidization}}\\
\midrule
\parbox[t]{4.0cm}{$d_{\mathrm{eq}} = d_{32}\psi$} & \parbox[t]{7.5cm}{Equivalent particle diameter} & \parbox[t]{1.5cm}{\si{m}} \\
\parbox[t]{4.0cm}{$d_{32}$} & \parbox[t]{7.5cm}{Sauter mean diameter} & \parbox[t]{1.5cm}{\si{m}} \\
\parbox[t]{4.0cm}{$\psi$} & \parbox[t]{7.5cm}{Particle sphericity} & \parbox[t]{1.5cm}{---} \\
\parbox[t]{4.0cm}{$\mathrm{Ar}=g\,d_{\mathrm{eq}}^{3}\rho_f(\rho_s-\rho_f)/\mu_g^{2}$} & \parbox[t]{7.5cm}{Archimedes number} & \parbox[t]{1.5cm}{---} \\
\parbox[t]{4.0cm}{$U$} & \parbox[t]{7.5cm}{Superficial gas velocity} & \parbox[t]{1.5cm}{\si{m.s^{-1}}} \\
\parbox[t]{4.0cm}{$U_{\mathrm{mf}}$} & \parbox[t]{7.5cm}{Minimum fluidization velocity} & \parbox[t]{1.5cm}{\si{m.s^{-1}}} \\
\parbox[t]{4.0cm}{$\Delta P$} & \parbox[t]{7.5cm}{Pressure drop across bed (Ergun context)} & \parbox[t]{1.5cm}{\si{Pa}} \\
\parbox[t]{4.0cm}{$L$} & \parbox[t]{7.5cm}{Bed length (Ergun context)} & \parbox[t]{1.5cm}{\si{m}} \\

\midrule
\multicolumn{3}{l}{\textit{1D diffusion--compaction governing equation}}\\
\midrule
\parbox[t]{4.0cm}{$D=k/(\mu\beta\varepsilon)$} & \parbox[t]{7.5cm}{Pressure diffusivity (linear limit)} & \parbox[t]{1.5cm}{\si{m^{2}.s^{-1}}} \\
\parbox[t]{4.0cm}{$D_0=k(\varepsilon_0)/(\mu\beta_0\varepsilon_0)$} & \parbox[t]{7.5cm}{Initial pressure diffusivity} & \parbox[t]{1.5cm}{\si{m^{2}.s^{-1}}} \\
\parbox[t]{4.0cm}{$S(t)=\dot{\varepsilon}/[\beta\varepsilon(1-\varepsilon)]$} & \parbox[t]{7.5cm}{Compaction/dilatation source term} & \parbox[t]{1.5cm}{\si{Pa.s^{-1}}} \\
\parbox[t]{4.0cm}{$R(t)=1/[\beta\varepsilon(1-\varepsilon)]$} & \parbox[t]{7.5cm}{Compaction coefficient} & \parbox[t]{1.5cm}{\si{Pa}} \\
\parbox[t]{4.0cm}{$\alpha=\pi^{2}D/(4H^{2})$} & \parbox[t]{7.5cm}{Fundamental modal decay rate} & \parbox[t]{1.5cm}{\si{s^{-1}}} \\
\parbox[t]{4.0cm}{$\alpha_{m,0}=\pi^{2}D_0/(4H_0^{2})$} & \parbox[t]{7.5cm}{Initial fundamental decay rate} & \parbox[t]{1.5cm}{\si{s^{-1}}} \\

\midrule
\multicolumn{3}{l}{\textit{Grain Deborah number (scale analysis)}}\\
\midrule
\parbox[t]{4.0cm}{$\mathrm{De}_d = d^2/(D\,t_0)$} & \parbox[t]{7.5cm}{Grain Deborah number (ratio of grain diffusion time to deformation time)} & \parbox[t]{1.5cm}{---} \\
\parbox[t]{4.0cm}{$d$} & \parbox[t]{7.5cm}{Grain diameter (in $\mathrm{De}_d$ context)} & \parbox[t]{1.5cm}{\si{m}} \\
\parbox[t]{4.0cm}{$t_0$} & \parbox[t]{7.5cm}{Characteristic deformation timescale} & \parbox[t]{1.5cm}{\si{s}} \\

\midrule
\multicolumn{3}{l}{\textit{1D numerical solver}}\\
\midrule
\parbox[t]{4.0cm}{$\Delta z$} & \parbox[t]{7.5cm}{Uniform spatial grid step} & \parbox[t]{1.5cm}{\si{m}} \\
\parbox[t]{4.0cm}{$N_{\mathrm{active}}$} & \parbox[t]{7.5cm}{Number of active (in-bed) grid nodes at time $t$} & \parbox[t]{1.5cm}{---} \\
\parbox[t]{4.0cm}{$\beta_{\mathrm{loc}}=1/(P_{\mathrm{amb}}+P_i^n)$} & \parbox[t]{7.5cm}{Local gas compressibility at node $i$, time $n$} & \parbox[t]{1.5cm}{\si{Pa^{-1}}} \\
\parbox[t]{4.0cm}{$D_{\mathrm{loc}}=k(t_n)/(\mu\beta_{\mathrm{loc}}\varepsilon(t_n))$} & \parbox[t]{7.5cm}{Local pressure diffusivity} & \parbox[t]{1.5cm}{\si{m^{2}.s^{-1}}} \\

\midrule
\multicolumn{3}{l}{\textit{Modal (Neumann--Dirichlet) decomposition}}\\
\midrule
\parbox[t]{4.0cm}{$\psi_k(z)=\cos[(\tfrac{\pi}{2}+k\pi)z/H]$} & \parbox[t]{7.5cm}{Eigenfunction of mode $k$} & \parbox[t]{1.5cm}{---} \\
\parbox[t]{4.0cm}{$\lambda_k=(\tfrac{\pi}{2}+k\pi)^{2}$} & \parbox[t]{7.5cm}{Eigenvalue of mode $k$} & \parbox[t]{1.5cm}{---} \\
\parbox[t]{4.0cm}{$P_k(t)$} & \parbox[t]{7.5cm}{Modal amplitude (coefficient) at time $t$, mode $k$} & \parbox[t]{1.5cm}{\si{Pa}} \\
\parbox[t]{4.0cm}{$P_{0}^{\,k}$} & \parbox[t]{7.5cm}{Initial modal coefficient, mode $k$} & \parbox[t]{1.5cm}{\si{Pa}} \\

\midrule
\multicolumn{3}{l}{\textit{Timescales and dimensionless parameters}}\\
\midrule
\parbox[t]{4.0cm}{$T_{\mathrm{diff}}=4H^{2}/(\pi^{2}D)$} & \parbox[t]{7.5cm}{Diffusive (drainage) timescale} & \parbox[t]{1.5cm}{\si{s}} \\
\parbox[t]{4.0cm}{$T_{c}=\varepsilon(1-\varepsilon)/(-\dot{\varepsilon})$} & \parbox[t]{7.5cm}{Compaction timescale} & \parbox[t]{1.5cm}{\si{s}} \\
\parbox[t]{4.0cm}{$\mathrm{De}=T_{\mathrm{diff}}/T_{c}$} & \parbox[t]{7.5cm}{Deborah number (timescale ratio)} & \parbox[t]{1.5cm}{---} \\
\parbox[t]{4.0cm}{$\Psi(t)=|R(t)\dot{\varepsilon}(t)|/[\alpha_m(t)P_b(t)]$} & \parbox[t]{7.5cm}{Instantaneous source-to-diffusion ratio} & \parbox[t]{1.5cm}{---} \\
\parbox[t]{4.0cm}{$\Psi_0$} & \parbox[t]{7.5cm}{Initial source-to-diffusion ratio ($=\mathrm{De}_0/(\beta_0 P_{b0})$)} & \parbox[t]{1.5cm}{---} \\

\midrule
\multicolumn{3}{l}{\textit{Effective diffusivity analysis}}\\
\midrule
\parbox[t]{4.0cm}{$D_{\text{pred}}=k(\varepsilon_0)/(\mu\beta_0\varepsilon_0)$} & \parbox[t]{7.5cm}{Predicted Darcy--Carman--Kozeny diffusivity} & \parbox[t]{1.5cm}{\si{m^{2}.s^{-1}}} \\
\parbox[t]{4.0cm}{$D_{\text{fit}}$} & \parbox[t]{7.5cm}{Effective diffusivity fitted to basal pressure decay} & \parbox[t]{1.5cm}{\si{m^{2}.s^{-1}}} \\
\parbox[t]{4.0cm}{$\alpha_{m,\text{fit}}$} & \parbox[t]{7.5cm}{Fitted exponential decay rate} & \parbox[t]{1.5cm}{\si{s^{-1}}} \\
\parbox[t]{4.0cm}{$D_{\Psi}=D_{\text{pred}}/(1+c\,\Psi_0^{m})$} & \parbox[t]{7.5cm}{Compaction-corrected diffusivity} & \parbox[t]{1.5cm}{\si{m^{2}.s^{-1}}} \\
\parbox[t]{4.0cm}{$c,\,m_e$} & \parbox[t]{7.5cm}{Empirical fitting constants in algebraic correction (Eq.~\eqref{eq:alg_model}; $c=1.5637$, $m_e=1.0392$)} & \parbox[t]{1.5cm}{---} \\

\parbox[t]{4.0cm}{$\Delta\varepsilon=\varepsilon_{\mathrm{start}}-\varepsilon_{\mathrm{end}}$} & \parbox[t]{7.5cm}{Total porosity change during drainage} & \parbox[t]{1.5cm}{---} \\
\parbox[t]{4.0cm}{$\varepsilon_{\mathrm{start}},\,\varepsilon_{\mathrm{end}}$} & \parbox[t]{7.5cm}{Porosity at onset and end of drainage interval} & \parbox[t]{1.5cm}{---} \\
\parbox[t]{4.0cm}{$\Pi_H=\beta_0(1-\varepsilon_0)\rho_s g H_0$} & \parbox[t]{7.5cm}{Dimensionless bed-thickness parameter} & \parbox[t]{1.5cm}{---} \\
\parbox[t]{4.0cm}{$A=\Delta\varepsilon/[\varepsilon_0(1-\varepsilon_0)]$} & \parbox[t]{7.5cm}{Normalised compaction amplitude} & \parbox[t]{1.5cm}{---} \\

\midrule
\multicolumn{3}{l}{\textit{Two-Fluid MFIX-TFM: frictional stress model}}\\
\midrule
\parbox[t]{4.0cm}{$\varepsilon_g$} & \parbox[t]{7.5cm}{Gas volume fraction (porosity), MFIX notation} & \parbox[t]{1.5cm}{---} \\
\parbox[t]{4.0cm}{$\varepsilon_m$} & \parbox[t]{7.5cm}{Solids volume fraction of phase $m$} & \parbox[t]{1.5cm}{---} \\
\parbox[t]{4.0cm}{$\varepsilon_s=\sum_m\varepsilon_m$} & \parbox[t]{7.5cm}{Total solids fraction ($=\phi$)} & \parbox[t]{1.5cm}{---} \\
\parbox[t]{4.0cm}{$\varepsilon_f^{\min}$} & \parbox[t]{7.5cm}{Minimum gas fraction threshold below which frictional stress activates} & \parbox[t]{1.5cm}{---} \\
\parbox[t]{4.0cm}{$\varepsilon^{\ast}$} & \parbox[t]{7.5cm}{Gas volume fraction at maximum packing (close-packing)} & \parbox[t]{1.5cm}{---} \\
\parbox[t]{4.0cm}{$\delta$} & \parbox[t]{7.5cm}{Smoothing (regularisation) constant in $P_c$} & \parbox[t]{1.5cm}{---} \\
\parbox[t]{4.0cm}{$\mathbf{v}_m$} & \parbox[t]{7.5cm}{Solid-phase velocity in MFIX notation ($\equiv\mathbf{u}_s$)} & \parbox[t]{1.5cm}{\si{m.s^{-1}}} \\
\parbox[t]{4.0cm}{$\Theta_m$} & \parbox[t]{7.5cm}{Granular temperature of phase $m$} & \parbox[t]{1.5cm}{\si{m^{2}.s^{-2}}} \\
\parbox[t]{4.0cm}{$S_{mij}$} & \parbox[t]{7.5cm}{Deviatoric strain-rate tensor of phase $m$} & \parbox[t]{1.5cm}{\si{s^{-1}}} \\
\parbox[t]{4.0cm}{$D_{mkk}=\nabla\cdot\mathbf{v}_m$} & \parbox[t]{7.5cm}{Volumetric strain rate (trace of deformation tensor)} & \parbox[t]{1.5cm}{\si{s^{-1}}} \\
\parbox[t]{4.0cm}{$d_m$} & \parbox[t]{7.5cm}{Particle diameter of phase $m$} & \parbox[t]{1.5cm}{\si{m}} \\
\parbox[t]{4.0cm}{$\varphi$} & \parbox[t]{7.5cm}{Internal friction angle} & \parbox[t]{1.5cm}{\si{deg}} \\
\parbox[t]{4.0cm}{$\mu_m^{\mathrm{fric}}$} & \parbox[t]{7.5cm}{Frictional shear viscosity of solid phase $m$} & \parbox[t]{1.5cm}{\si{Pa.s}} \\
\parbox[t]{4.0cm}{$\lambda_m^{\mathrm{fric}}$} & \parbox[t]{7.5cm}{Frictional second (bulk) viscosity of solid phase $m$} & \parbox[t]{1.5cm}{\si{Pa.s}} \\
\parbox[t]{4.0cm}{$P_m^{\mathrm{fric}}$} & \parbox[t]{7.5cm}{Frictional pressure of solid phase $m$} & \parbox[t]{1.5cm}{\si{Pa}} \\
\parbox[t]{4.0cm}{$P_f$} & \parbox[t]{7.5cm}{Frictional pressure (shared across solid phases)} & \parbox[t]{1.5cm}{\si{Pa}} \\
\parbox[t]{4.0cm}{$P_c=P_c(\varepsilon_g)$} & \parbox[t]{7.5cm}{Critical-state pressure} & \parbox[t]{1.5cm}{\si{Pa}} \\
\parbox[t]{4.0cm}{$\mathcal{L}(\varepsilon_g)$} & \parbox[t]{7.5cm}{Smoothing limiter function in $P_c$ ensuring regularity as $\varepsilon_g\to\varepsilon^{\ast}$} & \parbox[t]{1.5cm}{\si{Pa}} \\
\parbox[t]{4.0cm}{$\nu_{SS}$} & \parbox[t]{7.5cm}{Yield-surface exponent in Srivastava model} & \parbox[t]{1.5cm}{---} \\
\parbox[t]{4.0cm}{$\mathrm{Fr}$} & \parbox[t]{7.5cm}{Pressure-scale constant in $P_c$ formula} & \parbox[t]{1.5cm}{\si{Pa}} \\
\parbox[t]{4.0cm}{$r,\,s$} & \parbox[t]{7.5cm}{Exponents in critical-state pressure formula} & \parbox[t]{1.5cm}{---} \\

\midrule
\multicolumn{3}{l}{\textit{Two-Fluid MFIX-TFM: drag model}}\\
\midrule
\parbox[t]{4.0cm}{$\beta'_{gm}$} & \parbox[t]{7.5cm}{Interphase momentum exchange coefficient (drag)} & \parbox[t]{1.5cm}{\si{kg.m^{-3}.s^{-1}}} \\
\parbox[t]{4.0cm}{$\phi_s=1-\varepsilon_g$} & \parbox[t]{7.5cm}{Local solids concentration (Gidaspow notation)} & \parbox[t]{1.5cm}{---} \\
\parbox[t]{4.0cm}{$\mathbf{U}_g,\,\mathbf{U}_m$} & \parbox[t]{7.5cm}{Gas and solid-phase velocities (Gidaspow notation)} & \parbox[t]{1.5cm}{\si{m.s^{-1}}} \\
\parbox[t]{4.0cm}{$|U_{gi}-U_{mi}|$} & \parbox[t]{7.5cm}{Slip speed} & \parbox[t]{1.5cm}{\si{m.s^{-1}}} \\
\parbox[t]{4.0cm}{$\mathrm{Re}_p=\rho_g\varepsilon_g|\mathbf{U}_g-\mathbf{U}_m|d_m/\mu_g$} & \parbox[t]{7.5cm}{Particle Reynolds number} & \parbox[t]{1.5cm}{---} \\
\parbox[t]{4.0cm}{$C_D$} & \parbox[t]{7.5cm}{Drag coefficient (Wen--Yu correlation)} & \parbox[t]{1.5cm}{---} \\
\parbox[t]{4.0cm}{$\rho_g\equiv\rho_f$} & \parbox[t]{7.5cm}{Gas density in drag-model notation} & \parbox[t]{1.5cm}{\si{kg.m^{-3}}} \\
\parbox[t]{4.0cm}{$\mu_g$} & \parbox[t]{7.5cm}{Gas dynamic viscosity in drag-model notation} & \parbox[t]{1.5cm}{\si{Pa.s}} \\

\midrule
\multicolumn{3}{l}{\textit{IMEX depth-averaged pore-pressure model}}\\
\midrule
\parbox[t]{4.0cm}{$P_{exc}$} & \parbox[t]{7.5cm}{Excess pore pressure $P-P_{\mathrm{atm}}$} & \parbox[t]{1.5cm}{\si{Pa}} \\
\parbox[t]{4.0cm}{$P_{b,exc}$} & \parbox[t]{7.5cm}{Basal excess pore pressure $P_b-P_{\mathrm{atm}}$} & \parbox[t]{1.5cm}{\si{Pa}} \\
\parbox[t]{4.0cm}{$h$} & \parbox[t]{7.5cm}{Local flow thickness (depth-averaged layer height)} & \parbox[t]{1.5cm}{\si{m}} \\
\parbox[t]{4.0cm}{$\phi_p$} & \parbox[t]{7.5cm}{Porosity used in hydraulic diffusivity} & \parbox[t]{1.5cm}{---} \\
\end{supertabular}

\newpage

\tablefirsthead{%
\toprule
\textbf{Symbol} & \textbf{Name} & \textbf{Units} \\
\midrule
}

\begin{supertabular}{lll}
\parbox[t]{4.0cm}{$\mu_g\equiv\mu_g$} & \parbox[t]{7.5cm}{Dynamic viscosity of pore fluid (IMEX notation)} & \parbox[t]{1.5cm}{\si{Pa.s}} \\
\parbox[t]{4.0cm}{$\beta_f\equiv\beta$} & \parbox[t]{7.5cm}{Compressibility of pore fluid (IMEX notation)} & \parbox[t]{1.5cm}{\si{Pa^{-1}}} \\
\parbox[t]{4.0cm}{$\mathbf{u}$} & \parbox[t]{7.5cm}{Depth-averaged mixture velocity} & \parbox[t]{1.5cm}{\si{m.s^{-1}}} \\
\parbox[t]{4.0cm}{$v_{\text{loss,gas}}$} & \parbox[t]{7.5cm}{Volumetric gas-loss flux at free surface (Darcy flux)} & \parbox[t]{1.5cm}{\si{m.s^{-1}}} \\
\parbox[t]{4.0cm}{$f_{\text{inhibit}}$} & \parbox[t]{7.5cm}{Inhibition factor for pore-pressure dissipation and gas loss} & \parbox[t]{1.5cm}{---} \\
\parbox[t]{4.0cm}{$S_n(x)$} & \parbox[t]{7.5cm}{Generalised smooth-step function of order $n$} & \parbox[t]{1.5cm}{---} \\
\parbox[t]{4.0cm}{$x=(\alpha_s-\alpha_{tr})/(\alpha_{\max}-\alpha_{tr})$} & \parbox[t]{7.5cm}{Normalised solid fraction in $S_n$} & \parbox[t]{1.5cm}{---} \\
\parbox[t]{4.0cm}{$\alpha_s=\sum\alpha_{s,i}$} & \parbox[t]{7.5cm}{Total solid volume fraction in IMEX layer} & \parbox[t]{1.5cm}{---} \\
\parbox[t]{4.0cm}{$\alpha_{tr}$} & \parbox[t]{7.5cm}{Solid fraction at onset of degassing inhibition} & \parbox[t]{1.5cm}{---} \\
\parbox[t]{4.0cm}{$\alpha_{\max}$} & \parbox[t]{7.5cm}{Maximum packing solid fraction ($\approx0.64$ for spheres)} & \parbox[t]{1.5cm}{---} \\
\parbox[t]{4.0cm}{$n$} & \parbox[t]{7.5cm}{Smooth-step order parameter (polynomial half-order)} & \parbox[t]{1.5cm}{---} \\
\parbox[t]{4.0cm}{$N=2n+1$} & \parbox[t]{7.5cm}{Polynomial degree in $S_n$} & \parbox[t]{1.5cm}{---} \\
\parbox[t]{4.0cm}{$\rho_m$} & \parbox[t]{7.5cm}{Bulk mixture density} & \parbox[t]{1.5cm}{\si{kg.m^{-3}}} \\
\parbox[t]{4.0cm}{$\rho_a$} & \parbox[t]{7.5cm}{Ambient fluid density} & \parbox[t]{1.5cm}{\si{kg.m^{-3}}} \\
\parbox[t]{4.0cm}{$\rho_c$} & \parbox[t]{7.5cm}{Carrier/ambient gas density used in pore-fluid loss term} & \parbox[t]{1.5cm}{\si{kg.m^{-3}}} \\
\end{supertabular}

\newpage

\tablefirsthead{%
\toprule
\textbf{Symbol} & \textbf{Name} & \textbf{Units} \\
\midrule
}

\begin{supertabular}{lll}
\parbox[t]{4.0cm}{$Q_p=\rho_m h P_{b,exc}$} & \parbox[t]{7.5cm}{Conserved pore-pressure variable} & \parbox[t]{1.5cm}{\si{Pa.kg.m^{-2}}} \\
\parbox[t]{4.0cm}{$q_1=\rho_m h$} & \parbox[t]{7.5cm}{Mixture mass per unit area} & \parbox[t]{1.5cm}{\si{kg.m^{-2}}} \\
\parbox[t]{4.0cm}{$S_{Q_p}$} & \parbox[t]{7.5cm}{Source term for $Q_p$ in conservative form} & \parbox[t]{1.5cm}{\si{Pa.kg.m^{-2}.s^{-1}}} \\
\parbox[t]{4.0cm}{$S_m$} & \parbox[t]{7.5cm}{Total mass source term (all processes)} & \parbox[t]{1.5cm}{\si{kg.m^{-2}.s^{-1}}} \\
\parbox[t]{4.0cm}{$S_{m,\text{other}}$} & \parbox[t]{7.5cm}{Mass source/sink from erosion, sedimentation, entrainment} & \parbox[t]{1.5cm}{\si{kg.m^{-2}.s^{-1}}} \\
\parbox[t]{4.0cm}{$S_{m,\text{pore}}$} & \parbox[t]{7.5cm}{Mass source/sink from pore-fluid escape} & \parbox[t]{1.5cm}{\si{kg.m^{-2}.s^{-1}}} \\
\parbox[t]{4.0cm}{$S_{P_{b,exc}}$} & \parbox[t]{7.5cm}{Source term in non-conservative $P_{b,exc}$ equation} & \parbox[t]{1.5cm}{\si{Pa.s^{-1}}} \\
\parbox[t]{4.0cm}{$\dot{h}$} & \parbox[t]{7.5cm}{Volumetric influx per unit area at source (entry velocity)} & \parbox[t]{1.5cm}{\si{m.s^{-1}}} \\
\parbox[t]{4.0cm}{$\rho_{\text{in}}$} & \parbox[t]{7.5cm}{Density of injected/source material} & \parbox[t]{1.5cm}{\si{kg.m^{-3}}} \\
\parbox[t]{4.0cm}{$C_p$} & \parbox[t]{7.5cm}{Dimensionless fraction of lithostatic stress borne by excess pore pressure} & \parbox[t]{1.5cm}{---} \\
\parbox[t]{4.0cm}{$g'=g(\rho_m-\rho_a)/\rho_m$} & \parbox[t]{7.5cm}{Reduced gravity} & \parbox[t]{1.5cm}{\si{m.s^{-2}}} \\
\parbox[t]{4.0cm}{$P_{b,exc,\text{source}}$} & \parbox[t]{7.5cm}{Basal excess pore pressure of incoming source material} & \parbox[t]{1.5cm}{\si{Pa}} \\
\parbox[t]{4.0cm}{$S_{Q_p,\text{source}}$} & \parbox[t]{7.5cm}{Source of $Q_p$ due to injected material} & \parbox[t]{1.5cm}{\si{Pa.kg.m^{-2}.s^{-1}}} \\
\parbox[t]{4.0cm}{$\sigma_{\text{vert}}$} & \parbox[t]{7.5cm}{Total vertical stress at bed} & \parbox[t]{1.5cm}{\si{Pa}} \\
\parbox[t]{4.0cm}{$\sigma'_{\text{vert}}$} & \parbox[t]{7.5cm}{Effective vertical stress at bed} & \parbox[t]{1.5cm}{\si{Pa}} \\
\parbox[t]{4.0cm}{$\sigma'_{n}$} & \parbox[t]{7.5cm}{Effective normal stress on bed} & \parbox[t]{1.5cm}{\si{Pa}} \\
\parbox[t]{4.0cm}{$\tau_c$} & \parbox[t]{7.5cm}{Coulomb basal shear stress magnitude} & \parbox[t]{1.5cm}{\si{Pa}} \\
\parbox[t]{4.0cm}{$S_f$} & \parbox[t]{7.5cm}{Basal-friction source-term magnitude (Eq.~\eqref{eq:source_term})} & \parbox[t]{1.5cm}{\si{Pa}} \\
\parbox[t]{4.0cm}{$\mu$} & \parbox[t]{7.5cm}{Basal Coulomb friction coefficient} & \parbox[t]{1.5cm}{---} \\
\parbox[t]{4.0cm}{$\alpha$} & \parbox[t]{7.5cm}{Local bed-slope angle} & \parbox[t]{1.5cm}{\si{rad}} \\
\parbox[t]{4.0cm}{$\phi_{\mathrm{topo}}=1/\cos\alpha$} & \parbox[t]{7.5cm}{Topographic projection factor} & \parbox[t]{1.5cm}{---} \\
\parbox[t]{4.0cm}{$I$} & \parbox[t]{7.5cm}{Inertial number} & \parbox[t]{1.5cm}{---} \\
\parbox[t]{4.0cm}{$\dot{\gamma}$} & \parbox[t]{7.5cm}{Shear rate} & \parbox[t]{1.5cm}{\si{s^{-1}}} \\
\parbox[t]{4.0cm}{$d$} & \parbox[t]{7.5cm}{Characteristic grain size in inertial number $I$} & \parbox[t]{1.5cm}{\si{m}} \\
\parbox[t]{4.0cm}{$P_s$} & \parbox[t]{7.5cm}{Granular isotropic pressure in $I=\dot{\gamma}d/\sqrt{P_s/\rho_s}$} & \parbox[t]{1.5cm}{\si{Pa}} \\
\parbox[t]{4.0cm}{$\mu_1$} & \parbox[t]{7.5cm}{Quasi-static friction limit in $\mu(I)$ ($I\to0$)} & \parbox[t]{1.5cm}{---} \\
\parbox[t]{4.0cm}{$\mu_2$} & \parbox[t]{7.5cm}{High-$I$ asymptote in classical $\mu(I)$ ($I\to\infty$)} & \parbox[t]{1.5cm}{---} \\
\parbox[t]{4.0cm}{$I_o$} & \parbox[t]{7.5cm}{Characteristic inertial number in $\mu(I)$} & \parbox[t]{1.5cm}{---} \\
\parbox[t]{4.0cm}{$\mu_{\mathrm{reg}}$} & \parbox[t]{7.5cm}{Partially regularised friction coefficient} & \parbox[t]{1.5cm}{---} \\
\parbox[t]{4.0cm}{$\mu_{\infty}$} & \parbox[t]{7.5cm}{Collisional dissipation parameter (regularisation)} & \parbox[t]{1.5cm}{---} \\
\parbox[t]{4.0cm}{$I_1^{N}$} & \parbox[t]{7.5cm}{Stability-neutral inertial number (regularisation threshold)} & \parbox[t]{1.5cm}{---} \\
\parbox[t]{4.0cm}{$\varepsilon_{\min}$} & \parbox[t]{7.5cm}{Minimum porosity (close-packing limit) used in IMEX inhibition} & \parbox[t]{1.5cm}{---} \\
\parbox[t]{4.0cm}{$\Delta\varepsilon_{\max}=\max(0,\varepsilon_0-\varepsilon_{\min})$} & \parbox[t]{7.5cm}{Maximum available porosity drop} & \parbox[t]{1.5cm}{---} \\
\parbox[t]{4.0cm}{$A_0=\Delta\varepsilon_{\max}/[\varepsilon_0(1-\varepsilon_0)]$} & \parbox[t]{7.5cm}{Normalised compaction amplitude (a~priori IMEX estimate)} & \parbox[t]{1.5cm}{---} \\
\parbox[t]{4.0cm}{$\Psi_0^{\text{ap}}$} & \parbox[t]{7.5cm}{A-priori source-to-diffusion ratio used in IMEX} & \parbox[t]{1.5cm}{---} \\
\parbox[t]{4.0cm}{$D_{\Psi}^{\text{ap}}$} & \parbox[t]{7.5cm}{A-priori compaction-corrected diffusivity used in IMEX} & \parbox[t]{1.5cm}{\si{m^{2}.s^{-1}}} \\
\end{supertabular}
}

\bibliographystyle{apsrev4-2}

\bibliography{bibliography}

@article{Roche_2012_depositional,
author = {Roche, Olivier},
year = {2012},
month = {10},
pages = {},
title = {Depositional processes and gas pore pressure in pyroclastic flows: An experimental perspective},
volume = {74},
journal = {Bulletin of Volcanology},
doi = {10.1007/s00445-012-0639-4}
}

@article{Aravena_2021_influence,
author = {Aravena, {\'A}lvaro and Chupin, Laurent and Dubois, Thierry and Roche, Olivier},
title = {The influence of gas pore pressure in dense granular flows: numerical simulations versus experiments and implications for pyroclastic density currents},
volume = {83},
journal = {Bulletin of Volcanology},
doi = {10.1007/s00445-021-01507-7},
year = {2021}
}

@article{de2023imex_sflow2d,
  title={IMEX\_SfloW2D v2: a depth-averaged numerical flow model for volcanic gas-particle flows over complex topographies and water},
  author={{de' Michieli Vitturi}, Mattia and Esposti Ongaro, Tomaso and Engwell, Samantha},
  journal={Geoscientific Model Development Discussions},
  volume={2023},
  pages={1--42},
  year={2023},
  publisher={G{\"o}ttingen, Germany}
}

@article{gueugneau2017effects,
  title={Effects of pore pressure in pyroclastic flows: numerical simulation and experimental validation},
  author={Gueugneau, Valentin and Kelfoun, Karim and Roche, Olivier and Chupin, Laurent},
  journal={Geophysical Research Letters},
  volume={44},
  number={5},
  pages={2194--2202},
  year={2017},
  publisher={Wiley Online Library}
}

@article{iverson1997physics,
  title={The physics of debris flows},
  author={Iverson, Richard M},
  journal={Reviews of geophysics},
  volume={35},
  number={3},
  pages={245--296},
  year={1997},
  publisher={Wiley Online Library}
}

@article{denlinger2001flow,
  title={Flow of variably fluidized granular masses across three-dimensional terrain: 2. Numerical predictions and experimental tests},
  author={Denlinger, Roger P and Iverson, Richard M},
  journal={Journal of Geophysical Research: Solid Earth},
  volume={106},
  number={B1},
  pages={553--566},
  year={2001},
  publisher={Wiley Online Library}
}

@article{iverson2010perfect,
  title={The perfect debris flow? Aggregated results from 28 large-scale experiments},
  author={Iverson, Richard M and Logan, Matthew and LaHusen, Richard G and Berti, Matteo},
  journal={Journal of Geophysical Research: Earth Surface},
  volume={115},
  number={F3},
  year={2010},
  publisher={Wiley Online Library}
}

@article{savage1989motion,
  title={The motion of a finite mass of granular material down a rough incline},
  author={Savage, Stuart B and Hutter, Kolumban},
  journal={Journal of fluid mechanics},
  volume={199},
  pages={177--215},
  year={1989},
  publisher={Cambridge University Press}
}

@article{xia2018new,
  title={A new depth-averaged model for flow-like landslides over complex terrains with curvatures and steep slopes},
  author={Xia, Xilin and Liang, Qiuhua},
  journal={Engineering Geology},
  volume={234},
  pages={174--191},
  year={2018},
  publisher={Elsevier}
}

@book{greenshieldsweller2022,
  title     = "Notes on Computational Fluid Dynamics: General Principles",
  author    = "Greenshields, Christopher and Weller, Henry",
  year      = 2022,
  publisher = "CFD Direct Ltd",
  address   = "Reading, UK"
}

@article{zhu2020high,
  title={High-speed confined granular flows down smooth inclines: scaling and wall friction laws},
  author={Zhu, Yajuan and Delannay, Renaud and Valance, Alexandre},
  journal={Granular Matter},
  volume={22},
  number={4},
  pages={82},
  year={2020},
  publisher={Springer}
}

@article{Montserrat2012_pore,
  author  = {Montserrat, S. and Tamburrino, A. and Roche, O. and Ni{\~n}o, Y.},
  title   = {Pore fluid pressure diffusion in defluidizing granular columns},
  journal = {Journal of Geophysical Research: Earth Surface},
  year    = {2012},
  volume  = {117},
  number  = {F2},
  pages   = {F02034},
  doi     = {10.1029/2011JF002164}
}

@article{Aravena_2024_runout,
  author  = {Aravena, {\'A}lvaro and Chupin, Laurent and Dubois, Thierry and Roche, Olivier},
  title   = {Run-out distance of initially fluidized, collapsing granular columns with different aspect ratios: constraints and volcanological implications from experiments and 2{D} incompressible simulations},
  journal = {Bulletin of Volcanology},
  year    = {2024},
  volume  = {86},
  number  = {90},
  pages   = {90},
  doi     = {10.1007/s00445-024-01778-w}
}

@article{Breard2019b,
  author  = {Breard, Eric C. P. and Jones, Jim R. and Fullard, Luke and Lube, Gert and Davies, Clive and Dufek, Josef},
  title   = {The Permeability of Volcanic Mixtures---Implications for Pyroclastic Currents},
  journal = {Journal of Geophysical Research: Solid Earth},
  year    = {2019},
  volume  = {124},
  number  = {2},
  pages   = {1343--1360},
  doi     = {10.1029/2018JB016544}
}

@Article{IMEX2019,
AUTHOR = {de' Michieli Vitturi, M. and Esposti Ongaro, T. and Lari, G. and Aravena, A.},
TITLE = {IMEX\_SfloW2D 1.0: a depth-averaged numerical flow model for pyroclastic avalanches},
JOURNAL = {Geoscientific Model Development},
VOLUME = {12},
YEAR = {2019},
NUMBER = {1},
PAGES = {581--595},
URL = {https://gmd.copernicus.org/articles/12/581/2019/},
DOI = {10.5194/gmd-12-581-2019}
}

@article{NEGLIA2021107146,
title = {Shallow-water models for volcanic granular flows: A review of strengths and weaknesses of TITAN2D and FLO2D numerical codes},
journal = {Journal of Volcanology and Geothermal Research},
volume = {410},
pages = {107146},
year = {2021},
issn = {0377-0273},
doi = {https://doi.org/10.1016/j.jvolgeores.2020.107146},
url = {https://www.sciencedirect.com/science/article/pii/S0377027320305825},
author = {Francesco Neglia and Roberto Sulpizio and Fabio Dioguardi and Lucia Capra and Damiano Sarocchi}
}

@article{Karim2011,
author = {Kelfoun, Karim},
title = {Suitability of simple rheological laws for the numerical simulation of dense pyroclastic flows and long-runout volcanic avalanches},
journal = {Journal of Geophysical Research: Solid Earth},
volume = {116},
number = {B8},
pages = {},
doi = {https://doi.org/10.1029/2010JB007622},
url = {https://agupubs.onlinelibrary.wiley.com/doi/abs/10.1029/2010JB007622},
eprint = {https://agupubs.onlinelibrary.wiley.com/doi/pdf/10.1029/2010JB007622},
year = {2011}
}

@article{zuccarello2025trigger,
  title={Trigger mechanism and propagation dynamics of pyroclastic density currents at basaltic volcanoes},
  author={Zuccarello, Francesco and Andronico, Daniele and Del Carlo, Paola and de’Michieli Vitturi, Mattia and Di Roberto, Alessio and Ganci, Gaetana and Behncke, Boris and Esposti Ongaro, Tomaso and Ciancitto, Francesco and Cappello, Annalisa},
  journal={Communications Earth \& Environment},
  volume={6},
  number={1},
  pages={495},
  year={2025},
  publisher={Nature Publishing Group UK London}
}

@ARTICLE{Ogburn2017,
  
AUTHOR={Ogburn, Sarah E.  and Calder, Eliza S. },
         
TITLE={The Relative Effectiveness of Empirical and Physical Models for Simulating the Dense Undercurrent of Pyroclastic Flows under Different Emplacement Conditions},
        
JOURNAL={Frontiers in Earth Science},
        
VOLUME={Volume 5 - 2017},

YEAR={2017},

URL={https://www.frontiersin.org/journals/earth-science/articles/10.3389/feart.2017.00083},

DOI={10.3389/feart.2017.00083},

ISSN={2296-6463}}

@article{Iverson1997,
  author = {Iverson, Richard M.},
  title = {The physics of debris flows},
  journal = {Reviews of Geophysics},
  volume = {35},
  number = {3},
  pages = {245--296},
  year = {1997},
  doi = {10.1029/97RG00426}
}

@article{Iverson2001,
  author = {Iverson, R. M. and Denlinger, R. P.},
  title = {Flow of variably fluidized granular masses across three-dimensional terrain: 1. Coulomb mixture theory},
  journal = {Journal of Geophysical Research: Solid Earth},
  volume = {106},
  number = {B1},
  pages = {537--552},
  year = {2001},
  doi = {10.1029/2000JB900329}
}

@article{Iverson2014,
  author = {Iverson, R. M. and George, D. L.},
  title = {A depth-averaged debris-flow model that includes the effects of evolving dilatancy. I. Physical basis},
  journal = {Proceedings of the Royal Society A},
  volume = {470},
  number = {2170},
  pages = {20130819},
  year = {2014},
  doi = {10.1098/rspa.2013.0819}
}

@article{Montserrat2012,
  author = {Montserrat, S. and Tamburrino, A. and Roche, O. and Ni{\~{n}}o, Y.},
  title = {Pore fluid pressure diffusion in defluidizing granular columns},
  journal = {Journal of Geophysical Research: Earth Surface},
  volume = {117},
  number = {F2},
  pages = {F02034},
  year = {2012},
  doi = {10.1029/2011JF002164}
}

@article{Roche_2010_pore,
author = {Roche, O. and Montserrat, S. and Ni{\~n}o, Y. and Tamburrino, A.},
title = {Pore fluid pressure and internal kinematics of gravitational laboratory air-particle flows: Insights into the emplacement dynamics of pyroclastic flows},
journal = {Journal of Geophysical Research: Solid Earth},
volume = {115},
number = {B9},
doi = {10.1029/2009JB007133},
year = {2010}
}

@article{Roche2012,
  author = {Roche, Olivier},
  title = {Depositional processes and gas pore pressure in pyroclastic flows: an experimental perspective},
  journal = {Bulletin of Volcanology},
  volume = {74},
  number = {8},
  pages = {1807--1820},
  year = {2012},
  doi = {10.1007/s00445-012-0639-4}
}

@article{Breard2016,
  author = {Breard, E. C. P. and Lube, G. and Jones, J. R. and Dufek, J. and Cronin, S. J. and Valentine, G. A. and Moebis, A.},
  title = {Coupling of turbulent and non-turbulent flow regimes within pyroclastic density currents},
  journal = {Nature Geoscience},
  volume = {9},
  number = {10},
  pages = {767--771},
  year = {2016},
  doi = {10.1038/ngeo2794}
}

@article{Breard2019,
  author = {Breard, E. C. P. and Dufek, J. and Roche, O.},
  title = {Continuum modeling of pressure-balanced and fluidized granular flows in 2-D: comparison with glass bead experiments and implications for concentrated pyroclastic density currents},
  journal = {Journal of Geophysical Research: Solid Earth},
  volume = {124},
  number = {6},
  pages = {5557--5583},
  year = {2019},
  doi = {10.1029/2018JB016874}
}

@article{Breard2023,
  author = {Breard, E. C. P. and Dufek, J. and Charbonnier, S. and Gueugneau, V. and Giachetti, T. and Walsh, B.},
  title = {The fragmentation-induced fluidisation of pyroclastic density currents},
  journal = {Nature Communications},
  volume = {14},
  number = {1},
  pages = {2079},
  year = {2023},
  doi = {10.1038/s41467-023-17663-1}
}

@article{Lube2020,
  author = {Lube, G. and Breard, E. C. P. and Esposti Ongaro, T. and Dufek, J. and Brand, B. D.},
  title = {Multiphase flow behaviour and hazard prediction of pyroclastic density currents},
  journal = {Nature Reviews Earth \& Environment},
  volume = {1},
  number = {7},
  pages = {348--365},
  year = {2020},
  doi = {10.1038/s43017-020-0064-8}
}

@article{Bouchut2016,
  author = {Bouchut, F. and Fern{\'a}ndez-Nieto, E. D. and Mangeney, A. and Narbona-Reina, G.},
  title = {A two-phase two-layer model for fluidized granular flows with dilatancy effects},
  journal = {Journal of Fluid Mechanics},
  volume = {801},
  pages = {166--221},
  year = {2016},
  doi = {10.1017/jfm.2016.417}
}

@article{GarresDiaz2020,
  author = {Garres-D{\'i}az, J. and Bouchut, F. and Fern{\'a}ndez-Nieto, E. D. and Mangeney, A. and Narbona-Reina, G.},
  title = {Multilayer models for shallow two-phase debris flows with dilatancy effects},
  journal = {Journal of Computational Physics},
  volume = {419},
  pages = {109699},
  year = {2020},
  doi = {10.1016/j.jcp.2020.109699}
}

@article{Goren2010,
author = {Goren, L. and Aharonov, E. and Sparks, D. and Toussaint, R.},
title = {Pore pressure evolution in deforming granular material: A general formulation and the infinitely stiff approximation},
journal = {Journal of Geophysical Research: Solid Earth},
volume = {115},
number = {B9},
pages = {},
doi = {https://doi.org/10.1029/2009JB007191},
url = {https://agupubs.onlinelibrary.wiley.com/doi/abs/10.1029/2009JB007191},
eprint = {https://agupubs.onlinelibrary.wiley.com/doi/pdf/10.1029/2009JB007191},
year = {2010}
}

@article{Iverson2005,
author = {Iverson, Richard M.},
title = {Regulation of landslide motion by dilatancy and pore pressure feedback},
journal = {Journal of Geophysical Research: Earth Surface},
volume = {110},
number = {F2},
pages = {},
doi = {https://doi.org/10.1029/2004JF000268},
url = {https://agupubs.onlinelibrary.wiley.com/doi/abs/10.1029/2004JF000268},
eprint = {https://agupubs.onlinelibrary.wiley.com/doi/pdf/10.1029/2004JF000268},
year = {2005}
}

@article{Lube2019,
  title={Generation of air lubrication within pyroclastic density currents},
  author={Lube, Gert and Breard, Eric CP and Jones, Jim and Fullard, Luke and Dufek, Josef and Cronin, Shane J and Wang, Ting},
  journal={Nature Geoscience},
  volume={12},
  number={5},
  pages={381--386},
  year={2019},
  publisher={Nature Publishing Group UK London}
}

@article{iverson2011positive,
  title={Positive feedback and momentum growth during debris-flow entrainment of wet bed sediment},
  author={Iverson, Richard M and Reid, Mark E and Logan, Matthew and LaHusen, Richard G and Godt, Jonathan W and Griswold, Julia P},
  journal={Nature Geoscience},
  volume={4},
  number={2},
  pages={116--121},
  year={2011},
  publisher={Nature Publishing Group UK London}
}

@article{Barker_2017_partial,
title={Partial regularisation of the incompressible {\(\mu\)}(I)-rheology for granular flow},
volume={828},
DOI={10.1017/jfm.2017.428},
journal={Journal of Fluid Mechanics},
author={Barker, T. and Gray, J. M. N. T.},
year={2017},
pages={5--32}
}

@article{Jop_2006_constitutive,
  author  = {Jop, Pierre and Forterre, Yo{\"e}l and Pouliquen, Olivier},
  title   = {A constitutive law for dense granular flows},
  journal = {Nature},
  year    = {2006},
  volume  = {441},
  number  = {7094},
  pages   = {727--730},
  doi     = {10.1038/nature04801},
  pmid    = {16760972}
}

@article{Barker_2015_illposed,
title={Well-posed and ill-posed behaviour of the ${\it\mu}(I)$-rheology for granular flow},
volume={779},
DOI={10.1017/jfm.2015.412},
journal={Journal of Fluid Mechanics},
author={Barker, T. and Schaeffer, D.G. and Bohorquez, P. and Gray, J.M.N.T.},
year={2015},
pages={794--818}
}

@article{Lagree_2011_granulacollapse,
  author  = {Lagr{\'e}e, P.-Y. and Staron, L. and Popinet, S.},
  title   = {The granular column collapse as a continuum: validity of a two-dimensional Navier--Stokes model with a {\ensuremath{\mu}}(I)-rheology},
  journal = {Journal of Fluid Mechanics},
  volume  = {686},
  pages   = {378--408},
  year    = {2011}
}

\end{document}